\documentclass{aa}  

\usepackage{graphicx}
\usepackage{txfonts}
\usepackage{lscape}
\usepackage{array,multirow}
\usepackage{booktabs}
\usepackage{cleveref}
\usepackage{enumerate}
\begin{document}

   \title{30-micron sources in galaxies with different metallicities}

   %\subtitle{I. subtitle}

   \author{M. G{\l}adkowski \inst{1}
          \and
          R. Szczerba \inst{1}
          \and
          G. C. Sloan \inst{2}
          \and
          E. Lagadec \inst{3}
          \and
          K. Volk \inst{4}
          }

   \institute{Nicolaus Copernicus Astronomical Center, ul. Rabia{\'n}ska 8, 87-100 Toru{\'n}, Poland\\
              \email{seyfert@ncac.torun.pl}
         \and
         Astronomy Department, Cornell University, Ithaca, NY 14853-6801, USA
         \and
         Laboratoire Lagrange, UMR7293, Université C\^ote d'Azur, CNRS, Observatoire de la C\^ote d'Azur, Boulevard de l'Observatoire, F-06304 Nice Cedex 4, France
         \and
         Space Telescope Science Institute, 3700 San Martin Drive, Baltimore, Maryland, USA
             }

   \date{Received --; accepted --}

% \abstract{}{}{}{}{} 
% 5 {} token are mandatory
 
  \abstract
   {}
   {We present an analysis and comparison of the 
   30\,$\mu$m dust features seen in the Spitzer 
   Space Telescope spectra of 207 carbon-rich 
   asymptotic giant branch (AGB) stars, post-AGB 
   objects, and planetary nebulae (PNe) located in the 
   Milky Way, the Magellanic Clouds (MCs), or the 
   Sagittarius dwarf spheroidal galaxy (Sgr dSph), 
   which are characterised by different average 
   metallicities. We investigated whether the formation 
   of the 30\,$\mu$m feature carrier may be a function 
   of the metallicity. Through this study we expect to 
   better understand the late stages of stellar evolution 
   of carbon-rich stars in these galaxies.
   }
   {Our analysis uses the `Manchester method' as a 
   basis for estimating the temperature of dust for the 
   carbon-rich AGB stars and the PNe in our sample. 
   For post-AGB objects we changed the wavelength 
   ranges used for temperature estimation, because 
   of the presence of the 21\,$\mu$m feature on the 
   short wavelength edge of the 30\,$\mu$m feature. 
   We used a black-body function with a single 
   temperature deduced from the Manchester method 
   or its modification to approximate the continuum 
   under the 30\,$\mu$m feature.
   }
   {We find that the strength of the 30\,$\mu$m feature 
   increases until dust temperature drops below 400\,K. 
   Below this temperature, the large loss of mass 
   and probably the self-absorption effect reduces the 
   strength of the feature. During the post-AGB phase, 
   when the intense mass-loss has terminated, the 
   optical depth of the circumstellar envelope is smaller, 
   and the 30\,$\mu$m feature becomes visible again, 
   showing variety of values for post-AGB objects and 
   PNe, and being comparable with the strengths of 
   AGB stars. In addition, the AGB stars and post-AGB 
   objects show similar values of central wavelengths -- 
   usually between 28.5 and 29.5\,$\mu$m. However, 
   in case of PNe the shift of the central wavelength 
   towards longer wavelengths is visible.
   The normalised median profiles for AGB stars look 
   uniformly for various ranges of dust temperature, 
   and different galaxies. We analysed the profiles 
   of post-AGB objects and PNe only within one 
   dust temperature range (below 200\,K), and they 
   were also similar in different galaxies.
   In the spectra of 17 PNe and five post-AGB objects 
   we found the broad 16-24\,$\mu$m feature. Two 
   objects among the PNe group are the new 
   detections: SMP~LMC~51, and SMP~LMC~79, 
   whereas in the case of post-AGBs the new 
   detections are: IRAS~05370-7019, 
   IRAS~05537-7015, and IRAS~21546+4721.
   In addition, in the spectra of nine PNe we found the 
   new detections of 16-18\,$\mu$m feature. We also 
   find that the Galactic post-AGB object 
   IRAS~11339-6004 has a 21\,$\mu$m emission. 
   Finally, we have produced online catalogues of 
   photometric data and Spitzer IRS spectra for all 
   objects that show the 30\,$\mu$m feature. These 
   resources are available online for use by the 
   community.
   }
  {The most important conclusion of our work is the fact 
  that the formation of the 30\,$\mu$m feature is 
  affected by metallicity. Specifically that, as opposed to 
  more metal-poor samples of AGB stars in the MCs, 
  the feature is seen at lower mass-loss rates, higher 
  temperatures, and has seen to be more prominent in 
  Galactic carbon stars. The averaged feature (profile) 
  in the AGB, post-AGB objects, and PNe seems 
  unaffected by metallicity at least between a fifth and 
  solar metallicity, but in the case of PNe it is shifted 
  to significantly longer wavelengths.
  }


   \maketitle
%
%___________________________________________________

%-------------------------------------------------------------
%-------------------------------------------------------------
% ----------------- INTRODUCTION ------------------
%-------------------------------------------------------------
%-------------------------------------------------------------
\section{Introduction}
\label{sec:introduction}

   Low and intermediate-mass stars (from $\sim$0.8 to 
   $\sim$8 M$_\odot$) terminate their evolution on the 
   asymptotic giant branch (AGB) in a phase of intense 
   mass-loss. During this evolutionary phase, the star is 
   obscured by a thick envelope made of gas and dust, 
   and is mainly or even only visible in the infrared.
   
   AGB stars produce carbon in their helium-burning 
   shells, which is brought to the surface by a series of 
   dredge-ups \citep[see e.g.][]{Herwig:2005lr}. The 
   chemical composition of the outer layers of the star 
   depends mostly on their C to O abundance ratio. 
   When the C/O value exceeds unity, the star becomes 
   carbon-rich. Most of the available oxygen is locked in 
   the very stable CO molecule, and the leftover carbon 
   forms molecules typical for carbon stars such as 
   C$_{2}$, C$_{2}$H$_{2}$, and HCN, as well as 
   carbon-rich dust grains. \cite{Groenewegen:1995fk} 
   demonstrated that Galactic stars with initial masses 
   M$_{\ast}$ $\lesssim$ 1.55 M$_{\odot}$ never 
   become carbon stars. However, the formation of 
   carbon-rich stars also depends on the metallicity 
   of the host galaxy 
   \citep[see e.g. Fig. 3 in][]{Piovan:2003aa}. At lower 
   metallicity, stars with initial masses down to 
   1.1 M$_\odot$ could become carbon-rich. For 
   example, Z = 0.004, as for the Sagittarius dwarf 
   spheroidal galaxy (Sgr dSph) or the Small 
   Magellanic Cloud (SMC).
   
   During the post-AGB phase, when the high 
   mass-loss phase has terminated, the ejected 
   envelope moves away from the star, which can 
   become visible in the optical. When the mass of the 
   H-rich envelope decreases to about 0.001 of solar 
   mass for 0.6 M$_\odot$, the star moves quickly to 
   the left on the H-R diagram (increases its 
   temperature) at nearly constant luminosity 
   \citep[see e.g.][]{Frankowski:2003aa}. Afterwards, 
   the star can ionise the surrounding matter and a 
   planetary nebula (PN) is formed. The mass-loss 
   process in low- and intermediate-mass stars is one 
   of the most important mechanisms to provide 
   enriched gas and dust to the interstellar medium for 
   a new generations of stars.
   
   A broad-emission dust feature peaking at around 
   30\,$\mu$m is only seen in the spectra of some 
   carbon-rich AGB stars, post-AGB stars, and PNe. 
   This feature was observed for the first time by 
   \citet{Forrest:1981qy}, who discovered it in 
   CW~Leo, IC~418 and NGC~6572. Since its 
   discovery, it has been detected in many carbon-rich 
   objects. In carbon-rich post-AGB objects, this 
   feature can sometimes be observed along with the 
   still-unidentified 21\,$\mu$m feature.

   \cite{Goebel:1985uq} proposed solid magnesium 
   sulphide (MgS) as the possible carrier of the 
   30\,$\mu$m emission feature. Despite the MgS 
   identification being generally accepted by the 
   community \citep[e.g.][]{Hony:2002lr, Zijlstra:2006fj, 
   Lombaert:2012kx, Sloan:2014fj, Sloan:2016aa}, its 
   identification remains a matter of some debate. For 
   instance, \cite{Zhang:2009kx} have argued that the 
   abundance of MgS is not sufficient to explain the 
   strength of the feature in the post-AGB star 
   HD~56126. The amount of MgS mass required to 
   explain the power emitted by the 30\,$\mu$m feature 
   in this object would require ten times more atomic 
   sulphur than available in the ejected envelope.  
   %[I WOULD PREFER WE REFUTE THIS HERE WITH COATINGS - GREG]
   On the basis of a sample of extreme carbon stars, 
   \cite{Messenger:2013lr} found a correlation between 
   the 30\,$\mu$m feature and the 11.3\,$\mu$m silicon 
   carbide (SiC) one, which suggests that the 
   abundance of the carrier of the 30\,$\mu$m feature 
   is linked to the SiC abundance. On the basis of 
   fullerene-containing planetary nebulae, 
   \cite{Otsuka:2014eu} suggested that graphite can 
   also be the carrier of this feature, because it exhibits 
   a very broad, strong feature peaking around 
   30 -- 40\,$\mu$m \citep[see also][]{Speck:2009aa}. 
   % [NOTE: graphite has been proposed prior to Otsuka 
   % and remains doubleful to me because the wavelength 
   % of the peak is too long.  Kevin  Ditto - Greg, primarily 
   % because Graphite should have a strong feature at 11.5 um]
   Other materials have also been proposed as 
   carriers of the 30\,$\mu$m feature. \cite{Grishko:2001qy} 
   proposed hydrogenated amorphous carbon (HAC) 
   as a possible carrier. \cite{Duley:2000uq} suggested 
   that the 30\,$\mu$m feature may be indicative of 
   carbon-based linear molecules with specific side 
   groups. However, the hypothesis that MgS is 
   responsible for the 30\,$\mu$m feature is not ruled 
   out. This is partly because the optical properties of 
   MgS in the optical and ultraviolet range are still 
   unknown, preventing modellers from properly 
   estimating temperature inside the envelope. In 
   addition, the shape of MgS grains is important. For 
   example, \cite{Hony:2002lr} concluded that the 
   shape of the 30\,$\mu$m feature can be best 
   reproduced when the MgS grains are not perfect 
   spheres. In addition, \cite{Lombaert:2012kx} 
   suggested that MgS condenses as a coating on the 
   top of amorphous carbon and SiC grains. Using 
   their model it is possible to resolve the MgS mass 
   problem. However, \citet{Szczerba:1997fk}, showed 
   that for the simple spherical coating the feature has 
   two peaks, which seems not to be observed (see 
   their Fig. 3).
   
   \cite{Hony:2002lr} have analysed in a uniform way a 
   sample of 63 Galactic 30\,$\mu$m sources observed  
   by the Infrared Space Observatory (ISO). Before 
   \cite{Sloan:2016aa} this was the largest sample 
   analysed in a uniform way. The Spitzer Space 
   Telescope mission (hereafter Spitzer), thanks to its 
   sensitivity, was able to detect distant sources with 
   30\,$\mu$m features in our Galaxy (designated on 
   all the figures in this paper by the short name `GAL'), 
   but also in nearby galaxies such as the SMC, Large 
   Magellanic Cloud (LMC), and the Sgr dSph. On the 
   basis of the Spitzer spectra, \cite{Sloan:2016aa} 
   have analysed a sample of 184 carbon stars in the 
   Magellanic Clouds (MCs), and about 50\% of them 
   show the 30\,$\mu$m emission.
   
   In this paper we present a uniform analysis of the 
   30\,$\mu$m feature, on the basis of the spectra 
   from the InfraRed Spectrograph (IRS) on board 
   Spitzer for carbon-rich sources from the Milky Way 
   and the three nearby galaxies mentioned above. 
   All of these galaxies are characterised by the 
   different average metallicities. Therefore we can 
   compare the basic properties of this feature in 
   various environments.
   
   Our paper is organised as follows. In 
   Section~\ref{sec:the_spectral_sample}, we describe 
   the sample and our catalogues of 30\,$\mu$m 
   sources. In Section~\ref{sec:spectral_analysis} we 
   describe the method for fitting the continuum to the 
   spectra. We then explain how we determine the 
   properties of the 30\,$\mu$m feature and present 
   a description of the contents of the tables with the 
   results. In Section~\ref{sec:results_summary} we 
   show the correlation study between the properties 
   of the 30\,$\mu$m feature for the different 
   populations of the objects. The second part of this 
   section contains the summary of the results we 
   have obtained.

%-------------------------------------------------------------
%-------------------------------------------------------------
% ----------- THE SPECTRAL SAMPLE -----------
%-------------------------------------------------------------
%-------------------------------------------------------------
\section{Spectral sample}
\label{sec:the_spectral_sample}

%-------------------------------------------------------------
% -------------- TARGET SELECTION ---------------
%-------------------------------------------------------------
\subsection{Target selection and the full sample}
\label{subsec:target_selection}

   Our sample consists of archival data obtained by the 
   InfraRed Spectrograph (IRS), on board Spitzer. The 
   sample considered in this paper is quite similar to the 
   sample of 30\,$\mu$m sources described by 
   \cite{Sloan:2014fj, Sloan:2016aa} as far as the SMC 
   and LMC objects are concerned. In case of Sgr dSph 
   and Milky Way objects the samples were selected 
   from different authors. Details of our selection are 
   given in Appendix~\ref{sec:appendix_sample}.
   
   Our full sample consist of 207 objects in total: five 
   from the Sgr dSph (four AGB stars, one PN), 22 
   from the SMC (eight AGB stars, three post-AGB 
   objects, and 11 PNe), 121 from the LMC (83 AGB 
   stars, 17 post-AGB objects, and 21 PNe), and 59 
   objects from our Galaxy (17 AGB stars, 23 
   post-AGB objects, and 19 PNe).
   
   The majority of our objects were observed in the low 
   resolution (R $\sim$60--120) mode during various 
   observational programs. These spectra have two 
   short-low (SL) and two long-low (LL) modules, which 
   provide spectral coverage from $\sim$5--38\,$\mu$m, 
   and join together around 14\,$\mu$m. The SL and LL 
   modules each have their own two apertures, one for 
   the second order spectrum (5.2--7.7\,$\mu$m and 
   14--21.3\,$\mu$m) and one for the first order 
   spectrum (7.4--14.5\,$\mu$m and 19.5--38\,$\mu$m; 
   see \citealt{Lebouteiller:2011aa} for more details). 
   The telescope had to shift position to put the target in 
   each of the four apertures, which requires eight 
   pointings. Some of the objects were observed in the 
   high resolution modules 
   \citep[see e.g.][]{2015ApJS..218...21L}, short-high 
   (SH) and long-high (LH), which cover the spectral 
   ranges from $\sim$10--37\,$\mu$m (9.9--19.6\,$\mu$m 
   and 18.7--37.2\,$\mu$m, respectively) at the 
   resolution of $\sim$300.
   %\MGcomment{On the website of CASSIS database 
   % I have seen R = 600, whereas in the Greg paper 
   % there is 300. Do the CASSIS spectra have higher 
   % resolution? Put that before, AS YOU MENTION 
   % SL AND LL IN PREVIOUS PARAGRAPHS}   
   
   For both resolutions a jump between the short 
   segments (SL/SH) and long segments (LL/LH) of the 
   spectra is common. This jump is observed in partially 
   extended sources when reduced using the point-like 
   source flux calibration. For such objects, the LL/LH 
   spectrum contains more flux than the SL/SH, 
   because the LL/LH extraction aperture is larger than 
   the SL/SH (this problem is described by 
   \citealt{Lebouteiller:2011aa} in the case of low 
   resolution spectra). Discontinuities were removed 
   between orders by multiplying each spectral segment 
   by a scaling factor to align them in the regions of 
   overlap. In almost all cases the corrections are made 
   up to the brightest segment, on the grounds that it is 
   the one best centred in the slit.

   % There are four segments to the spectra, not two. SL and LL 
   % each have two apertures, one for the 2nd order spectrum 
   % (5.1-7.5 um and 14-19ish um), and one for the 1st order 
   % spectrum (7.5-14 um and 19ish-37 um). The telescope had to 
   % shift position to put the target in each of the four apertures, so 
   % they are separate pointings requiring individual corrections. 
   %
   % One undesirable effect is that, when using a point-like source 
   % flux calibration (whether in optimal extraction or tapered 
   % column extraction) on a partially extended source, the LL 
   % spectrum contains more flux than the SL spectrum because the
   % LL extraction aperture is larger than the SL aperture. A more 
   % advanced version of the flux calibration is currently under 
   % investigation and will be included in future CASSIS 
   % processings. It uses a theoretical PSF and computes the 
   % amount of light lost outside the slit as a function of a 
   % broadening parameter.
   
   The data reduction for the spectra in the MCs 
   followed the standard Cornell algorithm described in 
   detail by \citet{Sloan:2012aa}. The key element is 
   the optimal extraction of the spectra from the 
   background-subtracted and cleaned spectral images 
   \citep{Lebouteiller:2010aa}. The spectra of sources 
   in the Galaxy and the Sgr dSph were obtained from 
   the Combined Atlas of Sources with Spitzer IRS 
   Spectra 
   (CASSIS\footnote{CASSIS database is available 
   online at: \url{http://cassis.sirtf.com/atlas/welcome.shtml}}; 
   \citealt{Lebouteiller:2011aa, 2015ApJS..218...21L}), 
   which used a similar approach, including optimal 
   extraction.
      
   Table~\ref{tab:sgr_basic_info} lists basic information 
   about objects in the Sgr dSph. 
   Appendix~\ref{sec:appendix_tables} presents the 
   \Cref{app_tab:smc_basic_info,app_tab:lmc_basic_info,app_tab:gal_basic_info} 
   with basic information about objects from the SMC, 
   LMC, and Milky Way. The objects which are marked 
   by the bold face names in 
   Table~\ref{tab:sgr_basic_info}, and 
   \Cref{app_tab:smc_basic_info,app_tab:lmc_basic_info,app_tab:gal_basic_info}, 
   are excluded from further analysis for reasons 
   described in Section~\ref{subsec:manchester_method}.

   In each of these Tables we present the name of the 
   object, another common name if available, RA and 
   Dec coordinates in the J2000 system, its position 
   reference, classification of the object (Class), and 
   its main period if known with the corresponding 
   references. The position reference is IRAC for SMC 
   and LMC sources and 
   2MASS\footnote{2MASS: The Two Micron All Sky Survey} 
   for Galactic and Sgr dSph objects (see 
   Section~\ref{subsec:torun_catalogues} for details). 
   The column with the classification contains any 
   additional information about the sources and the 
   references (if needed). For the post-AGB objects 
   we put the name of the group for the object (see 
   Section~\ref{subsec:manchester_method} for 
   details), then we mark objects with 21\,$\mu$m 
   feature, C$_{2}$H$_{2}$ band, and objects showing 
   the broad 16-24\,$\mu$m feature. In a case of PNe 
   we put information about presence of the 
   16-18\,$\mu$m feature, broad 16-24\,$\mu$m 
   feature and the fullerenes. In the Galactic sample we 
   indicate the S-type stars. Finally, Program ID and 
   AOR keys (Spitzer Astronomical Observation 
   Request) for each source are given.

%*************************************************************************************************************************
%*************************************************************************************************************************
%
%_____________________________________________________________
%                                             Two column Table 
%_____________________________________________________________
%
\begin{table*}
\scriptsize
\caption{Sgr dSph sample: names, coordinates, reference 
position, classification of spectra (C-AGB: carbon-rich AGB star, 
C-pAGB: carbon-rich post-AGB star, C-PN: carbon-rich planetary 
nebula), period with reference, program identifier, and AOR key.}             
\label{tab:sgr_basic_info}      
\centering          
\begin{tabular}{llccclcccc} % 10 columns
\hline\hline                       
              			&						&	\multicolumn{2}{c}{(J2000.0)}   \\\cmidrule{3-4}
Name    			&	Other name			&	RA		&	Dec			&	Position	&	Class								&	Period	&	Period	&	Program ID& AOR key\\
				&						&	(deg)		&	(deg)			&	reference	&										&	(days)	&	reference	&			&\\
\hline
Sgr~3			&       IRAS~F18413-3040		&	281.129	&	$-$30.619		&	2MASS	&	C-AGB								&	446		&	(1)		&	30333	&	18054656\\
Sgr~7			&       IRAS~F18436-2849		&	281.715	&	$-$28.764		&	2MASS	&	C-AGB								&	512		&	(1)		&	30333	&	18054912\\
Sgr~15			&       IRAS~F18555-3001		&	284.683	&	$-$29.949		&	2MASS	&	C-AGB								&	417		&	(1)		&	30333	&	18055168\\
Sgr~18			&       IRAS~19013-3117		&	286.148	&	$-$31.216		&	2MASS	&	C-AGB								&	485		&	(1)		&	30333	&	18055424\\      
\textbf{Wray16-423}	&       PN~G006.8-19.8		&	290.544	&	$-$31.511		&	2MASS	&	C-PN, 16-24\,$\mu$m feat.\tablefootmark{a}	&	\ldots	&	\ldots	&	50261	&	25833216\\
\hline                  
\end{tabular}
\tablefoot{The tables with the Galactic and MCs 
objects are available in Appendix~\ref{sec:appendix_tables}.}
\tablebib{(1)~\citet{Lagadec:2009aa}.\\
(a)~\citet{Otsuka:2015aa}.
}
\end{table*}
%
%*************************************************************************************************************************
%*************************************************************************************************************************

%-------------------------------------------------------------
% ---------- THE TORUN CATALOGUES ----------
%-------------------------------------------------------------
\subsection{Toru\'n catalogues of 30\,$\mu$m objects}
\label{subsec:torun_catalogues}

   We have produced online catalogues of photometric 
   data and Spitzer IRS spectra for all objects that show 
   the 30\,$\mu$m feature. The 207 objects are 
   organised into three different catalogues: 22 objects 
   from the SMC, 121 from the LMC, and 59 Galactic 
   sources. The five Sgr dSph objects are listed along 
   with the Galactic catalogue and they are 
   distinguished there by the special comments. Each 
   part of our database is available online at:

\begin{itemize}
        \item \url{http://www.ncac.torun.pl/30smc}
        \item \url{http://www.ncac.torun.pl/30lmc}
        \item \url{http://www.ncac.torun.pl/30galactic}
\end{itemize}
   
   The structure of the catalogues is similar to the 
   evolutionary catalogue of Galactic post-AGB and 
   related objects, created 
   by~\cite{Szczerba:2007th, Szczerba:2012aa}. After 
   entering any of the catalogue links given above, the 
   home screen will appear. The user has to click on 
   the main button `30 micron SMC/LMC/Galactic' 
   depending on the version. After this action, a new 
   screen appears with the information about the total 
   number of objects in this section of the catalogue 
   and list of objects ordered according to increasing 
   right ascension (Col. 1). In addition, the other 
   names, if any (Col. 2), classification (Col. 3), 
   explanations to each object (e.g. information about 
   variability) and the Bibliography (Cols. 4 and 5, 
   respectively) are given. The bibliography contains 
   the link to SAO/NASA Astrophysical Data System 
   (ADS). After clicking it, the list of papers for a given 
   object appears.

   The main bar above the list of objects contains the 
   following buttons (starting from the left):
 
 \begin{itemize}
        \item \textbf{Home} -- allows the user to get back to the home screen;
        \item \textbf{Change catalogue} -- allows the user to change the sub-catalogue to a different one;
        \item \textbf{Info} -- show more information about the catalogue content;
        \item \textbf{Search} -- allows the user to search the sub-catalogue using different criteria (name, class, coordinates, etc.);
        \item \textbf{Export} -- allows the user to export the data (see below for details).
\end{itemize}

\begin{figure*}[h!]
   \resizebox{\hsize}{!}
            {\includegraphics[]{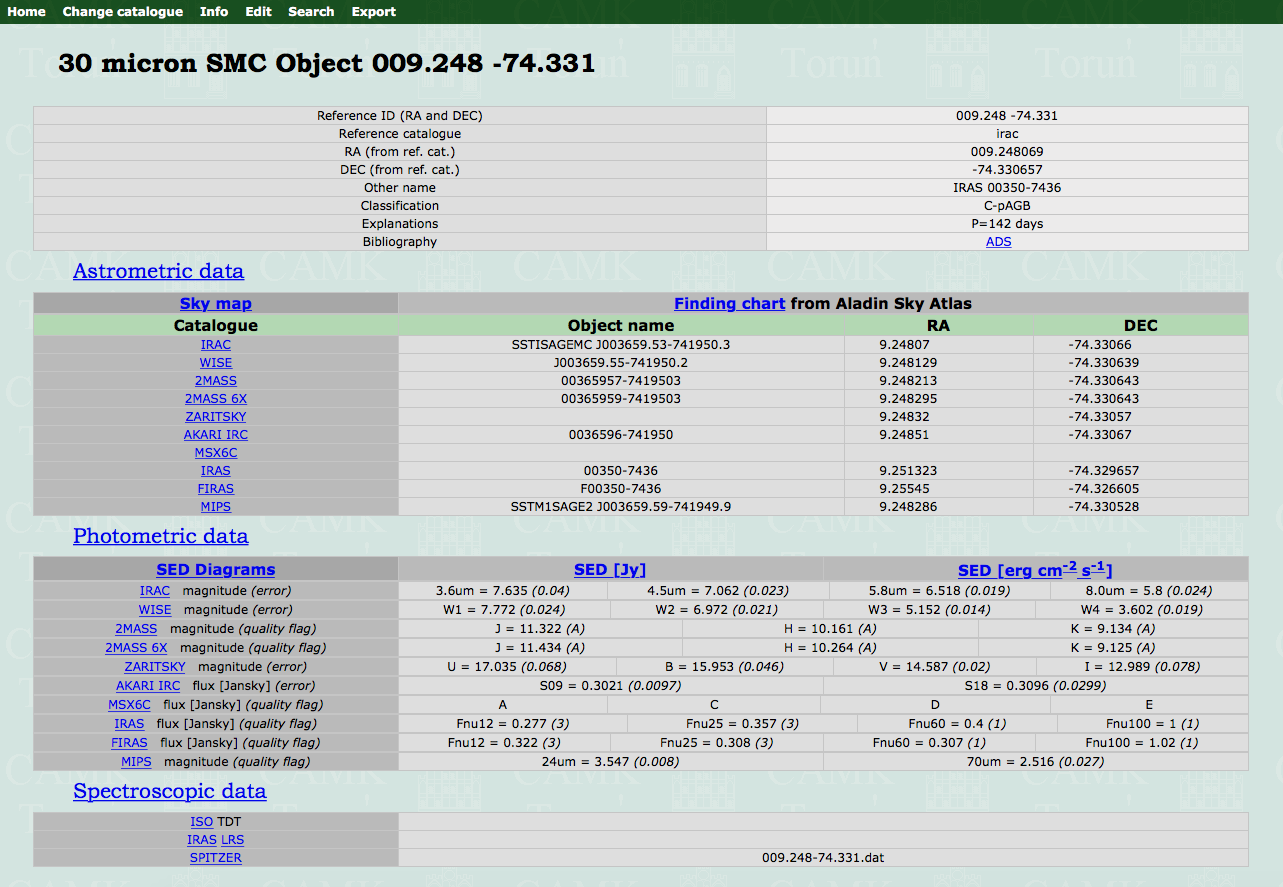}}
      \caption{Example of the screen shown for a selected 
      object from the SMC catalogue (IRAS~00350-7436).}
         \label{fig:catalogue_smc}
   \end{figure*}

   After clicking on the RA/Dec value of a given object, 
   a new screen appears with all the characteristics. 
   The main bar on the top of the screen remains 
   unchanged. These characteristics are divided into 
   three categories: `Astrometric data' with the names 
   and coordinates in degrees obtained for a few 
   photometric surveys; `Photometric data' with a 
   photometry provided by these surveys; and 
   `Spectroscopic data' containing the information 
   about available spectra for a given object. An 
   example of the screen showing all the information 
   for a given object is shown in 
   Figure~\ref{fig:catalogue_smc}. Inside the panel 
   with the photometric data, there are two `SED' 
   buttons, which allows the user to see the spectral 
   energy distribution (SED) in (Jy) or in 
   (${\rm erg}$ ${\rm cm}^{-2}$ ${\rm s}^{-1}$), as a 
   function of wavelength ($\mu$m). The photometry is 
   represented by the points on this diagram, whereas 
   the Spitzer spectrum is overplotted as the solid line 
   in the corresponding units. The data in our catalogue 
   can be exported by clicking on the `Export' button 
   (see Figure~\ref{fig:catalogue_smc}). The possible 
   format of the exported table is ascii or Excel.

   For the MCs objects we find the counterpart 
   sources in the SAGE\footnote{SAGE: Surveying the 
   Agents of Galaxy Evolution} survey of the LMC 
   \citep{Meixner:2006aa} and the SAGE-SMC survey 
   \citep{Gordon:2011aa}. The SAGE surveys give 
   photometry in the four filters (3.6, 4.5, 5.8, and 
   8\,$\mu$m) from the IRAC instrument as well as in 
   24 and 70\,$\mu$m (there is also the 160\,$\mu$m 
   filter but we have not noticed any detection) bands 
   from the MIPS\footnote{MIPS: Multiband Imaging 
   Photometer for Spitzer} photometer on board 
   Spitzer. Using the IRAC coordinates as the reference 
   positions, we searched several additional catalogues 
   for the photometric data. However, searching in the 
   radius of 5\arcsec we have not found the 
   counterparts in the SAGE catalogue for eight objects 
   in the LMC (IRAS~F04353-6702, IRAS~04375-7247, 
   IRAS~F04537-6509, SMP~LMC~61, SMP~LMC~79, 
   IRAS~F06108-7045, IRAS~06111-7023, and 
   SMP~LMC~99). Therefore we used 2MASS 
   coordinates to find the counterparts for them. In the 
   optical range we gathered information from the 
   MCPS\footnote{MCPS: Magellanic Cloud Photometric Survey} 
   of the LMC \citep{Zaritsky:2004aa} and SMC 
   \citep{Zaritsky:2002aa} in the U, B, V and I filters. 
   Near- and mid-IR photometry consist of a few surveys: 
   2MASS \citep{Skrutskie:2006aa} and 2MASS-6X 
   (J, H, and K$_{s}$ filters for both surveys) which 
   covered the MCs with longer exposure times 
   \citep{Cutri:2006}, WISE\footnote{WISE: Wide-field 
   Infrared Survey Experiment} \citep{Wright:2010aa}, 
   which provides observations in W1 (3.35\,$\mu$m), 
   W2 (4.6\,$\mu$m), W3 (11.6\,$\mu$m), and 
   W4 (22.1\,$\mu$m) filters, MSX6C\footnote{MSX6C: 
   The Midcourse Space Experiment (version 2.3); 6C 
   denotes data run using version 6.0 of convert 
   software with the names of the sources based on the 
   galactic positions} point source catalogue 
   \citep{2003AAS...203.5708E} from which we used 
   A, C, D, E filters (8.28, 12.13, 14.65, and 21.34, 
   respectively), and AKARI\footnote{AKARI was the first 
   Japanese astronomical satellite observing in the 
   infrared range.} IRC survey (9 and 18\,$\mu$m bands; 
   \citealt{Ishihara:2010aa}). We assembled additional 
   data from the IRAS\footnote{IRAS: Infrared 
   Astronomical Satellite} catalogue \citep{1988iras....7.....H} 
   of point sources and its deeper edition 
   FIRAS\footnote{FIRAS: IRAS Faint Source Catalogue} 
   (12, 25, 60, and 100\,$\mu$m bands for both surveys; 
   \citealt{Moshir:1990aa}). The photometry for the 
   MIPS 70\,$\mu$m band was available for only two 
   objects in the SMC and nine objects in the LMC, 
   whereas for the IRAS/FIRAS 60/100\,$\mu$m band 
   for only one object in the SMC, and four objects in 
   the LMC.
   
   In the case of Galactic objects (and Sgr dSph ones) 
   we have used the 2MASS coordinates as the 
   reference positions and added the data from a few 
   additional catalogues. However, in the case of 
   IRAS~15531-5704 we could not find the counterpart 
   in the 2MASS catalogue in the radius of 2\arcsec, 
   thus the WISE coordinates were used as the 
   reference. The third release of the 
   DENIS\footnote{DENIS: Deep Near Infrared Survey 
   of the Southern Sky} catalogue \citep{Denis:2005} 
   gives similar spectral coverage as 2MASS: I, J, and 
   K$_{\rm s}$. The first edition of the 
   GLIMPSE\footnote{GLIMPSE: Galactic Legacy 
   Infrared Mid-Plane Survey Extraordinaire} catalogue 
   \citep{Benjamin:2003aa, Churchwell:2009aa} 
   provided coverage at 3.6, 4.5, 5.8, and 8\,$\mu$m, 
   but only for two objects from our sample. The optical 
   photometry came from the GSC\footnote{GSC: The 
   Guide Star Catalogue} catalogue (version 2.3.2; 
   \citealt{Lasker:2008aa}) and contains the data in 
   Johnson B filter, and also B$_{\rm j}$, V, F, N 
   photographic filters. In a case of Galactic objects we 
   found counterparts in the AKARI-FIS catalogue 
   \citep{2010yCat.2298....0Y}, which has the four bands 
   at 65, 90, 140 (only two detections), and 160\,$\mu$m 
   (no detections). 
   
   The statistics in percentage of the collected 
   photometric data for each of the catalogues is 
   presented in Table~\ref{tab:cat_stat}. The first 
   column contains the information about the search 
   radius (\arcsec) used for each catalogue. The total 
   number of objects in each galaxy is given in the 
   square bracket. We note that the WISE survey 
   provides the photometry for almost all sources.

   All the catalogues consist of data that have been 
   obtained from various epochs. In some variable 
   sources the mismatch between the spectrum and 
   the photometry is visible in the SED diagrams. This 
   is clearly seen in the case of AGB stars, which are 
   known to be the strong pulsators.

%*************************************************************************************************************************
%*************************************************************************************************************************
%
%_____________________________________________________________
%                                             Two column Figure 
%-------------------------------------------------------------
% I added [ht] command to force LATEX to put the graph right here
% The valid options for the position specifier are:
% h : approximately here
% t : top of the page
% b : bottom of the page
% ! : to put emphasis on a specifier --> h! means "really here"
%
%
%*************************************************************************************************************************
%*************************************************************************************************************************

%*************************************************************************************************************************
%*************************************************************************************************************************
%
%______________________________________________________________
%
%_____________________________________________________________
%                                             Simple A&A Table
%_____________________________________________________________
%
\begin{table}[h!]
\scriptsize
\caption{Percentage of the collected photometric data for each of the catalogues.}             % title of Table
\label{tab:cat_stat}      % is used to refer this table in the text
\centering                          % used for centering table
\begin{tabular}{lccccc}        % 6 columns
\hline\hline                 % inserts double horizontal lines
				&       R		&	Sgr dSph	&       SMC&	LMC	&	Galactic\\    % table heading
				&			&	[5]		&       [22]	&	[121]	&	[59]\\
				&       (arcsec)	&	(\%)		&       (\%)	&	(\%)	&	(\%)\\
\hline                        % inserts single horizontal line
MCPS~(SMC)		&	5		&	0		&	77	&	0	&	0	\\      % inserting body of the table
MCPS~(LMC)		&	5		&	0		&	0	&	82	&	0	\\
GSC~2.3			&	2		&	80		&	0	&	0	&	86	\\
DENIS			&	3		&	100		&	0	&	0	&	49	\\
2MASS			&	2		&	100		&	95	&	86	&	98	\\
2MASS~6X		&	2		&	0		&	95	&	91	&	0	\\
SAGE-SMC~IRAC	&	5		&	0		&	100	&	0	&	0	\\
SAGE-LMC~IRAC	&	5		&	0		&	0	&	93	&	0	\\
GLIMPSE			&	5		&	0		&	0	&	0	&	3	\\
MSX6C			&	5		&	0		&	36	&	68	&	49	\\
WISE			&	5		&	100		&	95	&	96	&	100	\\ 
AKARI~IRC		&	5		&	100		&	50	&	92	&	100	\\
AKARI~FIS		&	10		&	0		&	0	&	0	&	56	\\
IRAS				&	10		&	20		&	14	&	55	&	100	\\
FIRAS			&	10		&	80		&	32	&	61	&	24	\\
MIPS~(24\,$\mu$m)	&	10		&	0		&	95	&	92	&	0	\\
MIPS~(70\,$\mu$m)	&	10		&	0		&	9	&	11	&	0	\\
\hline                                   %inserts single line
\end{tabular}
%\tablefoot{ND -- no data}
\end{table}
%
%*************************************************************************************************************************
%*************************************************************************************************************************

%-------------------------------------------------------------
%-------------------------------------------------------------
% -------------- SPECTRAL ANALYSIS --------------
%-------------------------------------------------------------
%-------------------------------------------------------------
\section{Spectral analysis}
\label{sec:spectral_analysis}

%-------------------------------------------------------------
% ------------ MANCHESTER METHOD ------------
%-------------------------------------------------------------
\subsection{Manchester method}
\label{subsec:manchester_method}

   Our analysis is based on the `Manchester method', 
   which was introduced by \cite{Zijlstra:2006fj} for 
   mass-losing carbon stars in the LMC and by 
   \cite{Sloan:2006kx} for mass-losing carbon stars in 
   the SMC. It uses two colour indices, [6.4]$-$[9.3] 
   and [16.5]$-$[21.5], from which the former correlates 
   with optical depth, while the latter serves for 
   determination of the colour temperature, which we 
   considered as the derived dust temperature 
   (T$_{\rm d}$). The [6.4]$-$[9.3] colour is calculated 
   from the Spitzer spectra using separate flux 
   integrations from 6.25 to 6.55\,$\mu$m and from 9.1 
   to 9.5\,$\mu$m. We summed over the regions 
   16--17\,$\mu$m and 21--22\,$\mu$m to simulate the 
   second colour value. This method was also used by 
   \citet{Sloan:2016aa} to model the continua for the 
   carbon-rich stars in the MCs. We 
   followed their recipe to get the continuum level in 
   each spectrum (with a modification for the post-AGB 
   objects and a few PNe).
   
   We used this method in its original form for the 
   carbon-rich AGB objects and PNe. However, the 
   range between 21 and 22\,$\mu$m in the Spitzer 
   spectra of PNe is not completely free from nebular 
   lines. In this spectral region the [Ar III] forbidden line 
   at 21.83\,$\mu$m can appear. Another transition of 
   the [Ar III] line is visible at 8.99\,$\mu$m. We have 
   checked all the PNe spectra in our sample for 
   presence of both [Ar III] lines, but only a very weak 
   [Ar III] line at 8.99 has been found. 
   
\begin{figure}
   \centering
   \includegraphics[width=\hsize]{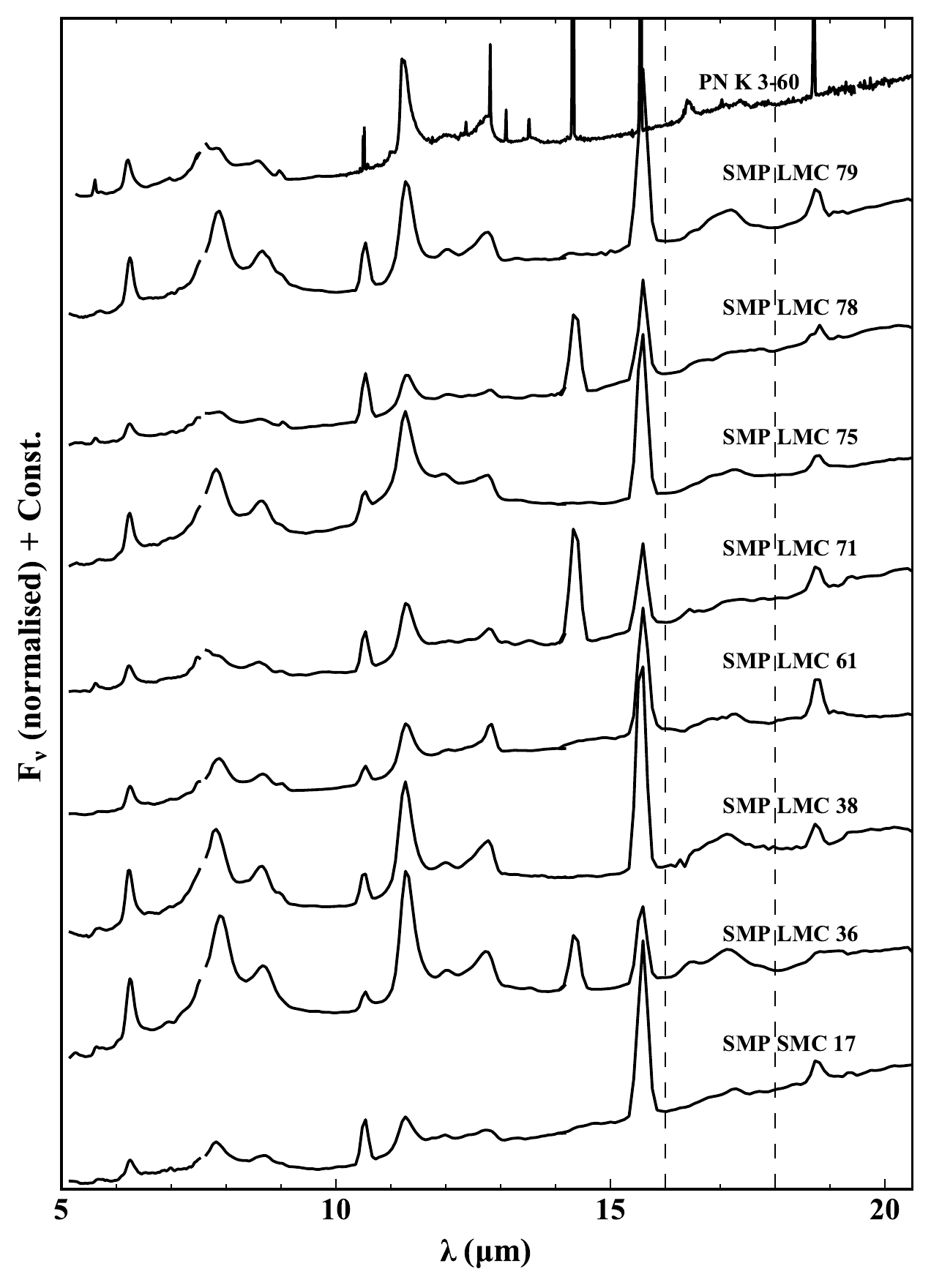}
      \caption{5--20.5\,$\mu$m spectra for objects 
      showing the 16-18\,$\mu$m emission feature 
      (solid lines). These spectra are normalised to the 
      flux density at 18\,$\mu$m. The names of objects 
      are shown above the spectra. The spectrum of 
      PN~K3-60 was partially obtained using low (SL) 
      and high-resolution (SH/LH) modes. The region 
      containing this feature is shown by the dashed 
      vertical lines.}
         \label{fig:16-18_feature}
   \end{figure}

   Moreover, in the spectra of nine PNe with strong 
   polycyclic aromatic hydrocarbons (PAHs), we found 
   the new detections of an unidentified emission 
   feature between about 16 and 18\,$\mu$m 
   (hereafter the 16-18\,$\mu$m feature), which might 
   be attributed to PAHs 
   \citep{Peeters:2004aa, Boersma:2010aa}. The 
   spectra in the 5-20.5\,$\mu$m range for those 
   objects are shown in Figure~\ref{fig:16-18_feature}. 
   %by solid lines. The region in which this feature 
   %appears is shown by a dashed line. 
   For this group of 
   objects we changed the shorter side of 
   [16.5]$-$[21.5] colour and integrated flux from 14.8 
   to 15.1\,$\mu$m. We note that the values of this new 
   [14.95]$-$[21.5] colour are given in 
   Table~\ref{tab:14-21_colour}. 
   
   However, in the spectra of SMP~LMC~78 and 
   SMP~LMC~79 the 16-24\,$\mu$m feature is also 
   visible (see Fig.~\ref{fig:pne_broad_emission} and 
   the related discussion below), and the 
   [14.95]$-$[21.5] colour is affected by this feature. 
   Therefore, we removed these objects from further 
   analysis, and did not report their [14.95]$-$[21.5] 
   colour. The names of those objects are given in 
   bold face in Table~\ref{tab:14-21_colour}.

%*************************************************************************************************************************
%*************************************************************************************************************************
% I added [h!] command to force LATEX to put the graph right here
% The valid options for the position specifier are:
% h : approximately here
% t : top of the page
% b : bottom of the page
% ! : to put emphasis on a specifier --> h! means "really here"
%
%*************************************************************************************************************************
%*************************************************************************************************************************

%*************************************************************************************************************************
%*************************************************************************************************************************
%
%______________________________________________________________
%
%_____________________________________________________________
%                                             Simple A&A Table
%_____________________________________________________________
%
\begin{table}[h!]
\scriptsize
\caption{List of [14.95]$-$[21.5] colour values for 
the PNe with the 16-18\,$\mu$m feature.}             % title of Table
\label{tab:14-21_colour}      % is used to refer this table in the text
\centering                          % used for centering table
\begin{tabular}{lc}        % 2 columns
\hline\hline                 % inserts double horizontal lines
Name				&		[14.95]$-$[21.5]			\\
					&		(mag)				\\
\hline
PN~K~3-60			&	1.618	$\pm$	0.005	\\
\textbf{SMP~LMC~79}	&			$\ldots$			\\
\textbf{SMP~LMC~78}	&			$\ldots$			\\
SMP~LMC~75			&	1.407	$\pm$	0.009	\\
SMP~LMC~71			&	1.722	$\pm$	0.018	\\
SMP~LMC~61			&	1.145	$\pm$	0.009	\\
SMP~LMC~38			&	1.453	$\pm$	0.015	\\
SMP~LMC~36			&	1.349	$\pm$	0.008	\\
SMP~SMC~17			&	1.611	$\pm$	0.009	\\
\hline                                   %inserts single line
\end{tabular}
\end{table}
%
%*************************************************************************************************************************
%*************************************************************************************************************************

   In the case of post-AGB stars we redefined the 
   colour indices, depending on the strength of the 
   21\,$\mu$m feature. Generally, if the 21\,$\mu$m 
   band is not visible or is weak (group I) we used the 
   [18.4]$-$[22.45] colour (summing the flux from 
   18.1--18.7 and 22.3--22.6\,$\mu$m). When the 
   21\,$\mu$m feature is relatively strong (group II) we 
   took the [18.4]$-$[22.75] colour (summing the flux 
   from 18.1--18.7 and 22.5--23\,$\mu$m). For objects 
   with a very strong 21\,$\mu$m feature (group III) we 
   used the [17.95]$-$[23.2] colour (summing the flux 
   from 17.8--18.1 and 23--23.4\,$\mu$m).
   
   The [6.4]$-$[9.3] colour has been defined in the 
   Manchester method on the basis that those spectral 
   ranges represent the continuum level of carbon-rich 
   AGB stars. However, for post-AGB objects and PNe 
   this colour overlaps with the PAH band at 
   6.2\,$\mu$m. Therefore we defined a new 
   [5.8]$-$[9.3] colour, which is similar to the 
   [6.4]$-$[9.3] one, but avoids the 6.2\,$\mu$m band, 
   and other spectral features such as the [Mg V] line 
   at 5.61\,$\mu$m. The short wavelength part of this 
   colour was constructed by summing the flux from 
   5.7 to 5.9\,$\mu$m, whereas we kept the long 
   wavelength part of the colour in its original form.
   
   The list of all colour indices that have been used in 
   this work is shown in Table~\ref{tab:all_colours}. It 
   contains the name of colour index, the short and 
   long wavelength integration ranges, the range of the 
   continuum fit, and number of objects for which such 
   a colour index was calculated (among AGBs, 
   post-AGBs, and PNe).

%*************************************************************************************************************************
%*************************************************************************************************************************
%
%______________________________________________________________
%
%_____________________________________________________________
%                                             Simple A&A Table
%_____________________________________________________________
%
\begin{table*}
\scriptsize
\caption{List of colour indices used in this work: name of 
the colour, left and right integration range, continuum fit 
range, and number of objects for which such the colour 
was obtained.}             % title of Table
\label{tab:all_colours}      % is used to refer this table in the text
\centering                          % used for centering table
\begin{tabular}{lcccccc}        % 7 columns
\hline\hline                 % inserts double horizontal lines
Colour index		&	Left int.		&	Right int.	&	Cont. fit. range	&	\#AGB	&	\#post-AGB		&	\#PNe		\\
				&	($\mu$m)		&	($\mu$m)	&	($\mu$m)		&			&					&				\\
\hline
[6.4]$-$[9.3]		&	6.25--6.55		&	9.1--9.5	&	\ldots		&	112		&	0				&	0			\\
{[5.8]}$-${[9.3]}		&	5.7--5.9		&	9.1--9.5	&	\ldots		&	0		&	35				&	50			\\
\hline
{[16.5]}$-${[21.5]}	&	16--17		&	21--22	&	16--22, 20--23	&	112		&	0				&	42			\\
{[14.95]}$-${[21.5]}	&	14.8--15.1		&	21--22	&	20--23		&	0		&	0				&	7\tablefootmark{a}\\
{[18.4]}$-${[22.45]}	&	18.1--18.7		&	22.3--22.6	&	20--23		&	0		&	22\tablefootmark{b}	&	0			\\
{[18.4]}$-${[22.75]}	&	18.1--18.7		&	22.5--23	&	18.5--19		&	0		&	10\tablefootmark{c}	&	0			\\
{[17.95]}$-${[23.2]}	&	17.8--18.1		&	23--23.4	&	18--18.5		&	0		&	11\tablefootmark{d}	&	0			\\
\hline                                   %inserts single line
\end{tabular}
\tablefoot{
\tablefoottext{a}{Sources with the 16-18\,$\mu$m feature.}
\tablefoottext{b}{Weak or not present 21\,$\mu$m feature.}
\tablefoottext{c}{Relatively strong 21\,$\mu$m feature.}
\tablefoottext{d}{Strong 21\,$\mu$m feature.}
}
\end{table*}
% the [] bracket after \\ causes the problem, thus I made {[]}
%*************************************************************************************************************************
%*************************************************************************************************************************

   To model the continuum under the 30\,$\mu$m 
   feature we used a black-body function with the 
   derived dust temperature and fitted its continuum 
   level to the selected wavelength ranges that depend 
   on the kind of the object. For the AGB stars, we 
   applied the same fitting range as 
   \citet{Sloan:2016aa}: 16--22\,$\mu$m, whereas for 
   the PNe we used 20--23\,$\mu$m. The fitting range 
   for the post-AGB objects is different for each of the 
   three groups above: 20--23 for the group I, 
   18.5--19.0 for the group II, and 18--18.5\,$\mu$m for 
   the group III. There are 22 carbon-rich post-AGB 
   objects in the group I (one SMC, 11 LMC, and ten 
   Galactic objects), ten in the group II (one SMC, two 
   LMC, and seven Galactic objects) and 11 in the 
   group III (one SMC, four LMC, and six Galactic 
   objects).
   
\begin{figure}
   \centering
   \includegraphics[width=\hsize]{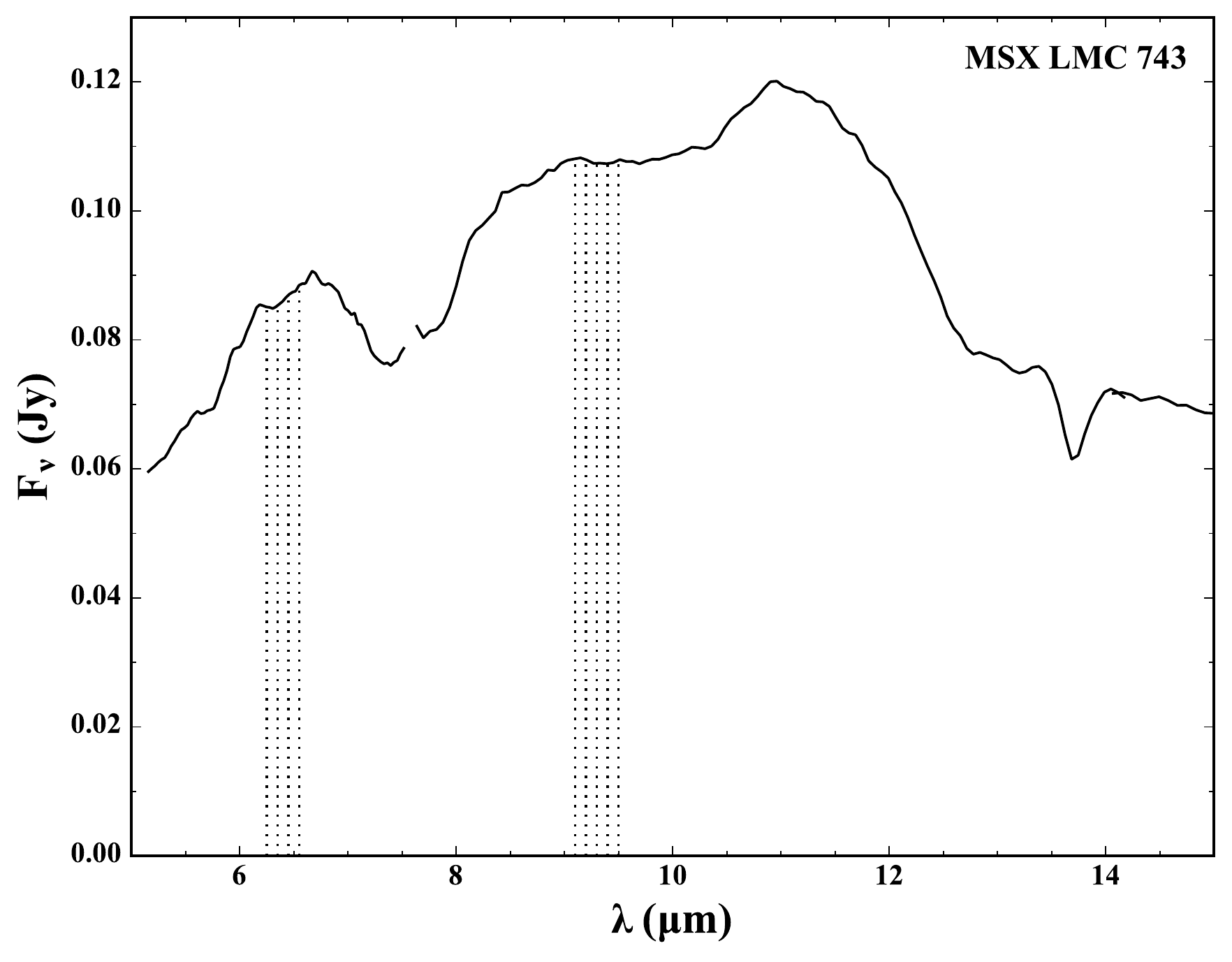}
      \caption{Manchester method applied to the shorter 
      part of the spectrum of MSX~LMC~743 (solid line). 
      The dotted areas show the ranges used to determine 
      the [6.4]$-$[9.3] colour.}
         \label{fig:manchester_method_6-9}
   \end{figure}

\begin{figure}
   \centering
   \includegraphics[width=\hsize]{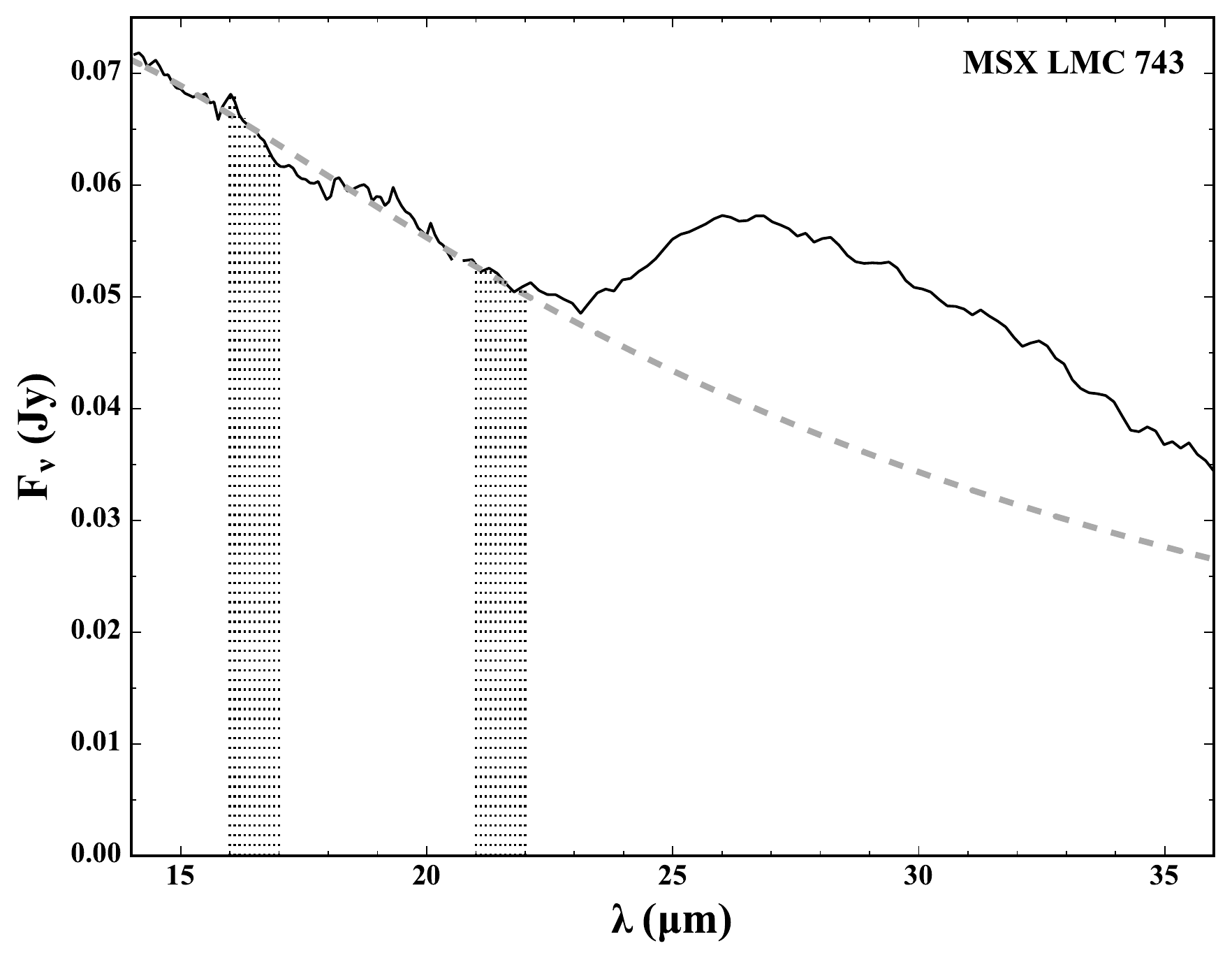}
      \caption{Manchester method applied to the longer 
      part of the spectrum of MSX~LMC~743 (solid line). 
      The dotted areas show the ranges used to 
      determine the [16.5]$-$[21.5] colour, and serves 
      for estimation of the colour temperature. The 
      dashed grey line represents the continuum under 
      the 30\,$\mu$m feature.}
         \label{fig:manchester_method_16-21}
   \end{figure}

\begin{figure}[h!]
   \centering
   \includegraphics[width=\hsize]{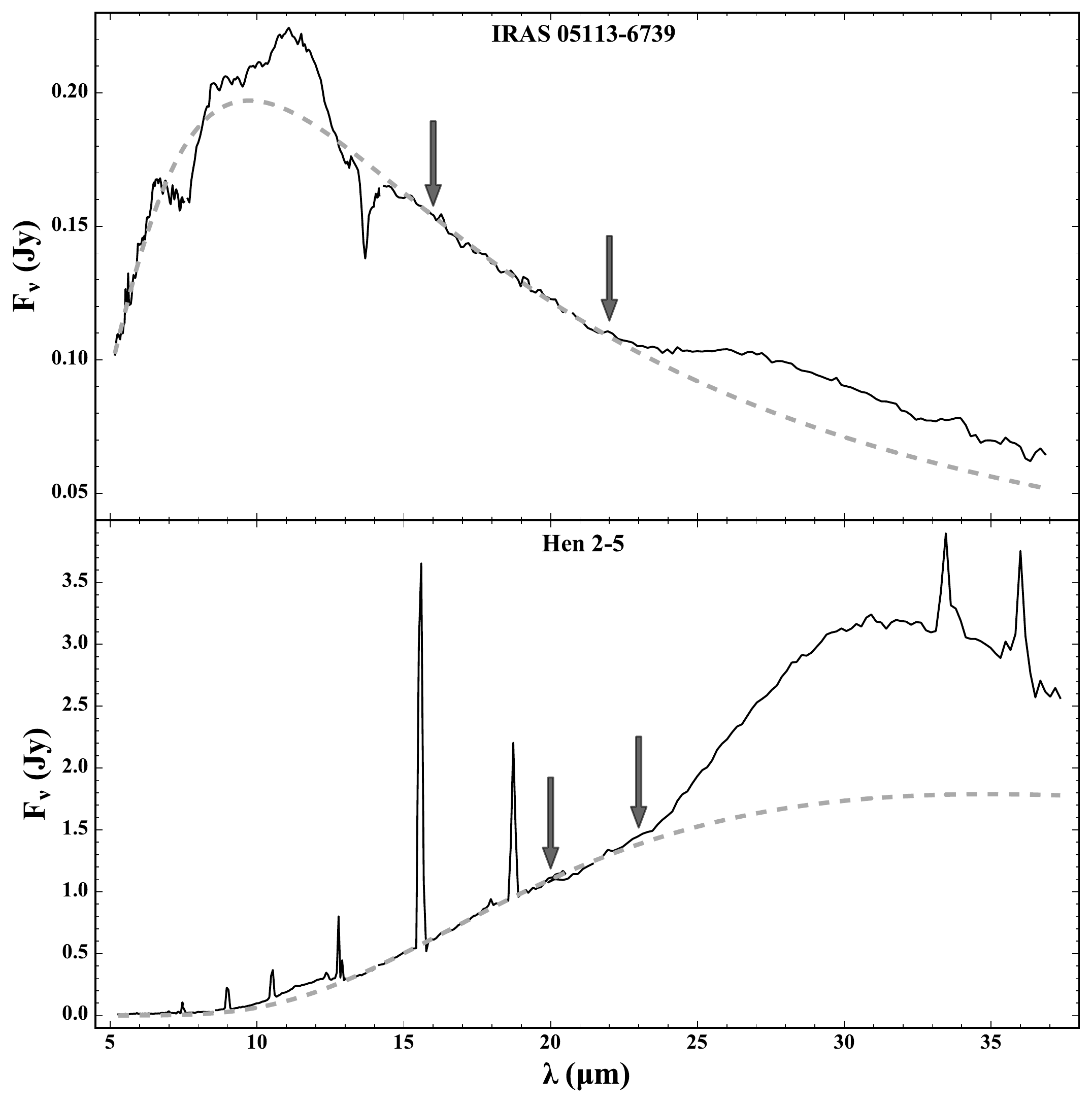}
      \caption{Examples of the fitted continuum for a 
      carbon-rich AGB star (top) and a PN (bottom). 
      Spectra are drawn in black solid lines, while the grey 
      dashed lines show the fitted black-body with a single 
      temperature derived from the [16.5]$-$[21.5] colour 
      (grey dashed lines). The arrows indicate the spectral 
      ranges to which the blackbodies are fitted. The 
      names of the objects are shown in the upper-central 
      part of each panel.}
         \label{fig:cont_fit_examples_agb_pn}
   \end{figure}

\begin{figure}[h!]
   \centering
   \includegraphics[width=\hsize]{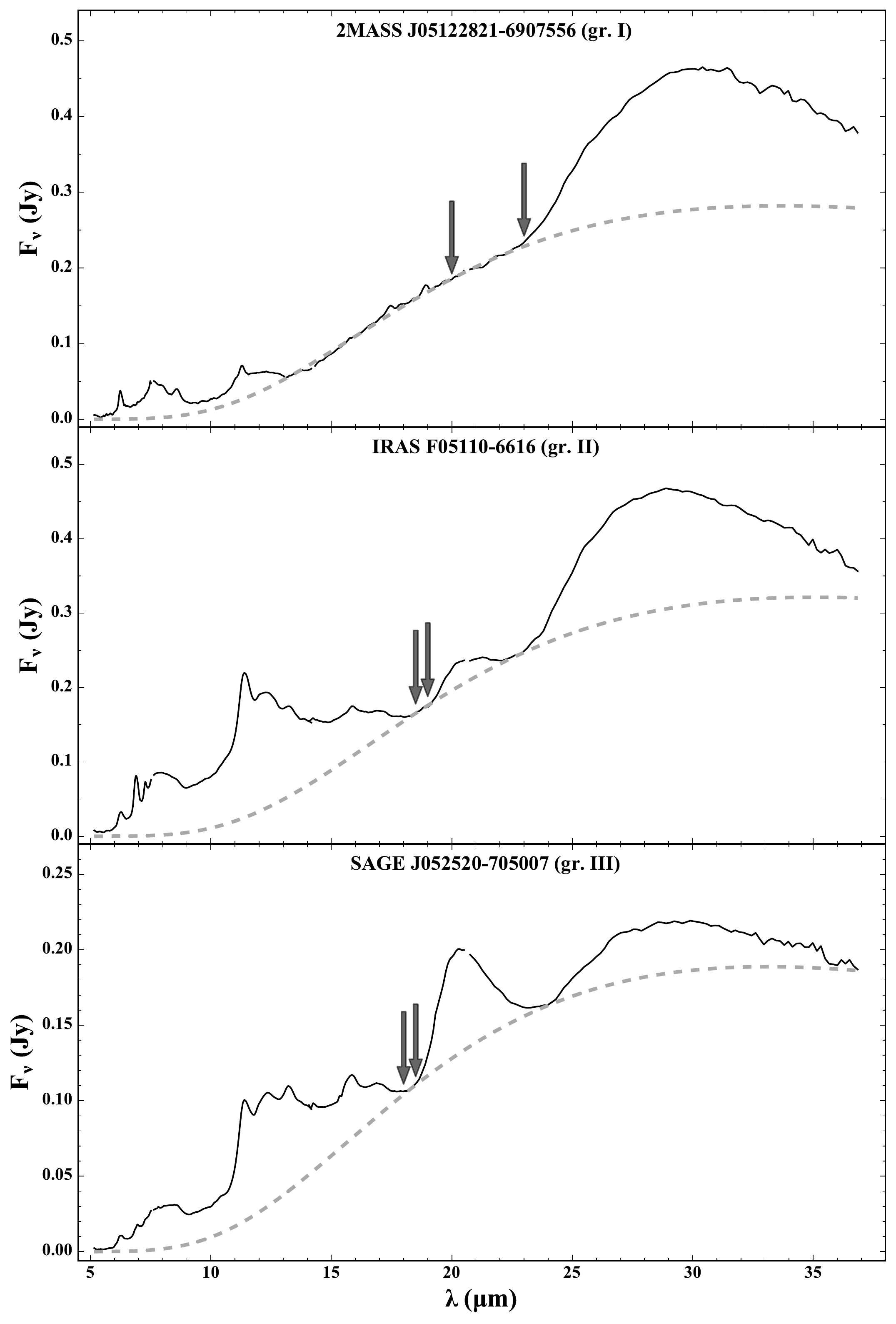}
      \caption{Examples of the fitted continuum for the 
      post-AGB objects from the group I to III (starting 
      from the top). Spectra are drawn in black solid lines, 
      while the grey dashed lines show the fitted 
      black-body with a single temperature. The arrows 
      indicate the spectral ranges to which the 
      blackbodies are fitted. The names of the objects are 
      shown in the upper-central part of each panel.}
         \label{fig:cont_fit_examples_pagb}
   \end{figure}

   Figures~\ref{fig:manchester_method_6-9} 
   and~\ref{fig:manchester_method_16-21} show how 
   the Manchester method works in its original form for 
   the shorter and longer part of the spectrum, 
   respectively. As an example we used the carbon star 
   MSX~LMC~743.
   %The spectra are drawn in
   %solid lines. The grey dashed line in 
   %Figure~\ref{fig:manchester_method_16-21} represents the 
   %black-body continuum fitted to the spectrum. The dotted areas 
   %mark ranges, which were used for flux integration and 
   %determination of Manchester method colours. 
   Figure~\ref{fig:cont_fit_examples_agb_pn} shows 
   examples of the fitted continuum for a carbon-rich 
   AGB star (IRAS~05113-6739) at the top panel and 
   a carbon-rich PN (Hen~2-5) on the bottom panel. 
   %The spectra are drawn in 
   %solid lines. The black-body fits with a single temperature are 
   %shown in grey dashed lines. The vertical arrows indicate the spectral 
   %ranges to which the blackbodies are fit. 
   In Figure~\ref{fig:cont_fit_examples_pagb} we show 
   examples of the fitted continuum for each of three 
   groups of post-AGB objects: 
   2MASS~J05122821-6907556 from group I at the 
   top panel, IRAS~F05110-6616 from group II in the 
   middle panel, and SAGE~J052520-705007 from 
   group III in the bottom panel. We note how the 
   strength of the 21\,$\mu$m feature is increasing 
   from the group I to III.
   %The spectra are drawn in solid lines, while 
   %the grey dashed lines represent the fitted black-body continua. 
   %Again, the vertical arrows indicate the spectral ranges to which 
   %the blackbodies are fitted.

%*************************************************************************************************************************
%*************************************************************************************************************************
%*************************************************************************************************************************
%*************************************************************************************************************************

%*************************************************************************************************************************
%*************************************************************************************************************************
%*************************************************************************************************************************
%*************************************************************************************************************************

%*************************************************************************************************************************
%*************************************************************************************************************************
%*************************************************************************************************************************
%*************************************************************************************************************************

%*************************************************************************************************************************
%*************************************************************************************************************************
%*************************************************************************************************************************
%*************************************************************************************************************************

   Despite the Manchester method being a very useful 
   tool for a uniform analysis of big samples of  
   carbon-rich stars, we noticed that its usage is limited 
   for some AGBs and post-AGBs with 21 $\mu$m 
   feature. We were unable to obtain black-body fits for 
   a total 33 objects (18 AGB stars and 15 post-AGB 
   objects). The objects with bad fits were removed 
   from the further analysis, and only the values 
   of the Manchester colours are presented in 
   \Cref{tab:sgr_basic_info,tab:sgr_spectral_results,%
   app_tab:smc_basic_info,app_tab:lmc_basic_info,%
   app_tab:gal_basic_info,app_tab:smc_spectral_results,%
   app_tab:lmc_spectral_results,app_tab:gal_spectral_results}.
   They are also distinguished by names in bold face.

\begin{figure}
   \centering
   \includegraphics[width=\hsize]{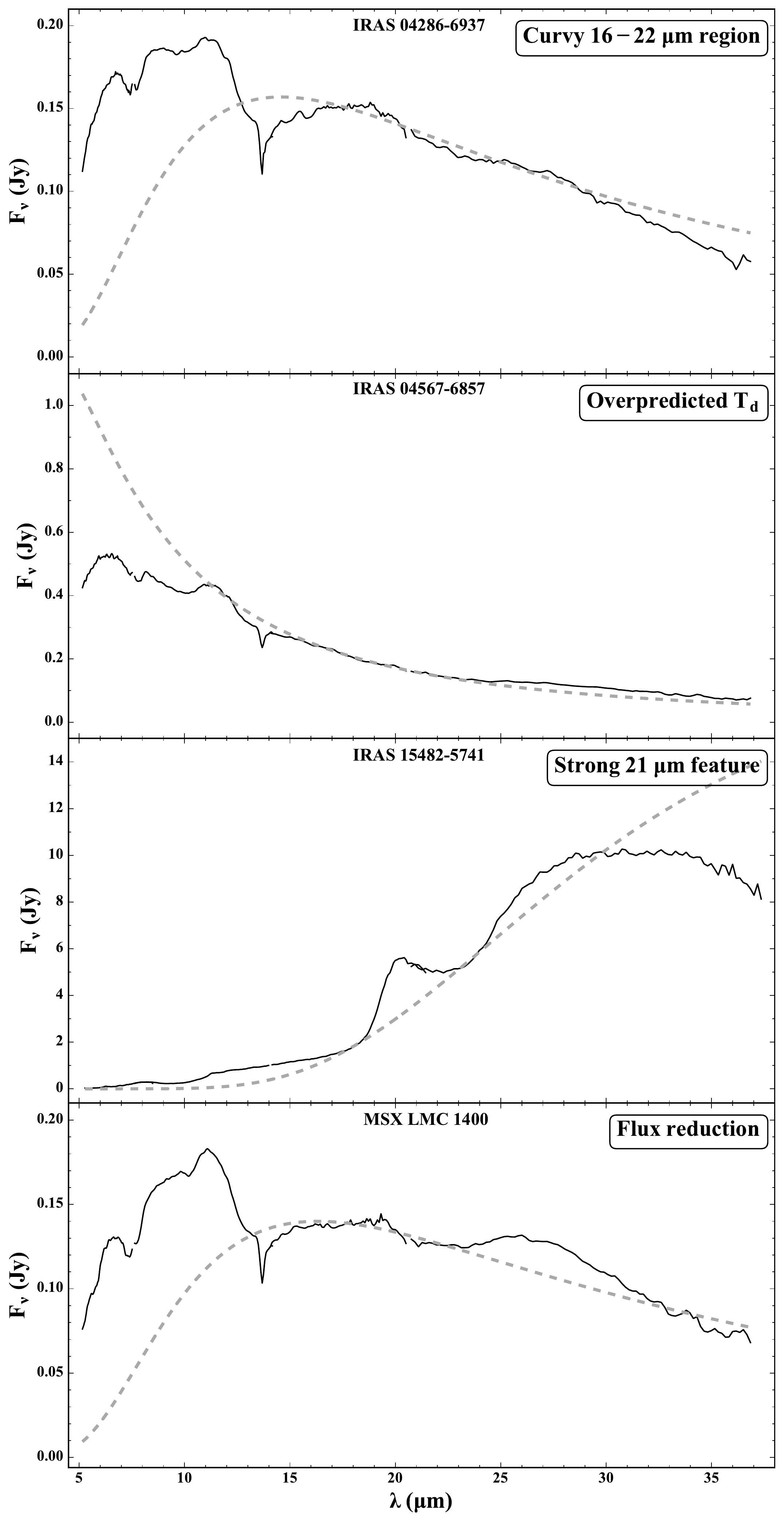}
      \caption{Examples of the bad continuum fits to the 
      spectra showing the constraints of the Manchester 
      method. The spectra are drawn in black solid lines, 
      while the grey dashed lines show the fitted 
      black-body with a single temperature. The names 
      of the objects are shown in the upper-central part 
      of each panel.}
         \label{fig:manchester_constraints}
   \end{figure}

\begin{figure}
   \centering
   \includegraphics[width=\hsize]{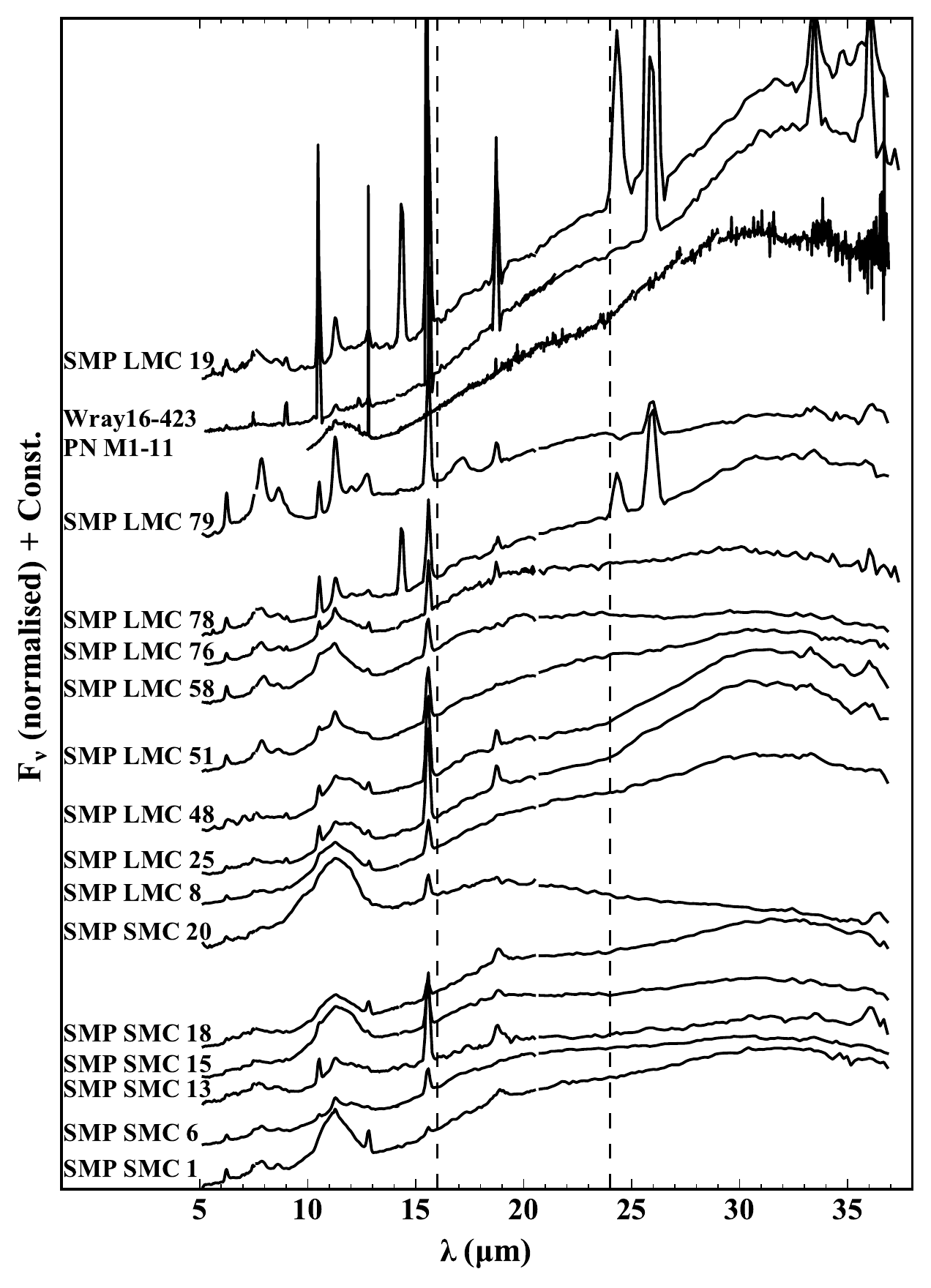}
      \caption{PNe excluded from analysed sample, 
      because of presence of the broad emission feature, 
      which is typically visible around 16 -- 24\,$\mu$m. 
      The spectra (solid lines) are normalised to the flux 
      density at 16\,$\mu$m and offset for clarity. The 
      names of the objects are shown above the spectra.}
         \label{fig:pne_broad_emission}
   \end{figure}

   The problems with modelling the continuum can be 
   divided into four categories, which are presented in 
   Figure~\ref{fig:manchester_constraints}. 
   %On each of four panels 
   %the spectra are shown by the solid lines. Single temperature 
   %black-body fits are drawn in grey dashed lines. The names 
   %of the objects are shown in the upper-central part of each panel 
   %of the Figure. 
   \textbf{The top panel} of 
   Figure~\ref{fig:manchester_constraints} shows the 
   case in which the shape of the 16-22\,$\mu$m 
   continuum causes the wrong determination of the 
   T$_{\rm d}$. There are 11 such objects of which ten 
   are AGB stars (all in the LMC), and the remaining 
   star is a post-AGB object without the 21\,$\mu$m 
   feature (IRAS~00350-7436 from the SMC). 
   \textbf{The second panel} of 
   Figure~\ref{fig:manchester_constraints} presents an 
   example of an object for which the obtained value of 
   T$_{\rm d}$ is too high. For five objects we observe 
   such a problem (one in the SMC, three in the LMC, 
   and one in the Milky Way). \textbf{The third panel} 
   of Figure~\ref{fig:manchester_constraints} shows 
   the problem of modelling the continuum for some 
   post-AGB objects. The explanation of this behaviour 
   may lie in the fact that one of the ranges used for 
   obtaining T$_{\rm d}$ is taken from the region 
   between the 21 and 30\,$\mu$m features. When the 
   21\,$\mu$m feature is strong, this region might be 
   bumped up, and causes too low the T$_{\rm d}$ 
   value. There are 12 such objects (one in the LMC 
   and 11 in the Milky Way). \textbf{The bottom panel} 
   shows that in some cases the Manchester Method 
   may lead to the reduction of the obtained flux of the 
   30\,$\mu$m feature. We notice this for three objects 
   (all in the LMC): MSX~LMC~1400, 
   2MASS~J05102834-6844313 which serves as the 
   example, and IRAS~05568-6753. In the last object 
   the continuum bisects the profile of the 30\,$\mu$m 
   feature.
   
   Two post-AGB objects with bad fits are not in these 
   four categories (both from the group I). In the Spitzer 
   spectrum of the first object, IRAS~19477+2401, the 
   SH part of the spectrum is very noisy, and a big jump 
   between the SH and LH segments is visible. These 
   two facts cause that the determination of the 
   continuum is uncertain. In a case of 
   IRAS~15531-5704, the 20 -- 23\,$\mu$m region of 
   spectrum where the continuum is fitted, contains a 
   curvy bump. This results in an overprediction of the 
   continuum.

%*************************************************************************************************************************
%*************************************************************************************************************************
%*************************************************************************************************************************
%*************************************************************************************************************************

   We excluded 17 additional PNe from the analysis, 
   because of the presence of an unidentified broad 
   emission feature between $\sim$16--24\,$\mu$m (in 
   some cases this feature can extends beyond 
   24\,$\mu$m), which affects the values of the 
   [16.5]$-$[21.5] colour. In these cases we only have 
   reported the values of the [5.8]$-$[9.3] colour. Six 
   out of these 17 objects (SMP~SMC~1, 6, and 
   SMP~LMC~8, 58, 76, 78) have been already 
   referred to by \citet{Bernard-Salas:2009aa} as 
   `Bump16' sources with a broad excess between 
   16-22\,$\mu$m. \citet{Garcia-Hernandez:2012qy} 
   found this feature in another six MCs 
   PNe that are in our sample (SMP~SMC~13, 15, 18, 
   20, and SMP~LMC~25 and 48). The carrier of this 
   unusual dust feature remains unclear. 
   \citet{Otsuka:2015aa} reported the presence of this 
   feature in the Sgr dSph planetary nebula 
   Wray-16-423, and called it the broad 
   `16-24\,$\mu$m feature'. In our work we follow their 
   nomenclature. In addition, the authors also 
   confirmed the presence of the 16-24\,$\mu$m 
   feature in SMP~LMC~19. In our Galaxy, 
   \citet{Otsuka:2013aa} found this feature in the 
   fullerene-containing planetary nebula PN~M1-11 
   and PN~M1-12. Nevertheless, the presence of this 
   feature in the spectrum of PN~M1-12 is not obvious 
   to us, thus we have not removed it from the further 
   analysis in our work. Moreover, the presence of this 
   spectral feature is indicated by the question mark 
   in Table~\ref{app_tab:gal_basic_info}. During our 
   analysis, we made the first detections of the 
   16-24\,$\mu$m feature in the Spitzer spectrum of 
   SMP~LMC~51 and 79. 
   % Three objects exhibiting the 16-24\,$\mu$m feature 
   % (SMP~SMC~1, SMP~LMC~25, and PN~M1-11) are also the 
   % sources with the fullerenes. 
   The spectra of all the PNe showing the 
   16-24\,$\mu$m feature are presented in 
   Figure~\ref{fig:pne_broad_emission}.
   %by the solid lines. By the dashed lines we 
   %marked the region in which this feature is 
   %visible (16 and 24\,$\mu$m). 
   Their names are also given in bold face in 
   \Cref{tab:sgr_basic_info,tab:sgr_spectral_results,%
   app_tab:smc_basic_info,app_tab:lmc_basic_info,%
   app_tab:gal_basic_info,app_tab:smc_spectral_results,%
   app_tab:lmc_spectral_results,app_tab:gal_spectral_results}.
   
\begin{figure}
   \centering
   \includegraphics[width=\hsize]{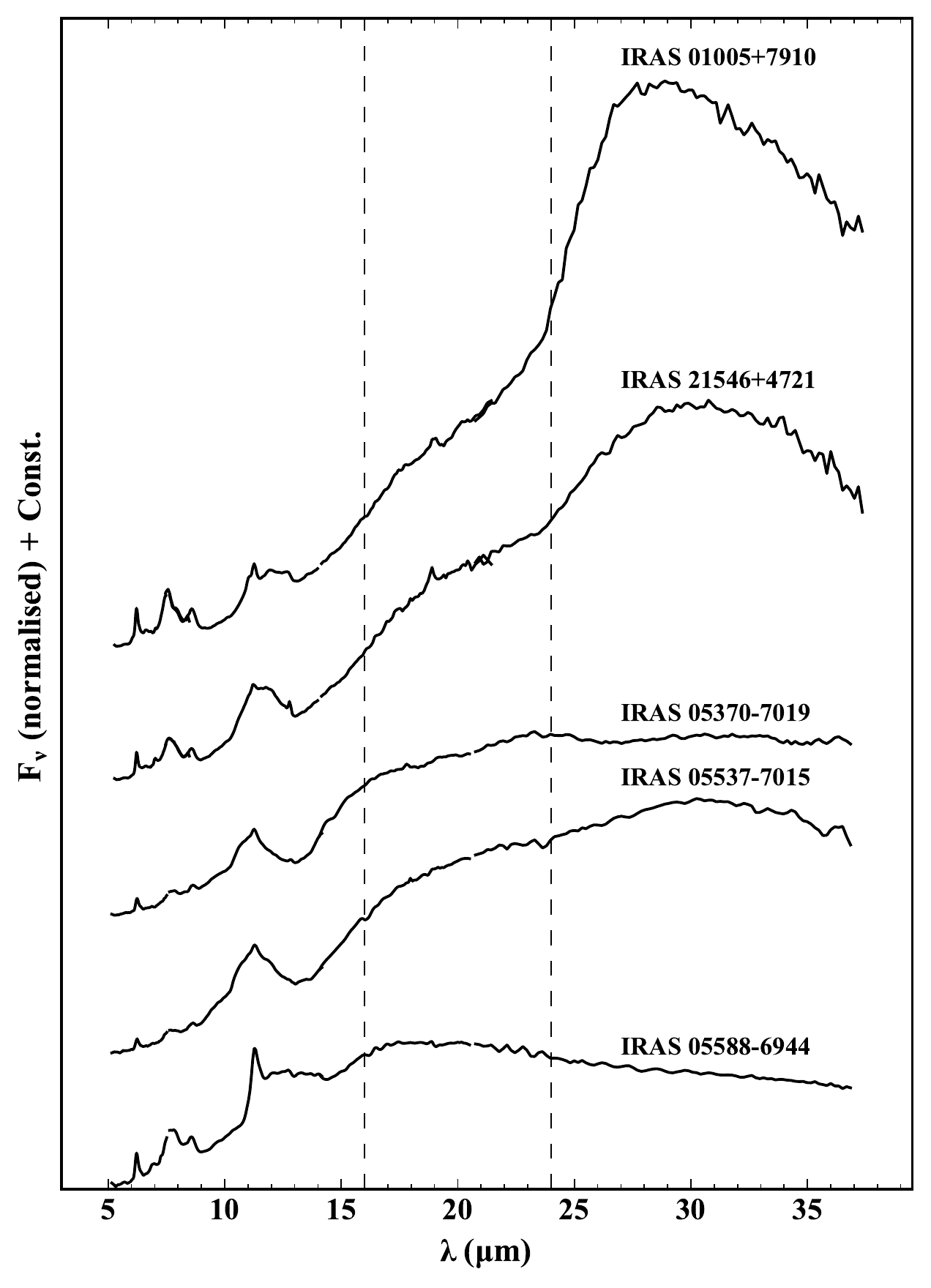}
      \caption{Post-AGB objects excluded from analysed 
      sample, because of the broad emission feature 
      presence, which is typically visible around 
      16--24\,$\mu$m. The spectra (solid lines) are 
      normalised to the flux density at 16\,$\mu$m and 
      offset for clarity. The names of the objects are 
      shown above the spectra.}
         \label{fig:pagb_broad_emission}
   \end{figure}

   The 16-24\,$\mu$m feature has been discovered in 
   the carbon-rich post-AGB objects as well. The first 
   detection was made by \citet{Zhang:2010aa} in the 
   Spitzer spectrum of IRAS~01005+7910. 
   \citet{Matsuura:2014fu} found this feature in the 
   LMC post-AGB object IRAS~05588-6944. We also 
   noticed very similar feature in the Spitzer spectra of 
   IRAS~05370-7019, IRAS~05537-7015, and 
   IRAS~21546+4721 for the first time. They are 
   analysed by \cite{Sloan:2014fj} and included to the 
   group called `big-11', because they show a strong 
   11.3\,$\mu$m feature. We excluded these five 
   post-AGB objects from the further analysis, and 
   marked them also by the bold face names in 
   \Cref{tab:sgr_basic_info,tab:sgr_spectral_results,%
   app_tab:smc_basic_info,app_tab:lmc_basic_info,%
   app_tab:gal_basic_info,app_tab:smc_spectral_results,%
   app_tab:lmc_spectral_results,app_tab:gal_spectral_results} 
   as in a case of the PNe with the 16-24\,$\mu$m 
   feature. The spectra of these post-AGBs are shown 
   in Figure~\ref{fig:pagb_broad_emission}. 
   %By the dashed lines we marked the region in 
   %which this feature is usually visible (between 
   %16 and 24\,$\mu$m).

%*************************************************************************************************************************
%*************************************************************************************************************************
%*************************************************************************************************************************
%*************************************************************************************************************************

%*************************************************************************************************************************
%*************************************************************************************************************************
%*************************************************************************************************************************
%*************************************************************************************************************************

%-------------------------------------------------------------
% -- THE STRENGTH AND CENTRAL WAV. ---
%-------------------------------------------------------------
\subsection{Strength and central wavelength of the 30\,$\mu$m feature}
\label{subsec:strength_central_wav}

   The 30\,$\mu$m band is present between 24 and 
   45\,$\mu$m \citep{Otsuka:2014eu}. The long 
   wavelength side of this feature is beyond the 
   end of the Spitzer spectral coverage, thus it is not 
   possible to estimate the full emission of the feature. 
   Therefore, we extrapolated the black-body 
   continuum on the short wavelength side of the 
   30\,$\mu$m feature only. We measured the strength 
   of the feature F/Cont (30\,$\mu$m), which is a ratio 
   of the integrated flux from 24 to 36\,$\mu$m above 
   the continuum, divided by the integrated underlying 
   continuum for the same spectral range. We applied 
   the same range of the integration for the AGB 
   objects as well as post-AGB stars and PNe to be in 
   agreement with \citet{Sloan:2016aa} results. For the 
   PNe we removed the forbidden emission lines of 
   [Ne V] at 24.32\,$\mu$m, [O IV] at 25.89\,$\mu$m, 
   [S III] at 33.48\,$\mu$m, and [Ne III] at 
   36.01\,$\mu$m, when they were visible, before 
   measuring the strength of the 30\,$\mu$m feature. 
   In addition, we determined the central wavelength 
   ($\lambda_{\rm c}$) of the 30\,$\mu$m feature. 
   This central wavelength is defined as the 
   wavelength at which (after the subtraction of the 
   continuum), the flux on the both sides is equal.

   Table~\ref{tab:sgr_spectral_results} lists the results 
   of the spectroscopic analysis for the objects in the 
   Sgr dSph. Appendix~\ref{sec:appendix_tables} 
   presents the 
   \Cref{app_tab:smc_spectral_results,%
   app_tab:lmc_spectral_results,app_tab:gal_spectral_results} 
   with the spectroscopic results for the SMC, LMC, 
   and Milky Way. In each of these Tables we present 
   the names of the objects, the colours derived from 
   the Manchester method and its modifications 
   ([5.8]$-$[9.3], [6.4]$-$[9.3], [16.5]$-$[21.5], 
   [18.4]$-$[22.45], [18.4]$-$[22.75], and 
   [17.95]$-$[23.2]) as well as the derived dust 
   temperatures (T$_{\rm d}$). 
   Table~\ref{tab:sgr_spectral_results} contains no 
   post-AGB objects, but we keep the columns with 
   the colours intended for those objects with an aim 
   of providing the same structure in all of the Tables. 
   Finally, we present the calculated central 
   wavelength ($\lambda_{\rm c}$), and the strength 
   of the 30\,$\mu$m feature, F/Cont. 
   Usually, the profile of the 30\,$\mu$m feature rises 
   starting from 24\,$\mu$m, and after reaching the 
   maximum it turns down at the long-wavelength 
   cut-off. If this decline is not visible in the spectrum, 
   then we assumed that the 30\,$\mu$m feature is 
   somehow contaminated or not present in the 
   spectrum, and we do not treat such an object as the 
   true 30\,$\mu$m source. Basically, we repeated the 
   values of the [6.4]$-$[9.3] colour, [16.5]$-$[21.5] 
   colour, and F/Cont together with their errors for the 
   carbon-rich AGB stars in the SMC and LMC from 
   \citet{Sloan:2016aa} (see their Table 5 in the online 
   content). However, Table~\ref{tab:sgr_spectral_results} 
   and \Cref{app_tab:smc_spectral_results,%
   app_tab:lmc_spectral_results,app_tab:gal_spectral_results} 
   were supplemented by the additional parameters 
   of the 30\,$\mu$m feature like the T$_{\rm d}$ and 
   $\lambda_{\rm c}$. On the other hand, in a few 
   cases the results given by \citet{Sloan:2016aa} 
   seemed to be unrealistic, and we have not include 
   them in our Tables with the results. Below we 
   present the arguments for these changes.
   
   We have not shown the results of the F/Cont 
   (30\,$\mu$m) for two objects from their SMC list: 
   MSX~SMC~036 and MSX~SMC~054. The values 
   of the T$_{\rm d}$ and $\lambda_{\rm c}$ (additional 
   parameters given by us) have not been shown for 
   them as well. In a case of MSX~SMC~036, the 
   authors reported the value of the F/Cont about 0.3. 
   We decided to remove it from our sample on the 
   grounds that the profile of the feature was not turning 
   down at the long-wavelength Spitzer cut-off in the 
   both used databases of Spitzer spectra. In a case of 
   the second object, MSX~SMC~054, the authors 
   reported the value of F/Cont $\sim$0.2, but the 
   maximum of the continuum was above the Spitzer 
   spectrum, which indicated that the obtained 
   T$_{\rm d}$ is too high (bad fit of continuum). We 
   marked this object in bold face in 
   \Cref{app_tab:smc_basic_info,app_tab:smc_spectral_results}. 
   The example of spectrum with such the problem is 
   shown in the second panel (from the top) of 
   Figure~\ref{fig:manchester_constraints}. 
   
   We find the differences in the profiles of the 
   30\,$\mu$m feature in the two databases of spectra 
   we have used (Spitzer \citep{Sloan:2016aa} and the 
   CASSIS) for three LMC objects: OGLE~J051306, 
   MSX~LMC~782, and MSX~LMC~787, as follows. 
   For the first object, OGLE~J051306, the authors 
   reported the value of the F/Cont $\sim$0.7, but the 
   profile of the feature was not turning down at the 
   end of the spectrum (like in a case of 
   MSX~SMC~036 in the SMC). On the other hand, 
   the CASSIS spectrum of this object has not shown 
   any rise behind 24\,$\mu$m, indicating that we were 
   not dealing with the true 30\,$\mu$m source. In the 
   end, this object was removed from the analysed 
   sample.
   
\begin{figure}
   \centering
   \includegraphics[width=\hsize]{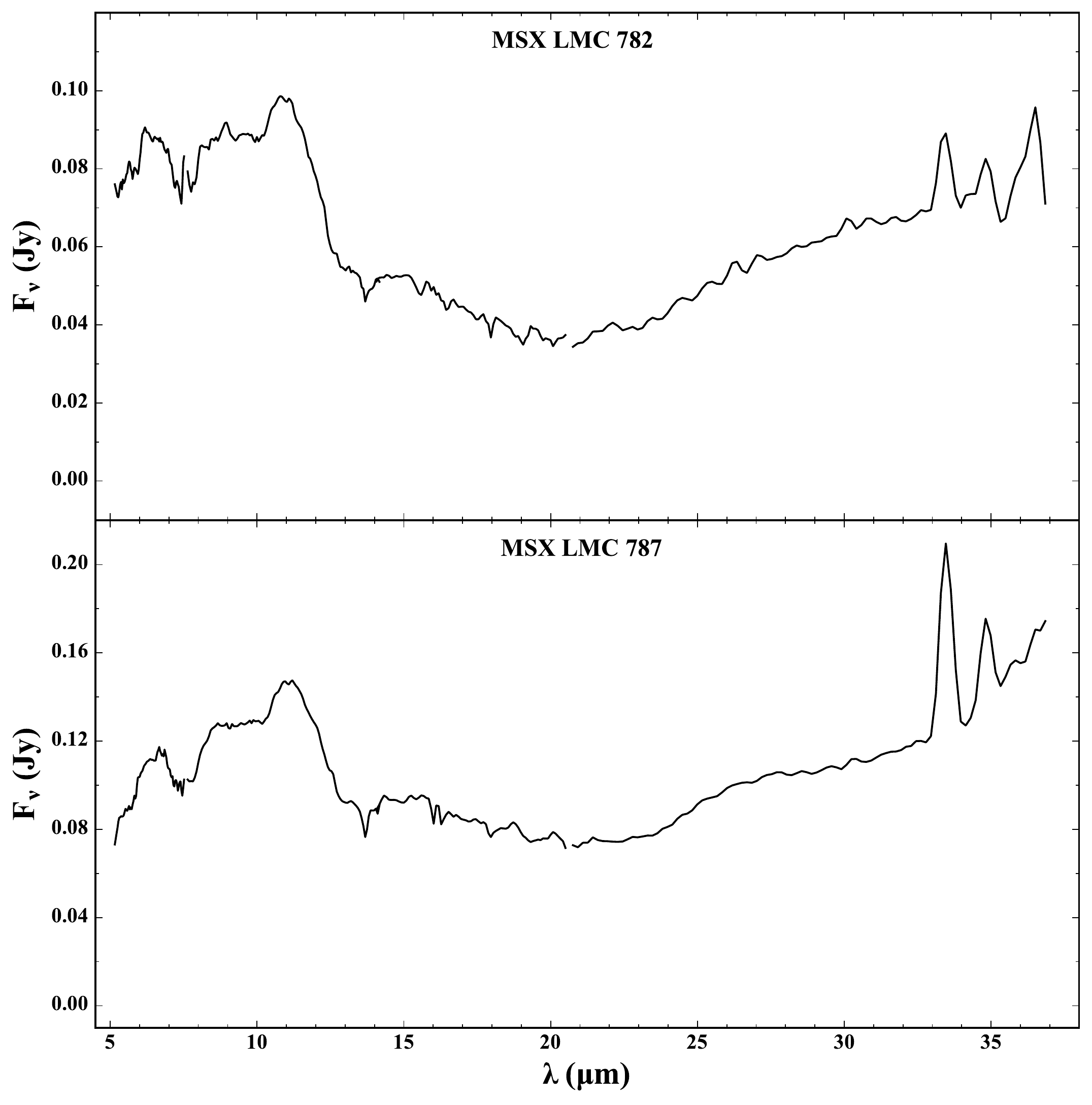}
      \caption{Spectra of MSX~LMC~782 (top), and 
      MSX~LMC~787 (bottom), showing the problem 
      with calibration. The Spitzer spectra are represented 
      by solid lines.}
         \label{fig:bad_calib_ex}
   \end{figure}

   The situation of MSX~LMC~782, MSX~LMC~787 
   was more difficult. Their Spitzer spectra from 
   \citet{Sloan:2016aa} are shown by the solid lines in 
   Figure~\ref{fig:bad_calib_ex}. The top panel 
   contains the spectrum of MSX~LMC~782, whereas 
   in the bottom panel we show the spectrum of 
   MSX~LMC~787. The authors reported the values 
   of the F/Cont of about 1.4 for MSX~LMC~782 and 
   1.1 for MSX~LMC~787. The spectra from 
   \citet{Sloan:2016aa} have shown the steep rise 
   starting from around 23\,$\mu$m, and no decline up 
   to the end of the spectra. From this point of view, 
   these objects should have been removed from the 
   analysed sample. However, the CASSIS spectra 
   have shown a clear decline, therefore, we have kept 
   them as the true 30\,$\mu$m sources. These spectra 
   are clearly affected by nearby background emission. 
   The extraction method for the CASSIS spectra was 
   more effective in removing that background than 
   our method, because it added the step of fitting the 
   background and separating it from the extracted 
   spectrum of the source. Because of this problem 
   the values of the $\lambda_{\rm c}$ and F/Cont 
   were not shown by us in 
   Table~\ref{app_tab:lmc_spectral_results}, but we 
   kept the T$_{\rm d}$. 
   
   In addition, we did not show results for another 18 
   objects from the \citet{Sloan:2016aa} LMC list (their 
   names are given in bold face in 
   \Cref{app_tab:lmc_basic_info,app_tab:lmc_spectral_results}, 
   except for one object), whereas in three cases the 
   different values have been reported by us 
   (IRAS~05416-6906, MSX LMC 494, and 
   IRAS~05026-6809). Below we describe the reasons 
   of these changes.
   
   The Spitzer spectrum of MSX~LMC~783 has not 
   shown any signs of decline behind 24\,$\mu$m. 
   The CASSIS spectrum has had the same trend, 
   thus we removed it from our sample. The authors 
   have reported the value of the F/Cont of about 0.3. 
   
   In a case of MSX~LMC~974 the weak 30\,$\mu$m 
   feature has been observed, whereas the authors 
   reported the value of the F/Cont $\sim$0.2. This 
   value is overestimated, just as value of 
   $\lambda_{\rm c}$ (> 33\,$\mu$m), because we 
   noticed that the Spitzer spectrum is very noisy over 
   32\,$\mu$m, giving the additional emission to the 
   end of spectrum. We report only the T$_{\rm d}$ 
   value for this object. 
   
   In the Spitzer spectra of ten LMC objects 
   (IRAS~04286-6937, IRAS~05010-6739, 
   IRAS~05103-6959, IRAS~05107-6953, 
   MSX~LMC~220, IRAS~05446-6945, and 
   LI-LMC~1758, IRAS~04374-6831, 
   IRAS~04473-6829, MSX~LMC~218), the 
   negative (or positive but close to 0) values of the 
   F/Cont were observed. The closer look into their 
   Spitzer spectra has shown they have the 
   30\,$\mu$m feature, but the continuum obtained 
   for them is not correct. It is caused by the curved 
   shape of the spectrum in the region from which the 
   Manchester colours were taken to determine the 
   T$_{\rm d}$. The exemplary spectrum showing this 
   kind of the bad continuum fit is shown in the top 
   panel of Figure~\ref{fig:manchester_constraints} 
   for IRAS~04286-6937. 
   
   In the Spitzer spectra of MSX~LMC~1400, 
   2MASS~J05102834, and IRAS~05568-6753 we 
   noticed the reduction of the obtained flux of the 
   30\,$\mu$m feature. The maximum of the 
   black-body continuum in IRAS~F04537-6509, 
   IRAS~04567-6857, and MSX~LMC~92 was above 
   the spectrum, which indicated that the value of the 
   T$_{\rm d}$ was overestimated (like for 
   MSX~SMC~054 in the SMC; see above).
   
   In the Spitzer spectrum of carbon-rich AGB object 
   IRAS~05416-6906, the [Ar III] at 21.83\,$\mu$m 
   and [S III] lines at 18.71/33.48\,$\mu$m were 
   observed. The classification of this object as 
   carbon-rich AGB star, was made by 
   \cite{Woods:2011aa}. \citet{Neugent:2012aa} 
   described IRAS~05416-6906 as the yellow 
   supergiant (YSG) with the spectral type of B2.5Ia, 
   which was determined by \citet{Fitzpatrick:1991aa}. 
   The presence of this emission lines may suggest 
   that this is a binary carbon-rich AGB star with a hot 
   companion. We deleted these lines before 
   measuring the value of the F/Cont. After removal 
   of the nebular lines we should have expected the 
   increase of the F/Cont. Meanwhile, the value of the 
   F/Cont, which has been obtained by us is about 
   6\% larger than the value reported by 
   \citet{Sloan:2016aa}. This was caused by the fact 
   that the value of the [16.5]$-$[21.5] colour has 
   changed after the removal of [Ar III] line, giving the 
   bigger value of the F/Cont.
   % IRAS 05416-6906 (SSID167). This object is variously classed 
   % as YSO (Whitney et al. 2008; Vijh et al. 2009), 
   % H ii region (Egan et al. 2001), and extreme AGB star 
   % (Srinivasan et al. 2009). We classify it as C-AGB.
   
   Finally, we decided to add two objects for which the 
   values of the F/Cont were not given by 
   \citet{Sloan:2016aa}: MSX~LMC~494 and 
   IRAS~05026-6809. This was made on the grounds 
   that the 30\,$\mu$m feature is well visible in their 
   Spitzer spectra, and the decline in the spectrum was 
   observed -- also in the CASSIS spectra.

%*************************************************************************************************************************
%*************************************************************************************************************************
%*************************************************************************************************************************
%*************************************************************************************************************************

%*************************************************************************************************************************
%*************************************************************************************************************************
\begin{table*}
\scriptsize
\caption{Spectroscopic results for the Sgr dSph objects: 
names, six colours, dust temperature (T$_{\rm d}$), 
central wavelength ($\lambda_{\rm c}$), and strength 
of the 30\,$\mu$m feature (F/Cont).}
\label{tab:sgr_spectral_results}
\centering
\begin{tabular}{lccccccccc} % 9 columns
\hline\hline
Target			&		[5.8]$-$[9.3]			&		[6.4]$-$[9.3]			&		[16.5]$-$[21.5]			&	[18.4]$-$[22.45]	&	[18.4]$-$[22.75]			&	[17.95]$-$[23.2]			&		T$_{\rm d}$			&	$\lambda_{\rm c}$ (30\,$\mu$m)	&	F/Cont (30\,$\mu$m)\\
				&			(mag)			&			(mag)			&			(mag)			&		(mag)	&		(mag)			&		(mag)			&			(K)				&			($\mu$m)				&\\
\hline
Sgr~3			&			\ldots			&	0.554	$\pm$	0.006	&	0.327	$\pm$	0.012	&		\ldots	&	\ldots	&	\ldots	&	436		$\pm$	13	&	28.978	$\pm$	0.111	&	0.390	$\pm$	0.014	\\
Sgr~7			&			\ldots			&	0.763	$\pm$	0.004	&	0.209	$\pm$	0.012	&		\ldots	&	\ldots	&	\ldots	&	645		$\pm$	31	&	29.326	$\pm$	0.228	&	0.195	$\pm$	0.014	\\
Sgr~15			&			\ldots			&	0.505	$\pm$	0.005	&	0.201	$\pm$	0.013	&		\ldots	&	\ldots	&	\ldots	&	669		$\pm$	36	&	29.338	$\pm$	0.194	&	0.165	$\pm$	0.015	\\
Sgr~18			&			\ldots			&	0.818	$\pm$	0.003	&	0.208	$\pm$	0.011	&		\ldots	&	\ldots	&	\ldots	&	650		$\pm$	28	&	29.175	$\pm$	0.136	&	0.243	$\pm$	0.013	\\      
\textbf{Wray16-423}	&	2.287	$\pm$	0.086	&			\ldots			&			\ldots			&		\ldots	&	\ldots	&	\ldots	&			\ldots		&			\ldots			&			\ldots	\\
\hline
\end{tabular}
\tablefoot{The tables with the Galactic and MCs objects 
are available in Appendix~\ref{sec:appendix_tables}.}
\end{table*}
%
%*************************************************************************************************************************
%*************************************************************************************************************************

%-------------------------------------------------------------
%-------------------------------------------------------------
% ---------- RESULTS AND SUMMARY -----------
%-------------------------------------------------------------
%-------------------------------------------------------------
\section{Results and summary}
\label{sec:results_summary}

%-------------------------------------------------------------
% -- RELATIONS BETWEEN PARAMETERS --
%-------------------------------------------------------------
\subsection{Relations between obtained parameters}
\label{subsec:relations_of_parameters}

\begin{figure}
   \centering
   \includegraphics[width=\hsize]{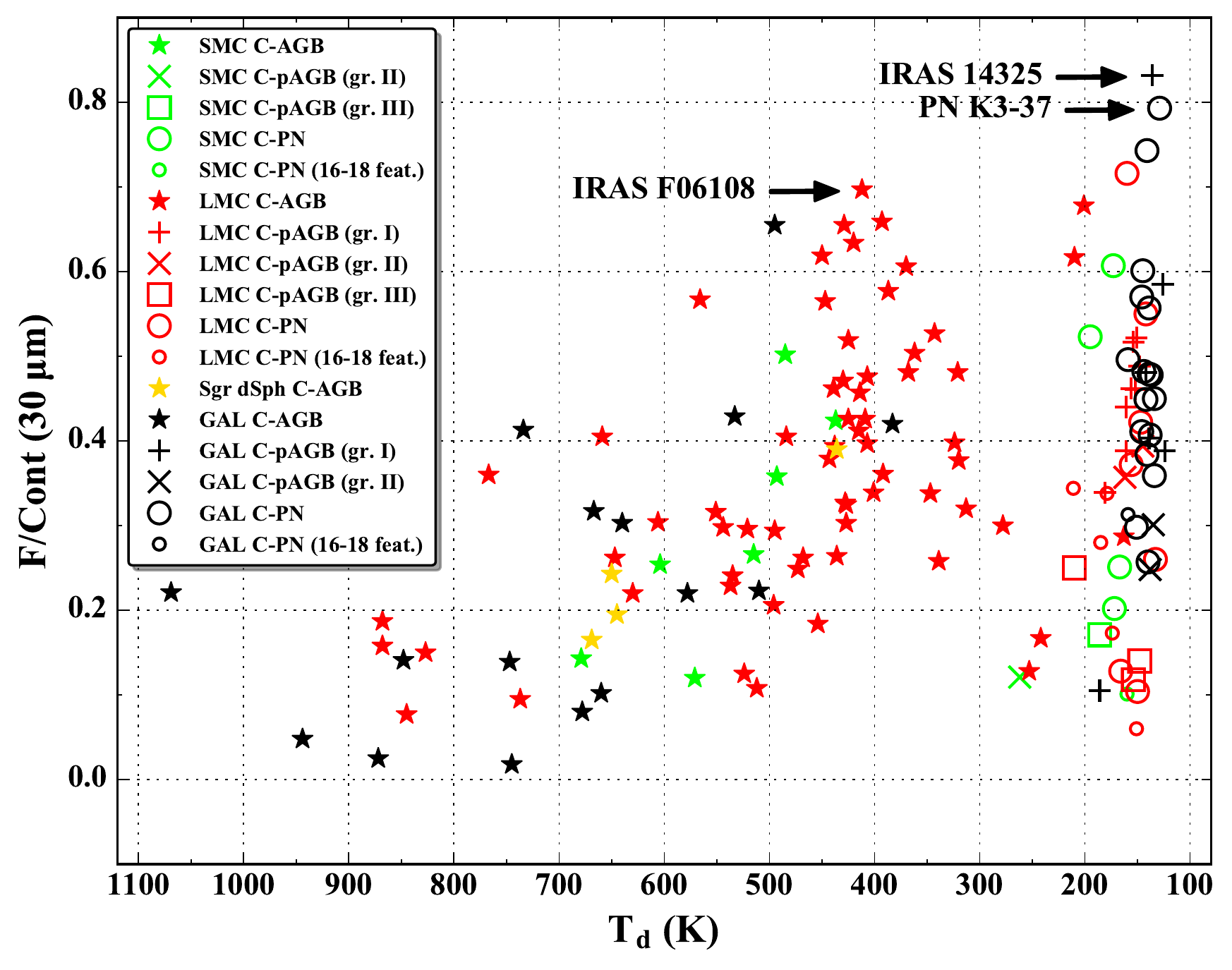}
      \caption{Strength of the feature as a function of the 
      dust temperature.}
         \label{fig:strength_t}
   \end{figure}

   Altogether, we are able to determine all 
   parameters of the 30\,$\mu$m feature for 42 
   Galactic objects (16 AGB stars, six post-AGB stars 
   in group I, two post-AGBs in gr. II, and 18 PNe); 
   89 LMC objects (64 AGB stars, eight post-AGB stars 
   in gr. I, two post-AGBs in gr. II, three post-AGBs in 
   gr. III, and 12 PNe); 14 SMC objects (seven AGB 
   stars, one post-AGB star in gr. II and III, and five 
   PNe); four Sgr dSph objects (AGB stars). In 
   Figure~\ref{fig:strength_t} we show the strength of 
   the 30\,$\mu$m feature as a function of the 
   T$_{\rm d}$. The horizontal axis is presented 
   reversely to show the evolution of C-rich objects 
   from the AGBs up to PNe. 
   %The explanation of the symbols on the graph is 
   %given in the legend, which is located in the 
   %upper-left corner. Generally, the different colours 
   %distinguish each of the analysed galaxies 
   %(lime -- SMC, red -- LMC, gold -- Sgr dSph, and 
   %black -- Milky Way). The stars represent the AGB 
   %stars, whereas the pluses, crosses and open squares 
   %mark the post-AGB objects from the group I to III, 
   %respectively. The open circles represent the PNe. 
   We also distinguish the group of the PNe with the 
   16-18\,$\mu$m feature, which has been discussed 
   in Section~\ref{subsec:manchester_method}. 
   These PNe are represented as small open circles. 
   
   The strength of the feature clearly increases as 
   T$_{\rm d}$ decreases to about 400\,K. After that, 
   the large mass-loss rate and probably the resulting 
   self-absorption reduces the strength of the 
   30\,$\mu$m feature. During the post-AGB phase, 
   when the intense mass-loss has terminated, the 
   optical depth of the circumstellar envelope is 
   smaller, and the feature is visible again, becoming 
   comparable in the strength to the strongest feature in 
   AGB stars with T$_{\rm d}$ between 300 and about 
   500\,K. In our sample AGB stars in this region of 
   T$_{\rm d}$ are dominated by the LMC objects. 
   However, the Galactic sample from 
   \cite{Sloan:2016aa}, which is based on the ISO 
   spectra, extends much further in the F/Cont than 
   our Galactic sample (see their Fig.~6). Their three 
   Galactic objects have F/Cont between 0.7 and 0.9. 
   Without unrealistic results for two LMC objects 
   discussed in 
   Section~\ref{subsec:strength_central_wav} 
   (MSX~LMC~782 and MSX~LMC~787), their 
   Galactic sources compose a group with the 
   strongest 30\,$\mu$m feature. The strongest 
   30\,$\mu$m feature in our sample was derived for 
   IRAS~F06108-7045 in case of AGB stars, 
   IRAS~14325-6428 among the post-AGB objects, 
   and PN~K3-37 in PNe. These objects are marked 
   by the black arrows, and abbreviations of their 
   names in Figure~\ref{fig:strength_t}.

%*************************************************************************************************************************
%*************************************************************************************************************************
%*************************************************************************************************************************
%*************************************************************************************************************************

\begin{figure}
   \centering
   \includegraphics[width=\hsize]{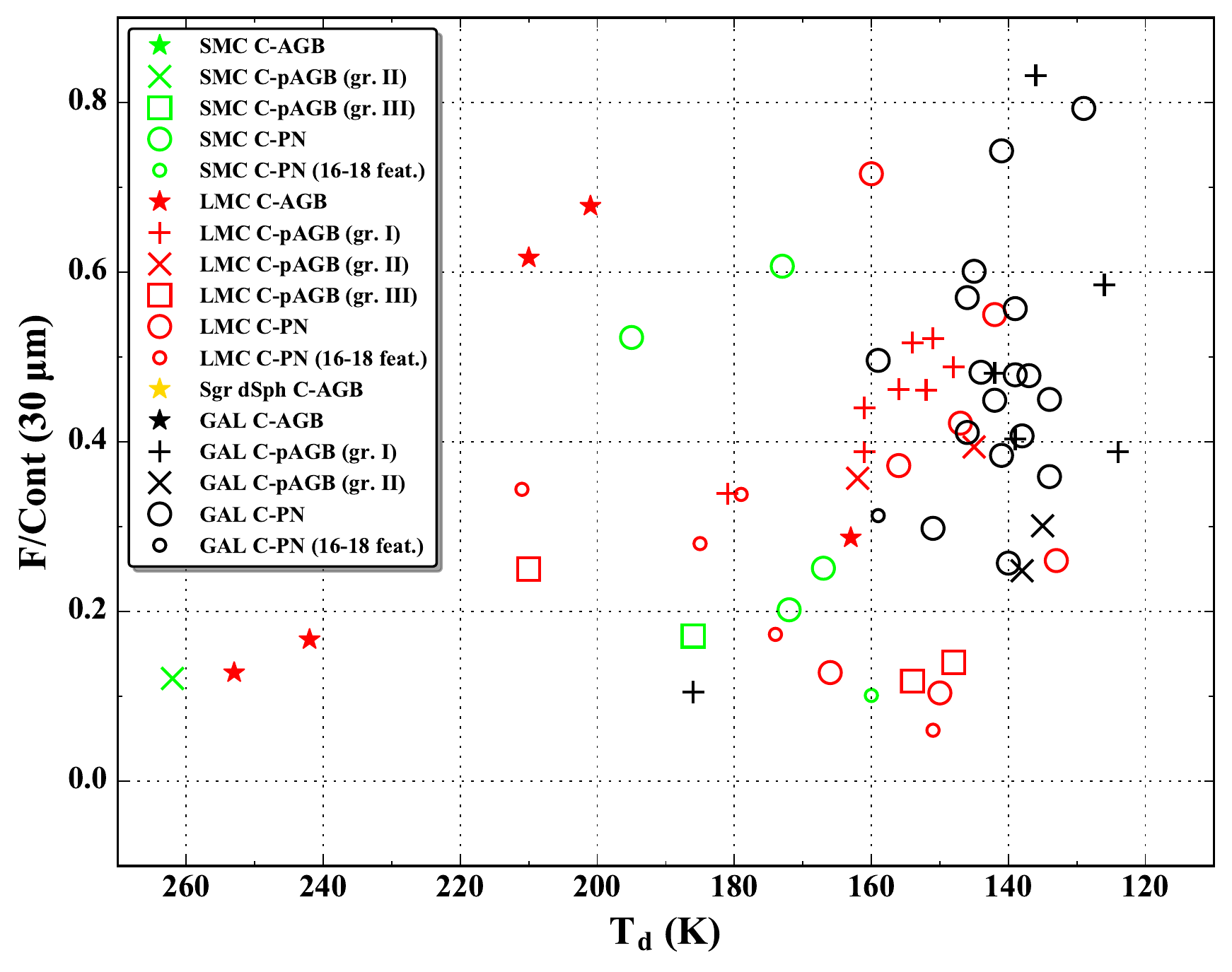}
      \caption{Strength of the feature as a function of the 
      dust temperature -- zoom for the region of the 
      objects with the coldest dust.}
         \label{fig:strength_t_zoom}
   \end{figure}

   In Figure~\ref{fig:strength_t_zoom} we present the 
   strength of the 30\,$\mu$m feature as the function 
   of the T$_{\rm d}$ between 110 and 270\,K, to show 
   the post-AGB objects and PNe more clearly. The 
   symbols on the graph are the same as in 
   Figure~\ref{fig:strength_t}. Despite the lack of 
   correlation between the strength of the feature and 
   T$_{\rm d}$, the Galactic post-AGB objects and 
   PNe seems to have a slightly smaller T$_{\rm d}$ 
   and larger strength of the feature in comparison 
   with their counterparts from other galaxies. There 
   are five carbon-rich AGB objects with 
   260\,K > T$_{\rm d}$ > 160\,K -- all of them are 
   the LMC members.

%*************************************************************************************************************************
%*************************************************************************************************************************
%*************************************************************************************************************************
%*************************************************************************************************************************

\begin{figure}
   \centering
   \includegraphics[width=\hsize]{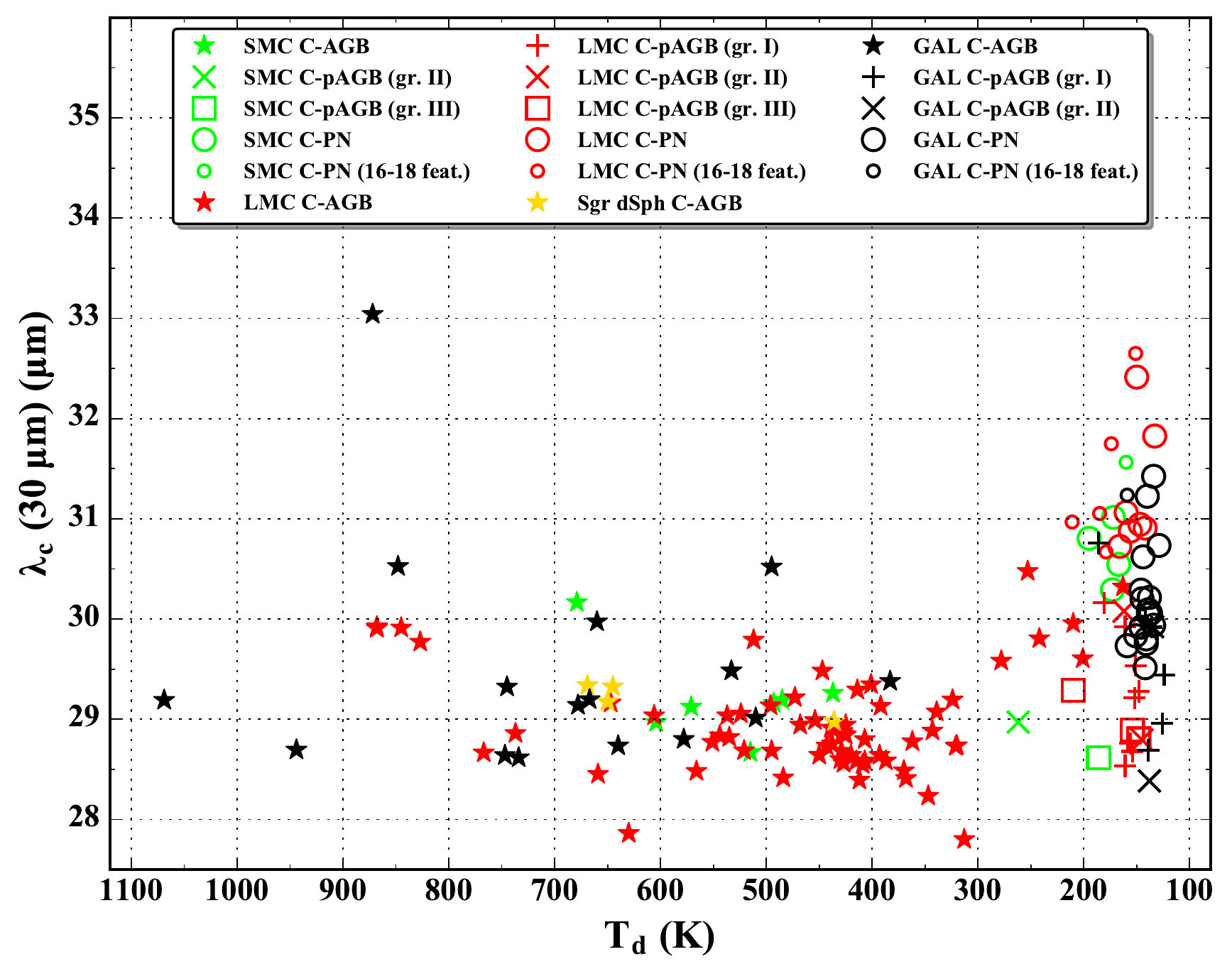}
      \caption{Central wavelength as a function of the 
      dust temperature.}
         \label{fig:central_t}
   \end{figure}

\begin{figure}
   \centering
   \includegraphics[width=\hsize]{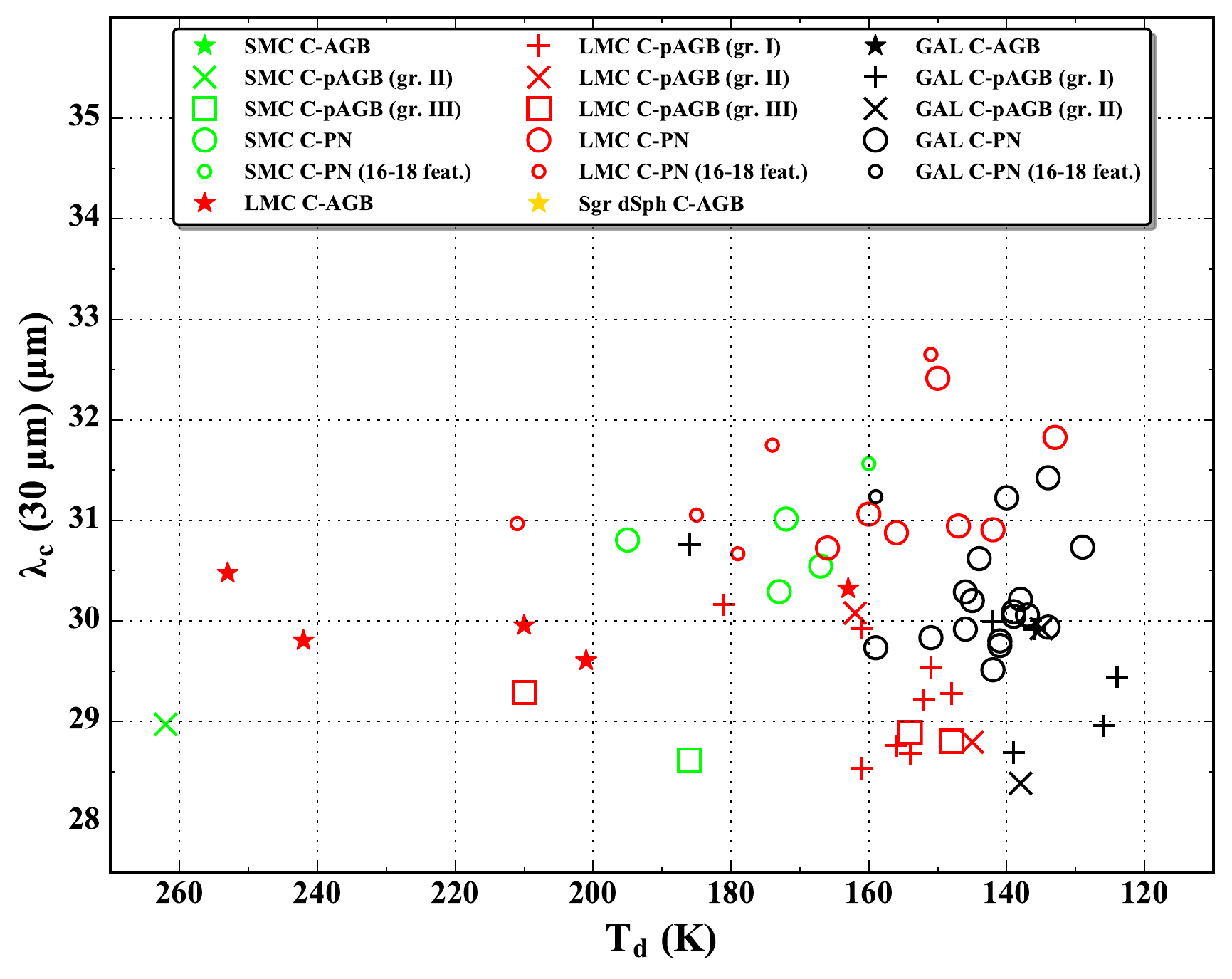}
      \caption{Central wavelength as a function of the 
      dust temperature -- zoom for the region containing 
      objects with the coldest dust.}
         \label{fig:central_t_zoom}
   \end{figure}

   Figure~\ref{fig:central_t} shows the 
   $\lambda_{\rm c}$ of the 30\,$\mu$m feature 
   as a function of the T$_{\rm d}$. The horizontal 
   axis is reversed to show the evolutionary sequence 
   from AGB stars, through post-AGBs up to PNe, and 
   the symbols on the graph are the same as in 
   Figure~\ref{fig:strength_t}. The $\lambda_{\rm c}$ 
   seems to be independent with the changing 
   T$_{\rm d}$. The majority of the carbon-rich AGB 
   objects occupy the region of the $\lambda_{\rm c}$ 
   between about 28.5 and 29.5\,$\mu$m. However, 
   some objects stick out from the main group. There 
   are two objects (MSX~LMC~950 and 
   IRAS~06025-6712 -- this one has the lower 
   T$_{\rm d}$), which are located in the region, where 
   the $\lambda_{\rm c}$ < 28\,$\mu$m. In the spectra 
   of these objects, the fitted continua are a little too 
   high at the end of the spectra (in comparison with 
   the other AGB objects), and thus result in the lower 
   value of the $\lambda_{\rm c}$. Another interesting 
   group of six LMC carbon-rich AGB objects occupies 
   the region between 280\,K > T$_{\rm d}$ > 160\,K and 
   30.5\,$\mu$m > $\lambda_{\rm c}$ > 29.5\,$\mu$m 
   (see Figure~\ref{fig:central_t_zoom} for more detailed 
   view). The obtained by us T$_{\rm d}$ for those 
   objects shows that their envelopes are cool. Four out 
   of these six objects were classified by 
   \citet{Gruendl:2008aa} as `extremely red objects' 
   (EROs). According to the authors, this class of 
   objects is mainly characterised by extremely red 
   mid-IR colours ([4.5] $-$ [8.0] > 4.0), and none of 
   them have counterparts in the 2MASS catalogue. 
   The mass-loss rates obtained by 
   \citet{Gruendl:2008aa} for these four objects are 
   between $8.9\times 10^{-5}$ and $2.3\times 10^{-4}\,{\rm M}_{\odot}\,{\rm yr}^{-1}$, 
   which shows that they experience an intense 
   mass-loss process. In 
   Table~\ref{tab:central_wav_EROs} we list these 
   objects in order of the decreasing T$_{\rm d}$. First, 
   the names of objects are presented, and then the 
   T$_{\rm d}$ and $\lambda_{\rm c}$ from 
   Table~\ref{app_tab:lmc_spectral_results} are shown. 
   After that the values of the mass-loss rates with the 
   corresponding reference are given.
   
   There are ten carbon-rich AGB sources in the range 
   of 900\,K > T$_{\rm d}$ > 300\,K and 
   $\lambda_{\rm c}$ > 29.7\,$\mu$m, which clearly 
   stick out from the main group of the carbon-rich AGB 
   stars. These objects are listed in 
   Table~\ref{tab:list_lambda>29.7} in order of the 
   decreasing T$_{\rm d}$. In the first column we 
   present the names of the objects, and then the 
   information about the host galaxy. After that the 
   values of the T$_{\rm d}$, $\lambda_{\rm c}$, and 
   F/Cont from 
   \Cref{app_tab:smc_spectral_results,%
   app_tab:lmc_spectral_results,app_tab:gal_spectral_results} 
   are repeated. In general, the 30\,$\mu$m feature in 
   these objects is weak, and then the 
   $\lambda_{\rm c}$ might be not certain and shifted 
   towards the longer wavelengths. The object showing 
   the largest $\lambda_{\rm c}$ (> 33\,$\mu$m) in the 
   analysed sample, IRAS~18120+4530, presents 
   simultaneously the weakest value of F/Cont 
   ($\sim$0.03). However, this is an artificial effect 
   caused by the weakness of the 30\,$\mu$m 
   feature, and not a manifestation of any physical 
   effect.

%*************************************************************************************************************************
%*************************************************************************************************************************
%*************************************************************************************************************************
%*************************************************************************************************************************

%*************************************************************************************************************************
%*************************************************************************************************************************
%
%______________________________________________________________
%
%_____________________________________________________________
%                                             Simple A&A Table
%_____________________________________________________________
%
\begin{table}[h!]
\scriptsize
\caption{
%The list of EROs with the high mass-loss rates, which are distinguished in Figure~\ref{fig:central_t}. 
List of AGB objects from LMC with the coolest dust.
The values of the T$_{\rm d}$ and $\lambda_{\rm c}$ are taken 
from Table~\ref{app_tab:lmc_spectral_results}.}             % title of Table
\label{tab:central_wav_EROs}      % is used to refer this table in the text
\centering                          % used for centering table
\begin{tabular}{lcccc}        % centered columns (4 columns)
\hline\hline                 % inserts double horizontal lines
Name				&	T$_{\rm d}$		&	$\lambda_{\rm c}$ (30\,$\mu$m)&	Mass-loss rate		&	Mass-loss rate\\
					&		(K)			&			($\mu$m)			&	($M_{\odot}\,yr^{-1}$)	&	reference\\
\hline
IRAS~05509-6956		&	278	$\pm$	3	&	29.58	$\pm$	0.08	&	$9.0\times 10^{-5}$	&	(1)		\\
IRAS~05026-6809		&	253	$\pm$	3	&	30.48	$\pm$	0.14	&		$\ldots$		&	$\ldots$	\\
IRAS~05133-6937		&	242	$\pm$	2	&	29.81	$\pm$	0.22	&	$8.9\times 10^{-5}$	&	(1)		\\
IRAS~05315-7145		&	210	$\pm$	4	&	29.96	$\pm$	0.04	&	$1.7\times 10^{-4}$	&	(1)		\\
SHV~0528350-701014	&	201	$\pm$	3	&	29.61	$\pm$	0.10	&		$\ldots$		&	$\ldots$	\\
IRAS~05495-7034		&	163	$\pm$	2	&	30.32	$\pm$	0.04	&	$2.3\times 10^{-4}$	&	(1)		\\
\hline                                   %inserts single line
\end{tabular}
\tablebib{(1)~\citet{Gruendl:2008aa}.
}
\end{table}
%
%*************************************************************************************************************************
%*************************************************************************************************************************

%*************************************************************************************************************************
%*************************************************************************************************************************
%
%______________________________________________________________
%
%_____________________________________________________________
%                                             Simple A&A Table
%_____________________________________________________________
%
\begin{table}[h!]
\scriptsize
\caption{List of AGB objects from Figure~\ref{fig:central_t}, 
which are located in the region of 900\,K > T$_{\rm d}$ > 300\,K 
and 33.1\,$\mu$m > $\lambda_{\rm c}$ > 29.7\,$\mu$m. 
}             % title of Table
\label{tab:list_lambda>29.7}      % is used to refer this table in the text
\centering                          % used for centering table
\begin{tabular}{lcccc}        % centered columns (4 columns)
\hline\hline                 % inserts double horizontal lines
Name							&	Location		&	T$_{\rm d}$		&$\lambda_{\rm c}$ (30\,$\mu$m)&	F/Cont (30\,$\mu$m)\\
								&				&		(K)			&	($\mu$m)				&	\\
\hline
IRAS~18120+4530\tablefootmark{a, b}	&	Milky Way		&	872	$\pm$	57	&	33.04	$\pm$	0.69	&	0.03		$\pm$	0.01	\\
IRAS~04433-7018\tablefootmark{a}		&	LMC			&	868	$\pm$	78	&	29.93	$\pm$	0.70	&	0.19		$\pm$	0.02	\\
OGLE~J052242\tablefootmark{a, d}		&	LMC			&	868	$\pm$	90	&	29.91	$\pm$	0.99	&	0.16		$\pm$	0.03	\\
IRAS~16339-0317\tablefootmark{a}		&	Milky Way		&	848	$\pm$	74	&	30.53	$\pm$	0.19	&	0.14		$\pm$	0.02	\\
MSX~LMC~1205\tablefootmark{a}		&	LMC			&	845	$\pm$	62	&	29.91	$\pm$	0.54	&	0.08		$\pm$	0.02	\\
MSX~LMC~494\tablefootmark{a}		&	LMC			&	827	$\pm$	64	&	29.77	$\pm$	0.46	&	0.15		$\pm$	0.02	\\
MSX~SMC~163\tablefootmark{a}		&	SMC			&	679	$\pm$	31	&	30.17	$\pm$	0.35	&	0.14		$\pm$	0.01	\\
IRAS~04188+0122\tablefootmark{a}		&	Milky Way		&	660	$\pm$	26	&	29.97	$\pm$	0.26	&	0.10		$\pm$	0.01	\\
MSX~LMC~474\tablefootmark{a}		&	LMC			&	512	$\pm$	17	&	29.79	$\pm$	0.08	&	0.11		$\pm$	0.01	\\
IRAS~18384-3310\tablefootmark{c}		&	Milky Way		&	495	$\pm$	12	&	30.52	$\pm$	0.06	&	0.66		$\pm$	0.01	\\
\hline                                   %inserts single line
\end{tabular}
\tablefoot{The values of the T$_{\rm d}$, $\lambda_{\rm c}$, 
and F/Cont are taken from 
\Cref{app_tab:smc_spectral_results,app_tab:lmc_spectral_results,app_tab:gal_spectral_results}.\\
\tablefoottext{a}{The 30\,$\mu$m feature is weak.}
\tablefoottext{b}{The 30\,$\mu$m feature is very weak.}
\tablefoottext{c}{The decline of the 30\,$\mu$m feature after maximum is very slow.}
\tablefoottext{d}{Full name of object is OGLE~J052242.09-691526.2.}
}
\end{table}
%
%*************************************************************************************************************************
%*************************************************************************************************************************

   Figure~\ref{fig:central_t_zoom} presents the 
   $\lambda_{\rm c}$ of the 30\,$\mu$m feature as 
   the function of the T$_{\rm d}$ between 110 and 
   270\,K, where the post-AGB objects and PNe are 
   mostly visible. The symbols on the graph are the 
   same as in Figure~\ref{fig:strength_t}. The main 
   group of the post-AGB objects is located in the 
   same range of $\lambda_{\rm c}$ as carbon-rich 
   AGB stars -- between 28.5 and 29.5\,$\mu$m. 
   However, there are eight post-AGB objects, which 
   are located above this range and one Galactic 
   post-AGB object a bit below at about 28.4\,$\mu$m 
   (IRAS~11339-6004). In addition, one post-AGB star, 
   2MASS~J01054645-7147053 (gr. II, SMC) has the 
   T$_{\rm d}$ = 262\,K, and evidently sticks out from 
   the main group. In case of PNe, the 
   $\lambda_{\rm c}$ is clearly shifted towards the 
   longer wavelengths. \citet{Hony:2002lr} suggested 
   that this is the effect of the temperature or a shape 
   of the MgS dust grains. The most of the PNe are 
   located in the range of the $\lambda_{\rm c}$ 
   between about 29.5 and 31.5\,$\mu$m. However, 
   the Galactic PNe show rather the smaller values of 
   the $\lambda_{\rm c}$ in comparison with the MCs 
   PNe.

%*************************************************************************************************************************
%*************************************************************************************************************************
%*************************************************************************************************************************
%*************************************************************************************************************************

   In the spectra of the PNe, which are located in 
   Figure~\ref{fig:central_t_zoom} beyond the 
   $\lambda_{\rm c}$ = 31.5\,$\mu$m, the 
   30\,$\mu$m feature starts to rise not from 
   24\,$\mu$m as usual, but from 27\,$\mu$m. These 
   objects are (in order of increasing 
   $\lambda_{\rm c}$): SMP~SMC~17 from the SMC, 
   SMP~LMC~38, 52, 99, 71 from the LMC. Moreover, 
   the region between 24-27\,$\mu$m in the Spitzer 
   spectra of SMP~LMC~52, SMP~LMC~71 is 
   additionally disrupted by presence of very strong 
   [O IV] forbidden line at 25.89\,$\mu$m, and relatively 
   strong [Ne V] line at 24.32\,$\mu$m. In the spectrum 
   of SMP~LMC~71, the [Ne V] line is roughly as strong 
   as the [O IV] one. Their presence causes that the left 
   edge of profile of the 30\,$\mu$m feature may be 
   disrupted, especially after removal of nebular lines. 
   We note that in these cases, measuring values of 
   the F/Cont and $\lambda_{\rm c}$ by integrating the 
   flux from 24 to 36\,$\mu$m leads to underestimated 
   values of those parameters. Therefore, we have 
   made the additional measurements of the F/Cont 
   and $\lambda_{\rm c}$, assuming the 
   27--36\,$\mu$m integration range. This caused the 
   increase in values of the F/Cont by 36-52\%, 
   whereas the change in the $\lambda_{\rm c}$ was 
   no bigger than 0.8\%.

\begin{figure}
   \centering
   \includegraphics[width=\hsize]{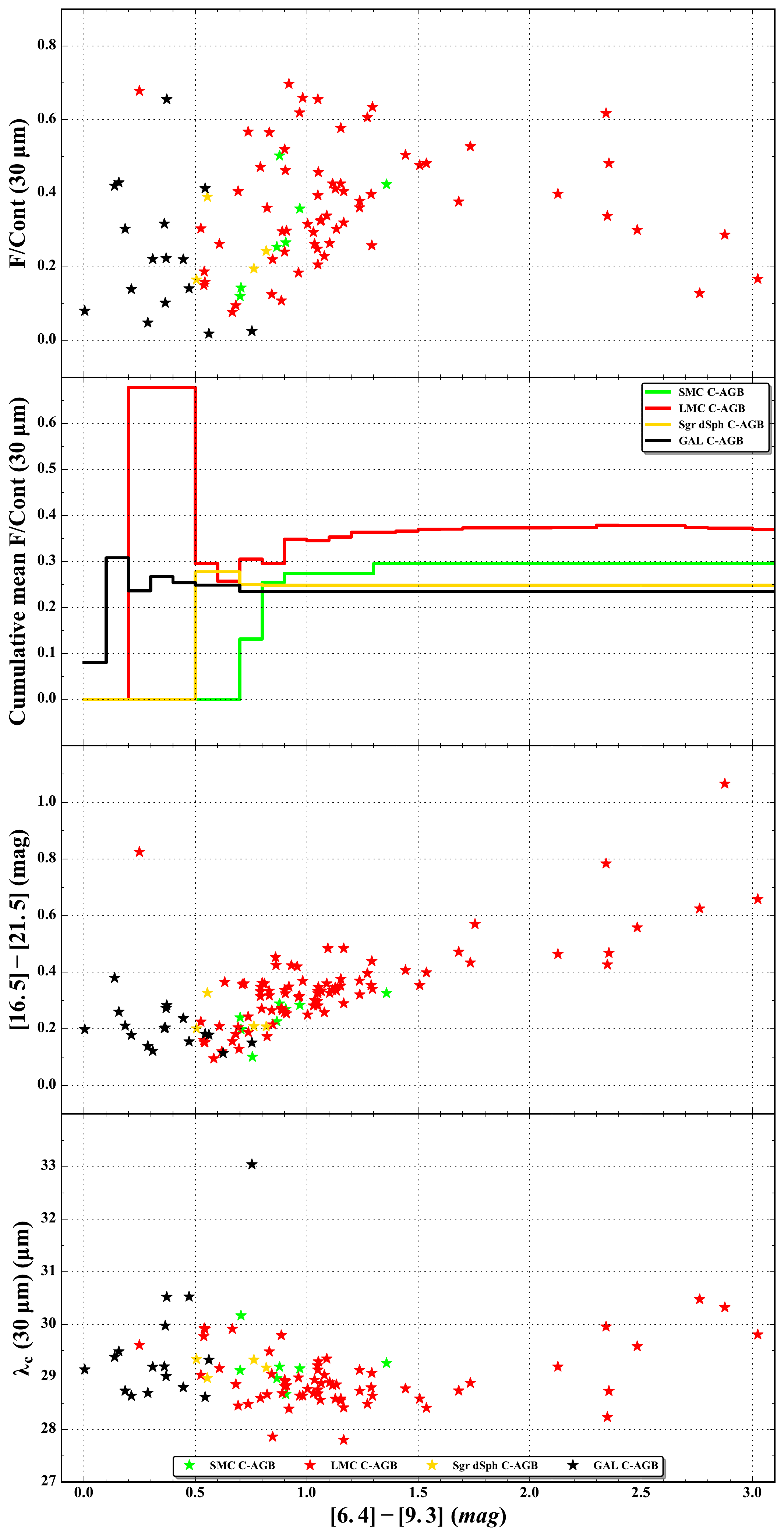}
      \caption{Top to bottom panels: The strength of the 
      feature (F/Cont), cumulative mean of the F/Cont, 
      [16.5]$-$[21.5] colour, and central wavelength 
      ($\lambda_{\rm c}$) as a function of the 
      [6.4]$-$[9.3] colour.}
         \label{fig:strength_cum_16-21_central_6-9}
   \end{figure}

   Figure~\ref{fig:strength_cum_16-21_central_6-9} 
   consists of four panels, and shows some parameters 
   as a function of the [6.4]$-$[9.3] colour for the 
   carbon-rich AGB stars. \citet{Zijlstra:2006fj} showed 
   that the [6.4]$-$[9.3] colour provides a good estimate 
   of the dust optical depth. In addition, the [6.4]$-$[9.3] 
   colour shows a linear correlation with the measured 
   mass-loss rates \citep{Groenewegen:2007yq}. 
   Symbols on the graph are in line with 
   Figure~\ref{fig:strength_t}. However, the second 
   panel (from the top) contains the cumulative 
   distributions, presented by solid lines. Their colours 
   are consistent with other panels.
   
   The upper panel of 
   Figure~\ref{fig:strength_cum_16-21_central_6-9} 
   shows the F/Cont (30\,$\mu$m) as a function of 
   the [6.4]$-$[9.3] colour for the different populations 
   of the carbon-rich AGB stars. The Galactic 
   carbon-rich AGB stars in our sample are visible 
   starting from about [6.4]$-$[9.3] $\sim$0 mag until 
   0.8 mag. However, our sample appears to be 
   truncated by a selection bias against the redder 
   objects. The \citet{Sloan:2016aa} sample of Galactic 
   objects based on the ISO is two times wider in the 
   [6.4]-[9.3] colour, and it spreads from about 0 to 1.7 
   mag (see their Fig. 6). For the LMC population of 
   objects, the first 30\,$\mu$m emission is visible at 
   the [6.4]$-$[9.3] = 0.25 mag (SHV~0528350-701014), 
   and then the feature is not visible until the 
   [6.4]$-$[9.3] colour is about 0.5 mag. SHV~0528350 
   is particularly red in the [16.5]$-$[21.5] colour. This 
   object is the carbon-rich star with cool and detached 
   dust shell (T$_{\rm d} = 201$ K). The Sgr dSph 
   objects also become visible from the [6.4]$-$[9.3] 
   colour $\sim$0.5 mag, whereas in the SMC the first 
   30\,$\mu$m emission is at about 0.7 mag. Basically, 
   the formation of the 30\,$\mu$m emission is clearly 
   related to the metallicity of the hosted galaxy. In the 
   metal-poor environments like the MCs, the 
   30\,$\mu$m emission becomes visible for much 
   redder [6.4]$-$[9.3] colours than in the Milky Way. 
   The location of the Sgr dSph carbon-rich stars on 
   this diagram indicates that the metallicity of this 
   galaxy is similar to the LMC, or between the LMC 
   and our Galaxy. 
   
   The second panel of 
   Figure~\ref{fig:strength_cum_16-21_central_6-9} 
   illustrates the cumulative mean of the F/Cont 
   (30\,$\mu$m) as a function of the [6.4]$-$[9.3] 
   colour for the carbon-rich stars in the different 
   galaxies. By solid lines we show the cumulative 
   distribution with a step of 0.1 mag in the [6.4]$-$[9.3] 
   colour. 
   %Their meaning is explained in the legend, which is 
   %located in the upper-right corner of this panel. 
   The cumulative distribution shows 
   clearly how the 30\,$\mu$m emission rises with the 
   increasing [6.4]$-$[9.3] colour for the carbon-rich 
   AGB stars in the different galaxies. In a case of the 
   metal-poor galaxies, the much higher dust 
   production rate and the cooler dust (see 
   also the next panel) is needed to make the 
   30\,$\mu$m emission visible. The plots for the 
   Galactic and Sgr dSph objects become flat past 
   the [6.4]$-$[9.3] colour of about 0.8 and 0.9 mag, 
   respectively. In case of the SMC objects it happens 
   past the [6.4]$-$[9.3] colour $\sim$1.4 mag. This is 
   caused by the lack of the redder objects in the 
   samples. \citet{Sloan:2016aa} presented a similar 
   cumulative plot, and their Galactic trace is above 
   the MCs rising up to the $\sim$1.7 mag in the 
   [6.4]$-$[9.3] colour, as has been already discussed 
   in the upper panel.
   
   The third panel of 
   Figure~\ref{fig:strength_cum_16-21_central_6-9} 
   presents the [16.5]$-$[21.5] colour as a function of 
   the [6.4]$-$[9.3] colour. \citet{Zijlstra:2006fj} showed 
   that the [16.5]$-$[21.5] colour serves as an indicator 
   of the dust temperature. The carbon-rich AGB stars 
   form a linear sequence. However, the points are 
   more dispersed for the [6.4]$-$[9.3] colour larger than 
   1.4 mag. There are 14 such objects, and five of them 
   are known EROs (their definition and selection criteria 
   is given above in this Section), which experience an 
   intense mass-loss process \citep{Gruendl:2008aa}. 
   Table~\ref{tab:high_mass-loss_rate_objects_LMC} 
   lists all these objects with their names in order of the 
   increasing values of the [6.4]$-$[9.3] colour. After 
   that, the values of the [16.5]$-$[21.5] colour, 
   T$_{\rm d}$, $\lambda_{\rm c}$, and F/Cont 
   (30\,$\mu$m) are given. All of these quantities are 
   repeated from Table~\ref{app_tab:lmc_spectral_results}. 
   Finally, the values of the mass-loss rates with the 
   corresponding reference are given. We note that 
   there is only one object in the SMC at 
   [6.4]$-$[9.3] = 1.36 mag, NGC~419~MIR~1, which 
   also might be counted as an object with an intense 
   mass-loss rate.
   
  In the bottom panel of 
  Figure~\ref{fig:strength_cum_16-21_central_6-9} we 
  show the $\lambda_{\rm c}$ (30\,$\mu$m) as a 
  function of the [6.4]$-$[9.3] colour. The 
  $\lambda_{\rm c}$ seems to be independent with 
  the increasing [6.4]$-$[9.3] colour. As we discussed 
  before, the large value of the $\lambda_{\rm c}$ for 
  nine objects with the central wavelength above 
  29.7\,$\mu$m may be result of the 30\,$\mu$m 
  feature weakness (see 
  Table~\ref{tab:list_lambda>29.7}).

%*************************************************************************************************************************
%*************************************************************************************************************************
%
%______________________________________________________________
%
%_____________________________________________________________
%                                             Simple A&A Table
%_____________________________________________________________
%
\begin{table*}[h!]
\scriptsize
\caption{List of AGB objects from the LMC, 
showing the intense mass-loss rates.}             % title of Table
\label{tab:high_mass-loss_rate_objects_LMC}      % is used to refer this table in the text
\centering                          % used for centering table
\begin{tabular}{lccccccc}        % centered columns (4 columns)
\hline\hline                 % inserts double horizontal lines
Name			&		[6.4]$-$[9.3]			&		[16.5]$-$[21.5]			&	T$_{\rm d}$		&	$\lambda_{\rm c}$ (30 $\mu$m	&	F/Cont (30\,$\mu$m)			&	Mass-loss rate		&	Mass-loss rate\\
				&			(mag)			&			(mag)			&		(K)			&			($\mu$m)			&							&	($M_{\odot}\,yr^{-1}$)	&	reference\\
\hline
IRAS~05053-6901	&	1.443	$\pm$	0.013	&	0.407	$\pm$	0.007	&	362	$\pm$	5	&	28.778	$\pm$	0.025	&	0.504	$\pm$	0.009	&		\ldots		&	\ldots	\\
IRAS~05125-7035	&	1.507	$\pm$	0.013	&	0.354	$\pm$	0.005	&	407	$\pm$	5	&	28.586	$\pm$	0.029	&	0.476	$\pm$	0.006	&		\ldots		&	\ldots	\\
IRAS~04535-6616	&	1.537	$\pm$	0.016	&	0.399	$\pm$	0.004	&	368	$\pm$	3	&	28.411	$\pm$	0.051	&	0.481	$\pm$	0.005	&		\ldots		&	\ldots	\\
IRAS~05416-6906	&	1.682	$\pm$	0.017	&	0.472	$\pm$	0.004	&	320	$\pm$	2	&	28.738	$\pm$	0.145	&	0.377	$\pm$	0.009	&		\ldots		&	\ldots	\\
IRAS~04518-6852	&	1.734	$\pm$	0.020	&	0.434	$\pm$	0.011	&	343	$\pm$	7	&	28.885	$\pm$	0.107	&	0.527	$\pm$	0.015	&		\ldots		&	\ldots	\\
IRAS~05568-6753	&	1.755	$\pm$	0.016	&	0.570	$\pm$	0.008	&		\ldots 		&			\ldots			&			\ldots			&		\ldots		&	\ldots	\\
IRAS~05305-7251	&	2.127	$\pm$	0.024	&	0.464	$\pm$	0.006	&	324	$\pm$	3	&	29.194	$\pm$	0.262	&	0.398	$\pm$	0.013	&	$4.2\times 10^{-5}$	&	(1)	\\
IRAS~05315-7145	&	2.343	$\pm$	0.031	&	0.784	$\pm$	0.017	&	210	$\pm$	4	&	29.956	$\pm$	0.037	&	0.617	$\pm$	0.020	&	$1.7\times 10^{-4}$	&	(1)	\\
IRAS~05306-7032	&	2.349	$\pm$	0.037	&	0.427	$\pm$	0.032	&	347	$\pm$	20	&	28.235	$\pm$	0.136	&	0.338	$\pm$	0.039	&		\ldots		&	\ldots	\\
IRAS~04589-6825	&	2.356	$\pm$	0.041	&	0.468	$\pm$	0.026	&	321	$\pm$	14	&	28.731	$\pm$	0.186	&	0.481	$\pm$	0.033	&		\ldots		&	\ldots	\\
IRAS~05509-6956	&	2.483	$\pm$	0.027	&	0.558	$\pm$	0.008	&	278	$\pm$	3	&	29.582	$\pm$	0.084	&	0.300	$\pm$	0.010	&	$9.0\times 10^{-5}$	&	(1)	\\
IRAS 05026-6809	&	2.763	$\pm$	0.031	&	0.625	$\pm$	0.008	&	253	$\pm$	3	&	30.477	$\pm$	0.137	&	0.128	$\pm$	0.009	&		\ldots		&	\ldots	\\
IRAS~05495-7034	&	2.876	$\pm$	0.041	&	1.066	$\pm$	0.017	&	163	$\pm$	2	&	30.322	$\pm$	0.040	&	0.287	$\pm$	0.020	&	$2.3\times 10^{-4}$	&	(1)	\\
IRAS~05133-6937	&	3.024	$\pm$	0.042	&	0.658	$\pm$	0.007	&	242	$\pm$	2	&	29.806	$\pm$	0.221	&	0.167	$\pm$	0.009	&	$8.9\times 10^{-5}$	&	(1)	\\
\hline                                   %inserts single line
\end{tabular}
\tablebib{(1)~\citet{Gruendl:2008aa}.
}
\tablefoot{The values of the T$_{\rm d}$, $\lambda_{\rm c}$, and 
F/Cont are taken from Table~\ref{app_tab:lmc_spectral_results}.}
\end{table*}
%
%*************************************************************************************************************************
%*************************************************************************************************************************

%*************************************************************************************************************************
%*************************************************************************************************************************
%*************************************************************************************************************************
%*************************************************************************************************************************

   The [6.4]$-$[9.3] colour is only useful for the 
   carbon-rich AGB objects, because the presence of 
   the PAH features between 6-9\,$\mu$m is interfering 
   with that colour in the spectra of the post-AGB stars 
   and PNe. The most offending is the PAH feature at 
   6.2\,$\mu$m, therefore, we modified only the left 
   side of the [6.4]$-$[9.3] colour so as to avoid this 
   spectral feature. This gave us the new [5.8]$-$[9.3] 
   colour, which was used for the post-AGB objects 
   and PNe.
   
   While, the [6.4]$-$[9.3] colour for the AGB stars 
   correlates well with mass-loss 
   \citep{Groenewegen:2007yq}, in case of the 
   post-AGB objects and PNe the [5.8]$-$[9.3] one 
   measures only optical depth of the detaching shell, 
   and cannot be easily related to the mass-loss 
   history. The geometrical effects (non-spherical 
   symmetry, existence of discs and/or tori), do not 
   allow for simple interpretation that decreasing 
   optical depth is the manifestation of longer 
   evolution after AGB and different mass-loss rates 
   during this phase of evolution.
   
\begin{figure}
   \centering
   \includegraphics[width=\hsize]{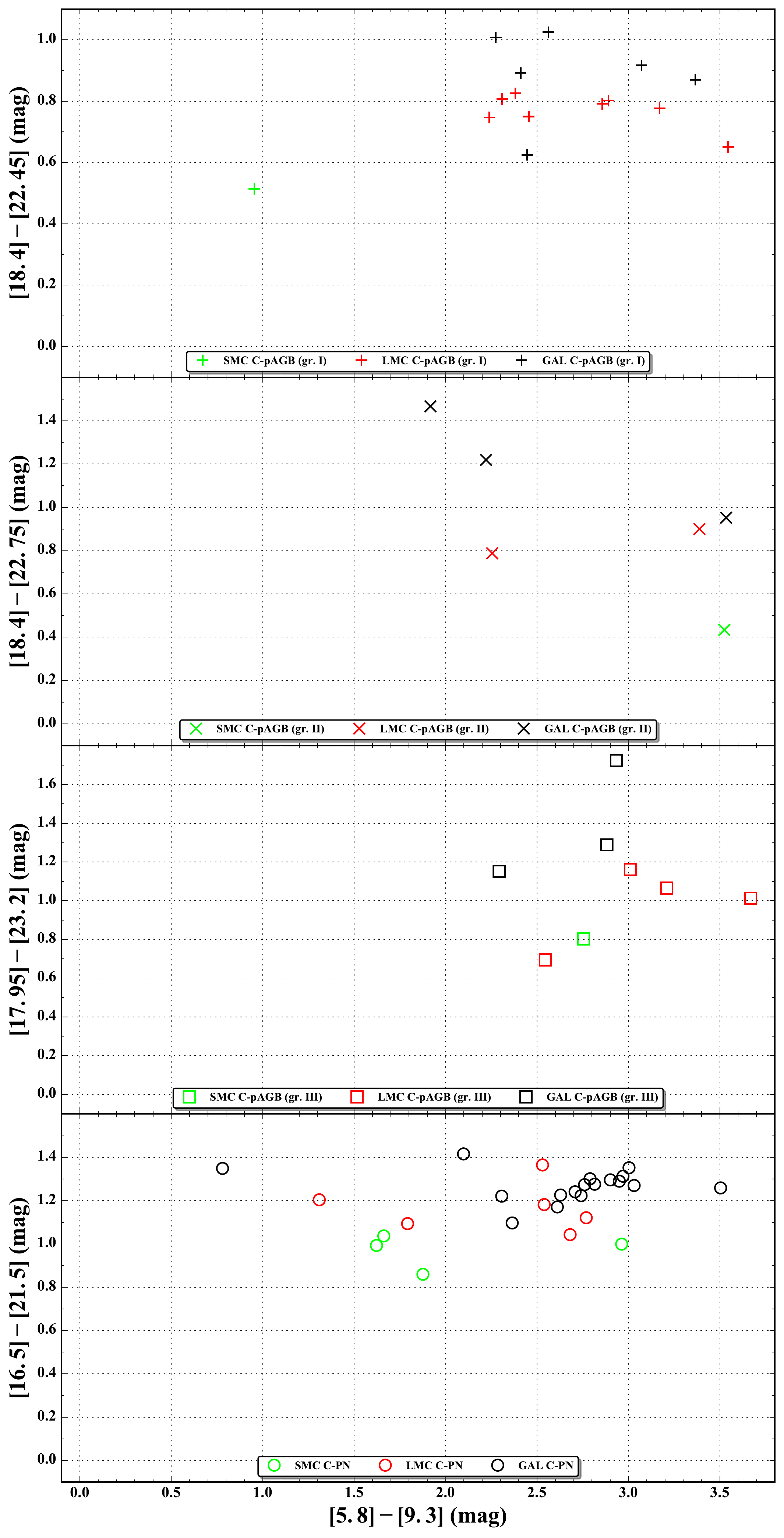}
      \caption{Colours used for determination of the 
      T$_{\rm d}$ in the different groups of the post-AGB 
      objects. From the top: [18.4]$-$[22.45] (gr. I), 
      [18.4]$-$[22.75] (gr. II), and [17.95]$-$[23.2] (gr. III) 
      colour as a function of the [5.8]$-$[9.3] one. The 
      bottom panel contains the [16.5]$-$[21.5] colour 
      used for the determination of the T$_{\rm d}$ in 
      PNe.}
         \label{fig:post-agb_PN_colours_5-9}
   \end{figure}

   In Figure~\ref{fig:post-agb_PN_colours_5-9} (four 
   panels) we present the relations between the 
   [5.8]$-$[9.3] and different colours used for 
   determination of the T$_{\rm d}$. 
   %The explanations of the symbols are given in the 
   %legends, which are located in the bottom-central 
   %part of every panel, and they are in line with 
   %Figure~\ref{fig:strength_t}. 
   Symbols on the graph are in line with 
   Figure~\ref{fig:strength_t}. In the case of the 
   post-AGB objects three different colours are used: 
   [18.4]$-$[22.45] for group I (top panel), 
   [18.4]$-$[22.75] for group II (second panel), and 
   [17.95]$-$[23.2] for group III (third panel). In the 
   bottom panel the PNe are shown with the 
   [16.5]$-$[21.5] colour determined for them.
   
   By comparing the positions of PNe in [5.8]$-$[9.3] 
   colour with those for the post-AGB objects we see, 
   that from statistical point of view the colours of PNe 
   are bluer (their optical depths are smaller) than 
   those for post-AGBs. The PNe colours concentrate 
   between 2 and 3 mag with significant extension to 
   the lower values of the [5.8]$-$[9.3] colour, while 
   post-AGBs have colours larger than about 2.5 mag 
   with significant contribution above 3 mag. The only 
   exception being IRAS~00350-7436 with the 
   [5.8]$-$[9.3] colour $\sim$1 mag on the top panel 
   and IRAS~22223+4327 ([5.8]$-$[9.3] $\sim$1.9 
   mag) on the second panel. The values of the 
   T$_{\rm d}$ are not known for them, because they 
   belong to the objects with bad fits of continuum. 
   However, their values of spectral indices are 
   obtained. Finally we note that, while colours used 
   for determination of the T$_{\rm d}$ are different for 
   various objects shown in 
   Figure~\ref{fig:post-agb_PN_colours_5-9}, there is 
   a clear sign that PNe have statistically lower 
   T$_{\rm d}$ (larger long-wavelength's colour) than 
   post-AGBs, and AGBs shown in the third panel of 
   Figure~\ref{fig:strength_cum_16-21_central_6-9}.

%*************************************************************************************************************************
%*************************************************************************************************************************
%*************************************************************************************************************************
%*************************************************************************************************************************

%-------------------------------------------------------------
% ---- PROFILES OF THE 30 UM FEATURE ---
%-------------------------------------------------------------
\subsection{Profiles of the 30\,$\mu$m feature}
\label{subsec:profiles_of_the_30_um_feature}

   The 30\,$\mu$m feature appears in plenty of 
   carbon-rich objects in many galaxies. As can be 
   seen in Figure~\ref{fig:strength_t}, it is also visible 
   in the wide range of the T$_{\rm d}$. Apart the 
   parameters of the feature, their profiles also might 
   provide information about the differences of dust 
   formation history in those galaxies. The profiles of 
   the 30\,$\mu$m feature were extracted from the 
   continuum subtracted spectra. However, it was 
   impossible to obtain the full profiles, because of the 
   spectral coverage of the Spitzer. Nevertheless, they 
   were long enough to contain the maxima of the 
   emission and the subsequent declines. After the 
   subtraction of the continuum, the profiles were 
   normalised to the region around the maximum of the 
   feature by dividing them by the average value of flux 
   density calculated for this region. The range used for 
   determination of the average value depended on the 
   kind of object: 28.5-29.5\,$\mu$m for the AGBs, and 
   28.5-30\,$\mu$m for the post-AGBs. In the case of 
   the PNe profiles we used the 30.5-32.5\,$\mu$m 
   normalisation range for low resolution spectra, and 
   30-31\,$\mu$m for high resolution Galactic profiles. 
   To determine those normalisation ranges, we also 
   took into consideration the behaviour of the 
   $\lambda_{\rm c}$, and presence of nebular lines in 
   a case of PNe. After normalisation, we determined 
   200\,K intervals in the T$_{\rm d}$ up to 800\,K, and 
   one range for objects with the T$_{\rm d}$ > 800\,K. 
   All the objects were split in terms of the kind of object 
   (AGB, post-AGB, and PN), hosted galaxy (SMC, 
   LMC, Sgr dSph, and GAL), and given T$_{\rm d}$ 
   range. The median profile was then calculated for 
   each of such a sample, but on the assumption that 
   there were at least three profiles. The spectra for 
   the Galactic post-AGBs and PNe are composed of 
   both resolutions. Because of this, we re-sampled 
   the low resolution spectra to the high resolution 
   wavelength grid using the linear interpolation 
   before calculating the median profile.

\begin{figure}[h!]
   \centering
   \includegraphics[width=\hsize]{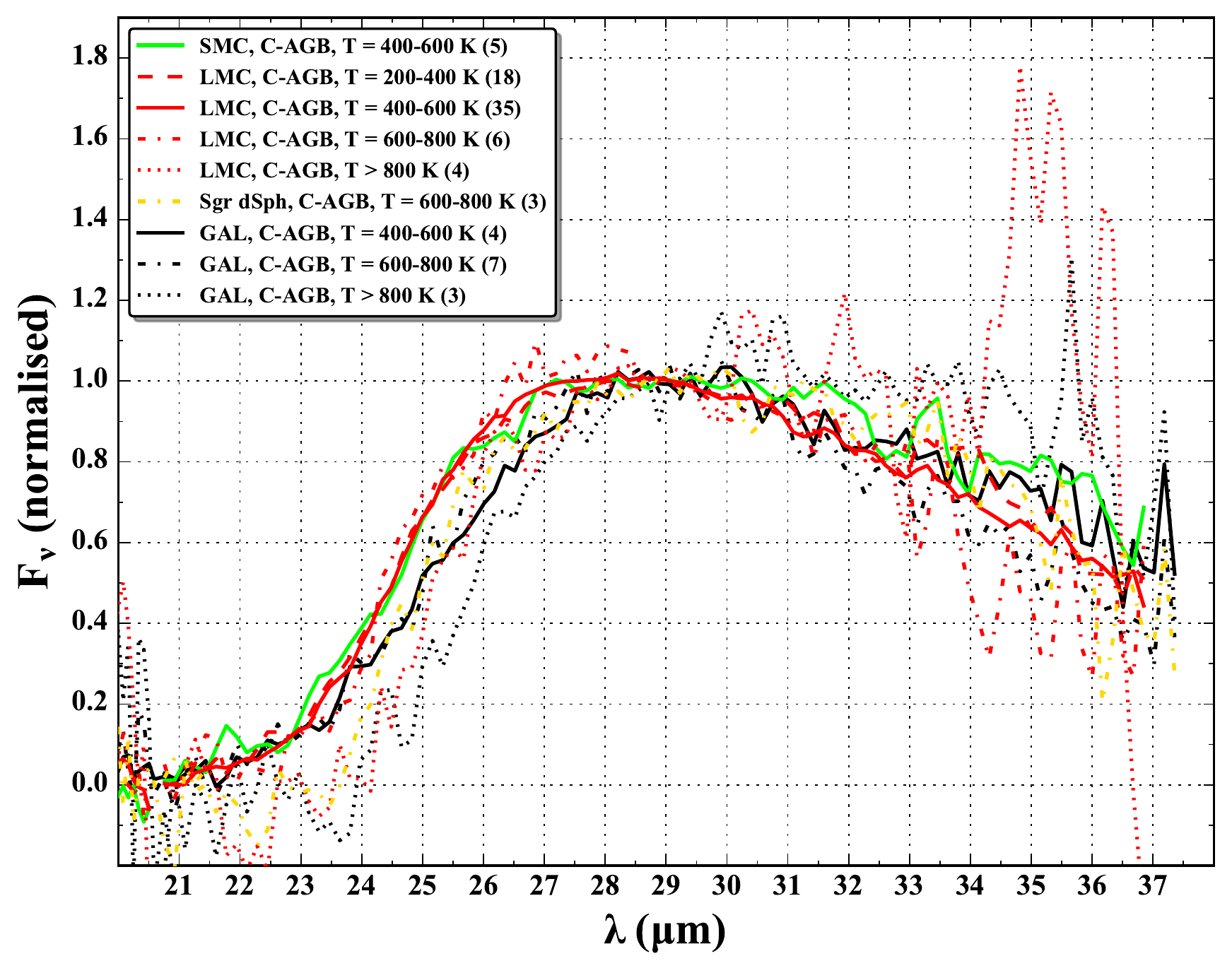}
      \caption{Normalised median profiles of the 
      30\,$\mu$m feature in the AGB stars calculated 
      for the different ranges of the T$_{\rm d}$. The 
      brackets in the legend show the number of 
      profiles taken to the calculation of a given median.}
         \label{fig:agb_norm_profiles_t}
   \end{figure}

   Figure~\ref{fig:agb_norm_profiles_t} shows the 
   normalised median profiles of the 30\,$\mu$m feature 
   in the carbon-rich AGB stars for the four ranges 
   of the T$_{\rm d}$. The colours of lines on the graph 
   are in line with the colours of the markers in 
   Figure~\ref{fig:strength_t}. 
   %The explanation of all the profiles on the graph is 
   %given in the legend, which is located in the 
   %upper-left corner. Generally, the different colours 
   %distinguish each of the analysed galaxies 
   %(lime -- SMC, red -- LMC, gold -- Sgr dSph, and 
   %black -- Milky Way). The various line styles 
   %represent the profiles for the different ranges of the 
   %T$_{\rm d}$: the dashed lines for 200-400\,K, solid 
   %lines for 400-600\,K, dotdash lines for 600-800\,K, 
   %and dotted lines for objects with the 
   %T$_{\rm d}$ > 800\,K. 
   There is only one object in the LMC with the 
   T$_{\rm d}$ below 200\,K, thus we do not present 
   any normalised median profile of the 30\,$\mu$m 
   feature in the AGB stars from this range. In addition, 
   one Galactic object (IRAS~18120+4530) is excluded 
   from median for T$_{\rm d}$ > 800\,K. The 
   $\lambda_{\rm c}$ for this object unusually 
   exceeds 33\,$\mu$m, therefore its normalised 
   profile completely sticks out from the others (see 
   Section~\ref{subsec:relations_of_parameters} for 
   details). 
   %The brackets in the legend contain the number 
   %of profiles taken to the calculation of a given median. 
   All of AGB profiles of the 30\,$\mu$m 
   feature were derived from the low resolution Spitzer 
   spectra. The normalised median profiles present 
   a relatively steep rise from about 23\,$\mu$m, and 
   after reaching a maximum at around 29\,$\mu$m, a 
   gradual decline is visible up to the long wavelength 
   Spitzer cut-off. The shape of particular median 
   profiles of the 30\,$\mu$m feature in the AGB stars 
   looks uniformly for the different galaxies and the 
   T$_{\rm d}$ ranges. Some changes in shape of the 
   median profiles are noticeable for the T$_{\rm d}$ 
   above 800\,K. However, the 30\,$\mu$m feature in 
   those objects is relatively weak, and their profiles 
   are noisy in general.

%*************************************************************************************************************************
%*************************************************************************************************************************
%*************************************************************************************************************************
%*************************************************************************************************************************

\begin{figure}[h!]
   \centering
   \includegraphics[width=\hsize]{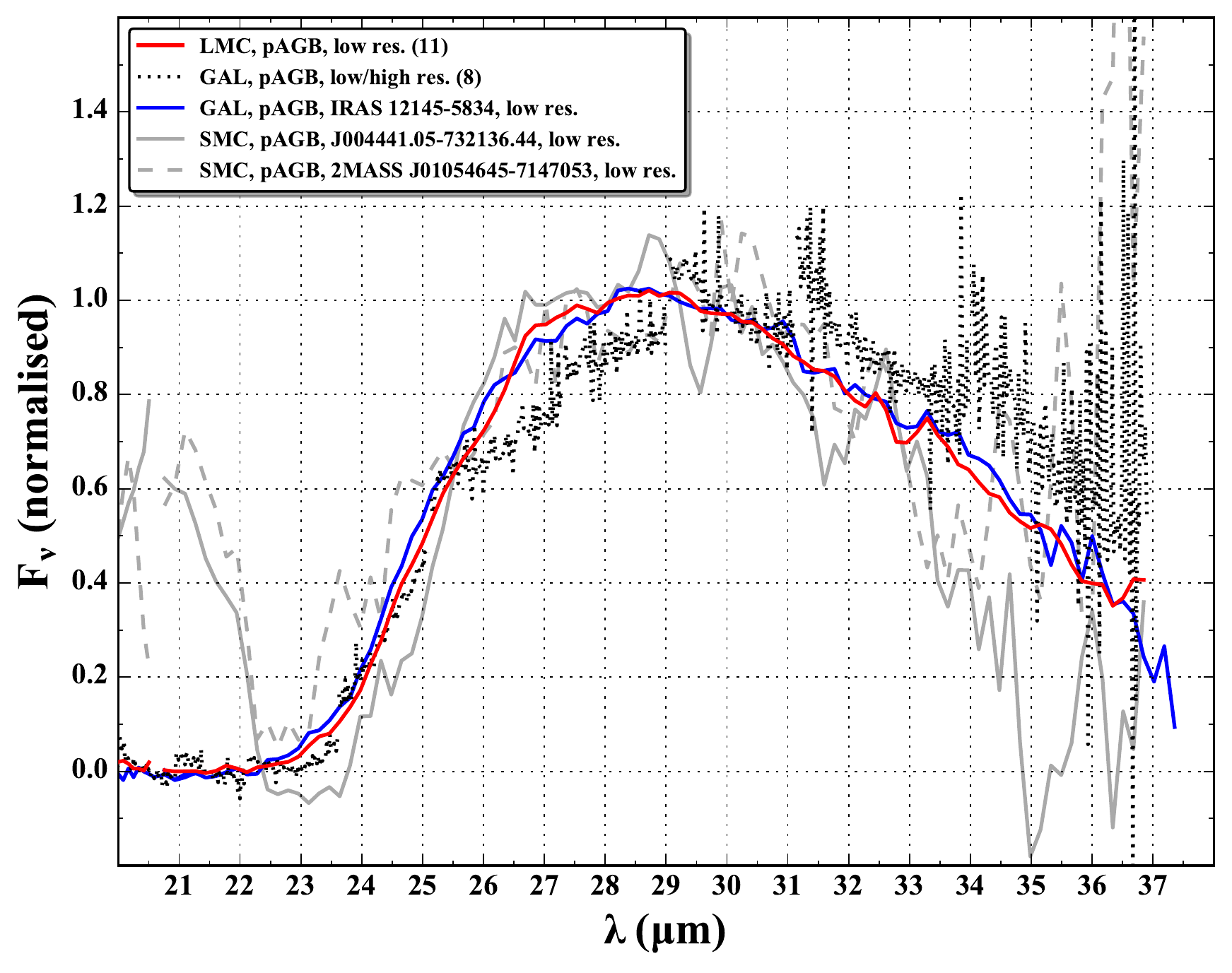}
      \caption{Normalised median profiles of the 
      30\,$\mu$m feature in the post-AGB stars. The 
      brackets in the legend show the number of 
      profiles taken to the calculation of a given median.}
         \label{fig:pagb_norm_profiles_t}
   \end{figure}

   Figure~\ref{fig:pagb_norm_profiles_t} shows the 
   normalised median profiles of the 30\,$\mu$m 
   feature in the carbon-rich post-AGB stars. All but two 
   of the post-AGB stars in our sample lie in only one 
   of the distinguished ranges of the T$_{\rm d}$ (below 
   200\,K). Accordingly, we have not compared the 
   median profiles as a function of the T$_{\rm d}$. 
   %The explanation of all the profiles on the graph is 
   %given in the legend, which is located in the 
   %upper-left corner. Generally, the different colours 
   %distinguish each of the analysed galaxies 
   %(red -- LMC, and black -- Milky Way). The brackets 
   %in the legend contain the number of profiles taken to 
   %the calculation of a given median. 
   By the black dotted line we present the normalised 
   median profile of the Galactic post-AGB objects. 
   This median profile consist of high resolution spectra 
   except one profile (shown by the blue solid line). 
   There are only two post-AGB objects in the SMC 
   with the derived normalised profiles of the 
   30\,$\mu$m feature, therefore we show them for 
   comparison instead of the median profile by the 
   grey solid and dashed lines. The individual spectra 
   for the SMC objects show the 21\,$\mu$m feature, 
   while the median spectra for the Milky Way and 
   LMC make the contribution from the 21\,$\mu$m 
   emitters unseen. Our sample does not contain any 
   of the post-AGB objects from the Sgr dSph galaxy. 
   
   The median profile for Galactic post-AGBs seems to 
   be a little bit different than other low resolution 
   spectra (median for the LMC and some individual 
   spectra). Nevertheless, the high resolution median 
   spectrum (Galactic) is quite noisy, so the difference 
   may not be significant. In addition, the low and high 
   resolution spectra from the CASSIS database were 
   processed using the different pipelines 
   \citep[see][]{Lebouteiller:2011aa, 2015ApJS..218...21L}, 
   which might led to the distortion of profiles of the 
   30\,$\mu$m feature. 
   
   The normalised median profiles of the 30\,$\mu$m 
   feature for the post-AGB objects look similar to the 
   AGB ones in general. However, the blue edge of the 
   individual normalised profiles might be affected by 
   the presence of the strong 21\,$\mu$m feature. The 
   low resolution median profiles for the Milky Way and 
   LMC objects show one common trace. The individual 
   normalised profiles of the SMC post-AGB objects 
   clearly confirm that shape.

%*************************************************************************************************************************
%*************************************************************************************************************************
%*************************************************************************************************************************
%*************************************************************************************************************************

\begin{figure}[h!]
   \centering
   \includegraphics[width=\hsize]{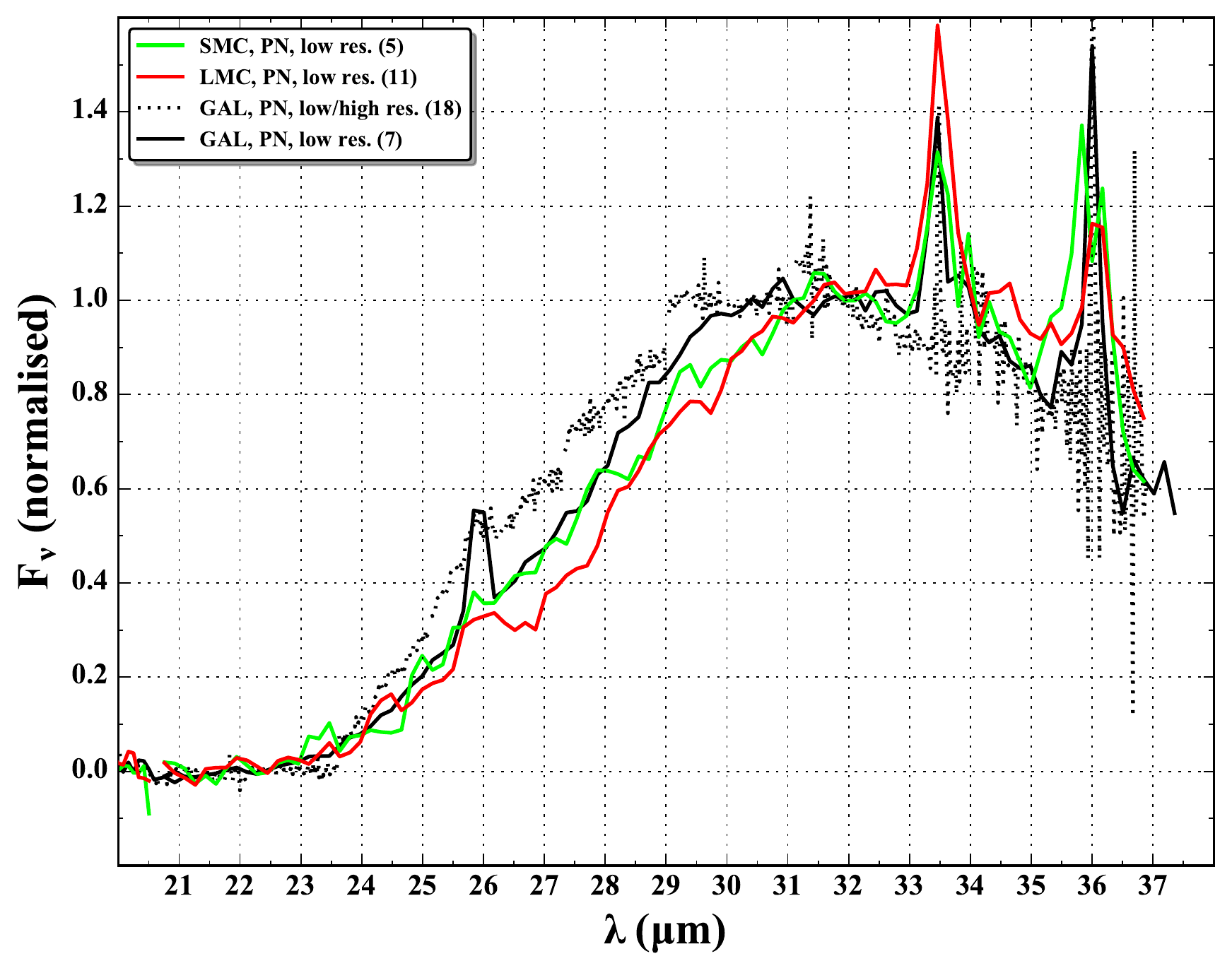}
      \caption{Normalised median profiles of the 
      30\,$\mu$m feature for the PNe. The brackets in 
      the legend show the number of profiles taken to 
      the calculation of a given median.}
         \label{fig:pne_norm_profiles_t}
   \end{figure}

   Figure~\ref{fig:pne_norm_profiles_t} illustrates the 
   normalised median profiles of the 30\,$\mu$m 
   feature in the carbon-rich PNe. All the PNe lie in one 
   of the distinguished ranges of the T$_{\rm d}$ (below 
   200\,K) except one object. The colours of lines on the 
   graph are in line with the colours of the markers in 
   Figure~\ref{fig:strength_t}. The Sgr dSph galaxy 
   contains only one PN with the 30\,$\mu$m feature 
   (Wray16-423), therefore the median profile is not 
   shown. We also do not show the individual 
   normalised profile for this object, because of the 
   presence of the 16-24\,$\mu$m feature (see 
   Section~\ref{subsec:manchester_method}). The 
   comparison between the Galactic median profiles 
   with low resolution spectra only, and median of all 
   the Galactic ones (seven low and 11 high resolution) 
   shows, that the difference in shape of the high 
   resolution median profile is a matter of the various 
   calibration of the spectra (different pipelines). 
   
\begin{figure}[h!]
   \centering
   \includegraphics[width=\hsize]{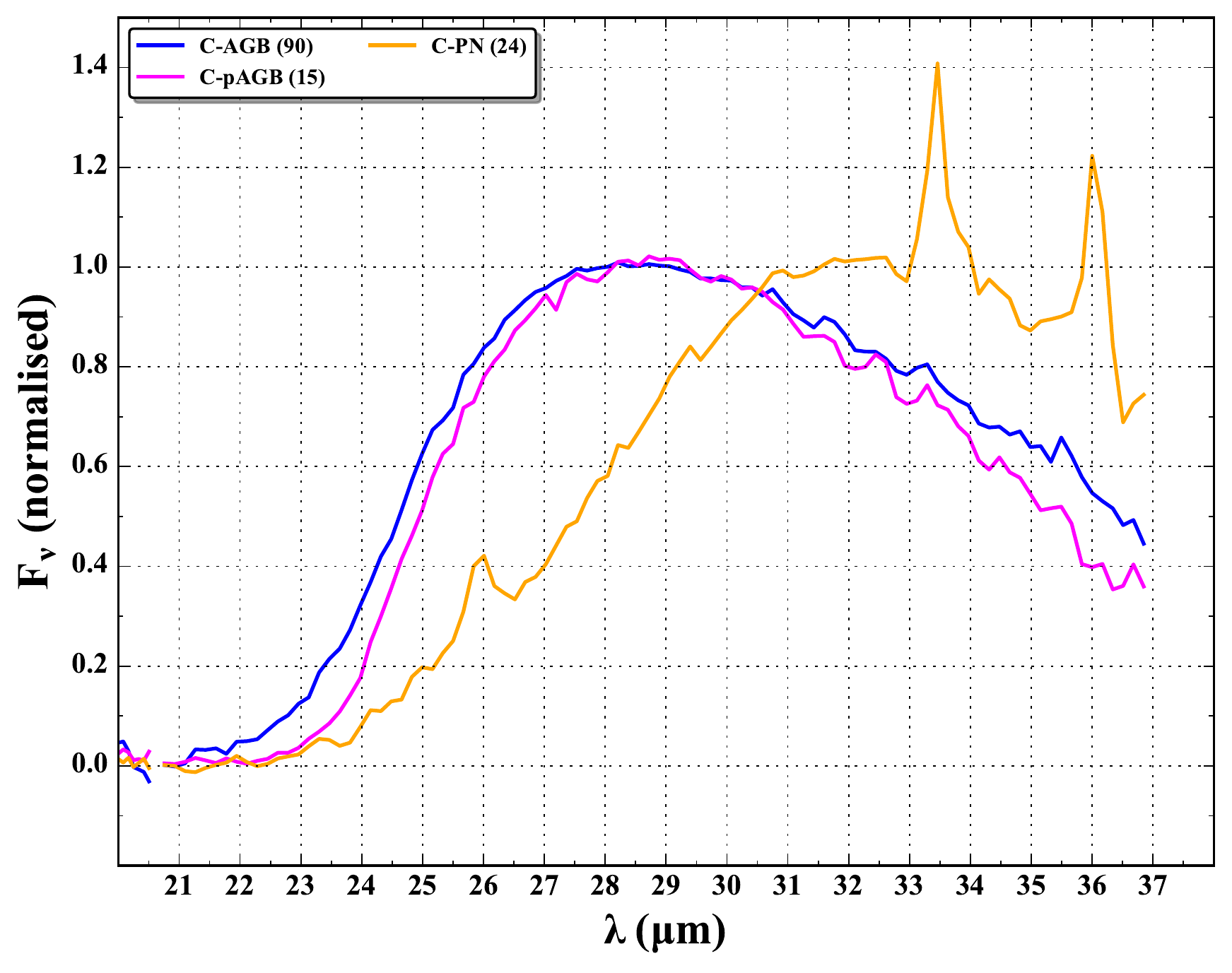}
      \caption{Normalised median profiles (only low 
      resolution) of the 30\,$\mu$m feature for all the 
      AGBs, post-AGBs, and PNe. The brackets in the 
      legend show the number of profiles taken to the 
      calculation of a given median.
      %The profiles are shown by the solid lines. Their colours 
      %distinguish the profiles for the different phase of the stellar 
      %evolution (blue -- AGBs, magenta -- post-AGBs, and 
      %orange -- PNe). The brackets in the legend contain the number 
      %of profiles taken to the calculation of a given median.
      }
         \label{fig:agb_pagb_pne_norm_profiles}
   \end{figure}

   The most remarkable, the normalised median profiles 
   of the PNe present clearly different shape than 
   the profiles of the AGBs and post-AGBs (see also 
   Figure~\ref{fig:agb_pagb_pne_norm_profiles}). The 
   30\,$\mu$m emission appears around 24\,$\mu$m 
   with more gradual rise up to the maximum, which is 
   significantly shifted towards the longer wavelengths 
   in comparison with the AGB and post-AGB profiles. 
   The shape of decline is more uncertain, because 
   the profiles of the PNe are deformed by the presence 
   of nebular lines -- [O IV] at 25.89\,$\mu$m, [S III] at 
   33.48\,$\mu$m, and [Ne III] at around 36\,$\mu$m are 
   the most noticeable. The low resolution median 
   profiles for the MCs and Galactic PNe show one 
   common trace, with no significant difference 
   between them. The individual median profiles from 
   \Cref{fig:agb_norm_profiles_t,fig:pagb_norm_profiles_t,fig:pne_norm_profiles_t} 
   together with every profile, which is a part of single 
   median are presented in 
   Appendix~\ref{sec:appendix_profiles}.

%*************************************************************************************************************************
%*************************************************************************************************************************
%*************************************************************************************************************************
%*************************************************************************************************************************

   The normalised median profiles of the 30\,$\mu$m 
   feature for the AGB stars look uniformly in the 
   various ranges of the T$_{\rm d}$ and hosted 
   galaxies. The low resolution median profiles for the 
   post-AGB objects and PNe show the same 
   behaviour. It allows us to collect the low resolution 
   median profiles from all the analysed galaxies and 
   create one common median profile of the 
   30\,$\mu$m feature for the carbon-rich AGB stars, 
   post-AGB objects, and PNe. However, the final 
   medians include more normalised profiles than the 
   previous ones, because this time we added profiles 
   of objects, which previously were not included (less 
   than three profiles, T$_{\rm d}$ < 200\,K for AGB 
   stars or T$_{\rm d}$ > 200\,K for post-AGB objects). 
   In case of AGB stars, five normalised profiles were 
   added: one Sgr dSph, two SMC, one LMC, and one 
   Galactic. In the case of post-AGB objects four 
   profiles were added: two SMC, one LMC, and one 
   Galactic, whereas in the case of PNe we added 
   one LMC profile.
   
   The comparison of the full median profiles is 
   shown in Figure~\ref{fig:agb_pagb_pne_norm_profiles}. 
   %The explanation of all the profiles on the graph 
   %is given in the legend, which is located in the 
   %upper-left corner. 
   We distinguish the median profiles for the different 
   phase of evolution by colour. 
   %(blue -- AGBs, magenta -- post-AGBs, and 
   %orange -- PNe). The brackets in the legend contain the number 
   %of profiles taken to the calculation of a given median. 
   The Figure clearly shows the similarity of the median 
   profiles for the AGB and post-AGB stars, but possibly 
   the narrower 30\,$\mu$m feature in the post-AGBs. 
   The normalised median profile calculated for all PNe 
   illustrates a distinctly different shape and a major 
   shift towards the longer wavelength in comparison 
   with the AGB and post-AGB ones, which is also 
   noticeable in the obtained values of the 
   $\lambda_{\rm c}$. The median profile for the PNe 
   is also deformed by the presence of [O IV], [S III], 
   and [Ne III] nebular lines. The second order 
   polynomial fit to the regions of maxima of the 
   30\,$\mu$m median profiles (between 
   26-32\,$\mu$m for the AGB and post-AGB objects, 
   and 27-35\,$\mu$m for the PNe -- but the [S III] line 
   at 33.48\,$\mu$m is removed from fit), shows the 
   maximum (with the standard error around 
   0.02\,$\mu$m) at 28.5\,$\mu$m for the AGB, 
   29\,$\mu$m for the post-AGB, and 32.5\,$\mu$m 
   for the PN profile. The shift of maximum between 
   the normalised median profiles of the 30\,$\mu$m 
   feature in AGB and post-AGB stars is 0.5\,$\mu$m 
   whereas between AGBs and PNe is 3.6\,$\mu$m.

%*************************************************************************************************************************
%*************************************************************************************************************************
%*************************************************************************************************************************
%*************************************************************************************************************************

%-------------------------------------------------------------
% ------- SYSTEMATICS OF THE METHOD ------
%-------------------------------------------------------------
\subsection{Systematics of the method}
\label{subsec:systematics}

   The key issue in the analysis of the 30 micron 
   feature is whether the shift in the $\lambda_{\rm c}$ 
   is an indication of changes in the feature itself, or it 
   is just an effect of the limited coverage of the Spitzer 
   spectra. To investigate that, we analysed some ISO 
   spectra from \citet{Hony:2002lr}. First, we made the 
   continuum fits the way that the Spitzer data were 
   fitted (24-36 $\mu$m), and then again with the full 
   wavelength coverage of the ISO spectra 
   (24-45 $\mu$m). This allowed us to see what 
   uncertainties are introduced by not having the 
   continuum sampled on the longer side of the feature. 
   This simple test showed that the values of the 
   $\lambda_{\rm c}$ obtained with the full range of 
   ISO spectra are always longer than those 
   corresponding to the Spitzer range. The median 
   of this difference is 3\% in case of AGBs, 7\% for 
   post-AGBs, and 5\% for PNe. It shows that, 
   despite introducing a systematic shift to the shorter 
   wavelength for the Spitzer range in all cases (AGBs, 
   post-AGBs, and PNe), our method of continuum fit
   demonstrates that the shift of the $\lambda_{\rm c}$ 
   in PNe is real and does not depend on the 
   environment 
   %%%%%(galaxy).\LEt{Please check I have retained your intended meaning. }
   
   In Table~\ref{tab:central_hony_vs_our} we list the 
   medians of the $\lambda_{\rm c}$ for the AGB stars, 
   post-AGB objects and PNe from \citet{Hony:2002lr} 
   and our work. The number of objects included into 
   a given median is shown in the square bracket. 
   The values in the last column of the Table are 
   achieved by multiplying the medians from our work 
   by the scaling factors discussed above. The scaled 
   values of the $\lambda_{\rm c}$ are almost the same 
   in case of AGB stars. On the other hand, for the 
   post-AGB objects they are larger by~1.1\,$\mu$m, 
   while in the case of PNe they are smaller by the 
   same value. These differences may be explained 
   on the ground that we use different method of fitting 
   the continuum. \citet{Hony:2002lr} use the modified 
   black-body continuum, while we use the black-body 
   with a single temperature. The increase of the 
   $\lambda_{\rm c}$ while going from 24-36 into 
   24-45 $\mu$m range concerns the Manchester 
   method itself. The values of the $\lambda_{\rm c}$  
   are close to each other for AGB stars, because the 
   continua of their spectra can be well defined by the 
   black-body with a single temperature (with the dust 
   emissivity index p = 0). When going into the 
   post-AGB objects and PNe, the continua are not so 
   well defined by a single black-body, and the 
   non-zero value of dust emissivity index is often 
   needed.

%*************************************************************************************************************************
%*************************************************************************************************************************
%
%______________________________________________________________
%
%_____________________________________________________________
%                                             Simple A&A Table
%_____________________________________________________________
%
\begin{table}[h!]
\scriptsize
\caption{Comparison of medians of the central wavelengths
obtained for ISO and Spitzer spectra. The square brackets 
contain the numbers of objects included into a given median.}             % title of Table
\label{tab:central_hony_vs_our}      % is used to refer this table in the text
\centering                          % used for centering table
\begin{tabular}{lccc}        % 4 columns
\hline\hline                 % inserts double horizontal lines
			&	$\lambda_{\rm c}$ 	&	$\lambda_{\rm c}$	&	$\lambda_{\rm c}$\tablefootmark{a}	\\
Source		&	\citep{Hony:2002lr}	&	(our work)			&	(scaled)	\\
			&	($\mu$m)			&	($\mu$m)			&	($\mu$m)	\\
\hline
AGBs		&	29.7 [36]	&	28.9 [91]	&	29.8	\\
post-AGBs	&	30.1 [14]	&	29.2 [35]	&	31.2	\\
PNe			&	33.3 [13]	&	30.7 [23]	&	32.2	\\
\hline                                   %inserts single line
\end{tabular}
\tablefoot{
\tablefoottext{a}{Result of multiplying the third column by the 
scaling factor obtained from the ISO spectra (AGBs: 1.03, 
post-AGBs: 1.07, and PNe: 1.05).}
}
\end{table}
% the [] bracket after \\ causes the problem, thus I made {[]}
%
%*************************************************************************************************************************
%*************************************************************************************************************************

%-------------------------------------------------------------
% ---------------------- SUMMARY ---------------------
%-------------------------------------------------------------
\subsection{Summary}
\label{subsec:summary}

   % SUMMARY - CONCLUSIONS, METHOD AND 
   % THE STRENGTH OF THE FEATURE
   In this paper we present an analysis of the 
   30\,$\mu$m dust feature as seen by Spitzer. 
   Our sample consisted of 207 objects being in 
   evolved stages of stellar evolution from four 
   galaxies: five from the Sgr dSph (four carbon-rich 
   AGB stars, one carbon-rich PN), 22 from the SMC 
   (eight carbon-rich AGB stars, three carbon-rich 
   post-AGB objects, and 11 PNe), 121 from the LMC 
   (83 carbon-rich AGBs, 17 carbon-rich post-AGBs, 
   and 21 carbon-rich PNe), and 59 from our Galaxy 
   (17 carbon-rich AGBs, 23 carbon-rich post-AGBs, 
   and 19 carbon-rich PNe). On the basis of this 
   sample, we have created three online catalogues 
   with photometric data and Spitzer IRS spectra for 
   the objects with 30\,$\mu$m feature from the SMC, 
   LMC, and the Milky Way, separately. Five objects 
   from the Sgr dSph were added to the Galactic 
   catalogue with a special comment.
   
   We applied the uniform method to fit the spectra 
   with a black-body of a single temperature deduced 
   from the Manchester method and its modifications. 
   However, use of this method does not guarantee 
   derivation of good continuum fits. We were unable 
   to obtain the correct fits for 33 objects (16\% of 
   whole sample). In addition, 17 PNe and five 
   post-AGB stars were also removed from the further 
   analysis, because of the presence of the 
   16-24$\,\mu$m feature in their Spitzer spectra. In 
   case of two planetary nebulae (SMP~LMC~51 and 
   79), and three post-AGB objects (IRAS~05370-7019, 
   IRAS~05537-7015, and IRAS~21546+4721), the 
   presence of the 16-24$\,\mu$m feature has been 
   recognised for the first time by us. The 16-24\,$\mu$m 
   feature affected the value of the [16.5]$-$[21.5] 
   colour and precluded determination of correct 
   continuum. Taking into account the above limitations 
   we were able to obtain a good fit to the continuum 
   for 152\footnote{This number includes three objects for which 
   only T$_{\rm d}$ is given in 
   Table~\ref{app_tab:lmc_spectral_results}: MSX~LMC~782, 
   MSX~LMC~787 showing the problem with calibration (see 
   Section~\ref{subsec:strength_central_wav} for details), and 
   MSX~LMC~974 with very noisy spectrum over 32\,$\mu$m.} 
   objects (73\%) from our initial sample. Then, 
   subtraction of continuum allowed us to determine the 
   parameters of the 30\,$\mu$m feature, like its central 
   wavelength, $\lambda_{\rm c}$, and the strength, 
   F/Cont. In addition, we determined the standard 
   Manchester colour indices, like the [6.4]$-$[9.3] 
   and [16.5]$-$[21.5], and some their modifications. 
   One of them was done because of the new 
   detections of the 16-18\,$\mu$m feature in the 
   spectra of nine PNe. We also found that the 
   Galactic post-AGB object IRAS~11339-6004 has a 
   21\,$\mu$m emission.
   
   Taking into account the Galactic sample of carbon 
   stars with the ISO spectra \citep[see][]{Sloan:2016aa}, 
   the strength of the 30\,$\mu$m feature is the highest 
   among Galactic objects. Moreover, this feature shows 
   up at the highest dust temperature for the Galactic 
   AGB stars. The first AGB objects with $30\,\mu$m 
   feature in the LMC are visible below 900\,K, whereas 
   such objects in the SMC and Sgr dSph objects do 
   not appear until T$_{\rm d}$ drops below 700\,K. 
   The strength of the feature increases until T$_{\rm d}$ 
   drops to about 400\,K, and then decreases to finally 
   show again variety of values for post-AGB objects 
   and PNe. The sample of the AGB stars is dominated 
   by the LMC objects. The explanation of this 
   behaviour lies in the fact, that it is much easier to 
   create a carbon-rich star in the metal-poor 
   environments \citep[see e.g.][]{Piovan:2003aa}. 
   However, such a regularity is not visible in the SMC 
   sample. The AGB objects with T$_{\rm d}$ < 400\,K, 
   seem to experience very large mass-loss rates, 
   which may be responsible for the drop in the strength 
   of the 30\,$\mu$m feature due to self-absorption. 
   During the post-AGB and PN phases, the strength 
   of the 30\,$\mu$m feature seems to not depend on 
   the metallicity of galaxy.

   % SUMMARY - CENTRAL WAVELENGTH 
   Our analysis of central wavelength of the 
   30\,$\mu$m feature shows that it is rather 
   independent of T$_{\rm d}$ for AGB and post-AGB 
   objects. The majority of the carbon-rich AGB and 
   post-AGB objects occupy the region of the 
   $\lambda_{\rm c}$ between about 28.5 and 
   29.5\,$\mu$m. There are, however, some exceptions. 
   These include: 
   a) AGB objects with 
   $\lambda_{\rm c} < 28$\,$\mu$m, which is probably 
   result of too high continuum derived, resulting in too 
   low $\lambda_{\rm c}$; 
   b) a group of ten AGB objects with 900\,K > T$_{\rm d}$ > 300\,K 
   and $\lambda_{\rm c}$ > 29.7\,$\mu$m, characterised 
   by rather low F/Cont, which may cause uncertain 
   determination of $\lambda_{\rm c}$ and its shift to the 
   longer wavelengths; and 
   c) an interesting group of six AGB stars with low dust 
   temperature and central wavelength larger than about 
   29.7\,$\mu$m. In the last case, the low dust 
   temperature suggests the large mass-loss rate, which 
   is confirmed by estimation of mass-loss rates by 
   \citet{Gruendl:2008aa}. In this sense, cold AGB dust 
   may have central wavelength of the 30\,$\mu$m 
   feature shifted towards longer values. On the other 
   hand even colder dust in post-AGB objects has a 
   $\lambda_{\rm c}$ similar to that of most AGB stars. 

   From our analysis, we see that the central 
   wavelengths shift towards longer values, typically 
   from 29.5 and 31.5\,$\mu$m, in case of PNe. 
   \cite{Hony:2002lr} suggested that such shift 
   is caused by the low temperature of the feature 
   carrier, or change in shape of dust particles. However, 
   we see several post-AGB objects with similar dust 
   temperature, but much lower central wavelength than 
   in PNe. Therefore, we expect that dust 
   processing (e.g. due to irradiation by UV photons from 
   central stars of PNe) may be more 
   important in shifting central wavelength of the 
   30\,$\mu$m feature towards larger values. Finally, we 
   note that it seems the Galactic PNe have rather 
   smaller $\lambda_{\rm c}$ values than their MCs 
   counterparts. If our suggestions are correct, this 
   could mean that processing of the 30\,$\mu$m 
   feature carrier is more efficient in the MCs than in 
   the Milky Way.

   % SUMMARY - PROFILES OF THE 30 MICRON FEATURE
   We have also searched for the median profiles of the 
   30\,$\mu$m feature in different galaxies and/or dust 
   temperature. Before determination of such profiles 
   spectra were normalised to the average flux in the 
   ranges typical for each class of objects and/or 
   resolution of the spectra. The average fluxes were 
   determined in range 28.5-29.5\,$\mu$m in case of 
   AGB stars, 28.5-30\,$\mu$m in case of post-AGB 
   objects, 30.5-32.5\,$\mu$m for low resolution 
   spectra of PNe, and 30-31\,$\mu$m for high 
   resolution spectra of PNe. We have shown that the 
   averaged shapes of the 30\,$\mu$m feature do 
   not change as a function of metallicity for 
   investigated objects. However, the shape of the 
   feature in PNe is clearly different 
   (not only the central wavelength) than shape of the 
   feature in AGB or post-AGB objects.

\begin{acknowledgements}
      This work was financially supported by the National 
      Science Center (NCN) of Poland through grant 
      No. 2014/15/N/ST9/04629. R. Sz. acknowledges 
      support from the NCN grant No. 2016/21/B/ST9/01626. 
      We have made extensive use of the SIMBAD and 
      Vizier databases operated at the Centre de 
      Donn\'ees Astronomiques de Strasbourg. We also 
      collected low and high resolution spectra from the 
      Combined Atlas of Sources with Spitzer IRS spectra 
      (CASSIS), which is a product of the IRS instrument 
      team, supported by NASA and JPL.
\end{acknowledgements}


%-------------------------------------------------------------
%-------------------------------------------------------------
% -------- THE BIBLIOGRAPHY - BIBTEX --------
%-------------------------------------------------------------
%-------------------------------------------------------------
\bibliographystyle{aa}
\bibliography{aa}
%
%__________________________________________________________________

%-------------------------------------------------------------
%-------------------------------------------------------------
%-------------------- APPENDICES --------------------
%-------------------------------------------------------------
%-------------------------------------------------------------

\begin{appendix}

%-------------------------------------------------------------
% ----------- APPENDIX - THE SAMPLE -----------
%-------------------------------------------------------------
\section{Full sample}
\label{sec:appendix_sample}

   In the case of post-AGB objects and PNe 
   from \cite{Sloan:2014fj}, we added one 
   PN (SMP~SMC~17) to the list of the 
   SMC objects, along with adding one post-AGB 
   object (SAGE~J051825-700532), and 11 PNe 
   (SMP~LMC~19, SMP~LMC~27, 
   SMP~LMC~34, SMP~LMC~36, SMP~LMC~38, 
   SMP~LMC~52, SMP~LMC~61, SMP~LMC~71, 
   SMP~LMC~75, SMP~LMC~78, and SMP~LMC~79) 
   to the list of the LMC objects. On the other hand, we 
   deleted four objects from the LMC (SMP~LMC~11 
   and 25; IRAS~05315-7145 and IRAS~05495-7034 
   were removed from the post-AGB list, because we 
   classified them as AGB stars). We present our 
   arguments for these changes below. 
   
   \textbf{SMP~SMC~17} is classified by 
   \cite{Stanghellini:2007aa} as an 
   intermediate-excitation PN ([Ar III], [S IV], and [Ne III] 
   nebular lines are visible) showing the PAH features 
   at 6.2, 7.7, and 8.6\,$\mu$m, which are blended with 
   a broader 8\,$\mu$m plateau feature, and 
   11.3\,$\mu$m at the top of a broader 12\,$\mu$m 
   plateau feature \citep[see e.g.][]{Kwok:2013aa}. The 
   Spitzer spectrum also shows a feature around 
   19\,$\mu$m, which could be attributed to the 
   fullerenes (C$_{\rm 60}$), but its identification seems 
   to be doubtful \citep{Garcia-Hernandez:2012qy}. We 
   also find the 16-18\,$\mu$m feature in the spectrum 
   of this object (see 
   Section~\ref{subsec:manchester_method} for details).
   
   \textbf{SAGE~J051825-700532} is classified by 
   \cite{Matsuura:2014fu} as a carbon-rich AGB star. 
   The Spitzer spectrum shows only weak SiC emission, 
   with no sign of the PAH bands, and an easily visible 
   30\,$\mu$m feature. We find an optical counterpart 
   0\arcsec.89 away (using 
   IRAC\footnote{IRAC: Infrared Array Camera on board Spitzer} 
   coordinates as the reference position), and if it is a 
   good match the SED suggests that we are dealing 
   with a post-AGB object having a detached dust shell. 
   We also find a matching entry in the OGLE-III 
   catalogue of long-period variables in the LMC 
   \citep{Soszynski:2009aa}. This object is identified 
   as an `OGLE small amplitude red giant' with two 
   pulsation periods of 16.8 and 15.9 days (I-band) 
   and the corresponding amplitudes of 0.006 and 
   0.008 mag. These pulsation periods are too short 
   for an AGB star, but would be consistent with 
   variability during the post-AGB phase. Taking all 
   these arguments into account, we classify 
   SAGE~J051825-700532 as a carbon-rich 
   post-AGB object.
   
   % FROM STANGHELLINI ET AL. 2007:
   % We designate the PNe to be low excitation if only the [S iii] 
   % lines are visible, intermediate excitation when the spectra 
   % shows [Ar iii], [S iv], and [Ne iii] nebular emission, high 
   % excitation if we also detect [O iv], and very high excitation if 
   % we also see the [Ne iv] or [Ne v] emission. Several PNe show 
   % both low- and high-excitation lines, likely due to different 
   % excitation zones in the nebulae.
   For the additional LMC PNe, five were taken 
   from \cite{Stanghellini:2007aa}. 
   \textbf{SMP~LMC~19} is classified by them as a 
   carbon-rich PN. The Spitzer spectrum shows PAHs, 
   especially the 11.3 band which is clearly visible on 
   top of the 12\,$\mu$m plateau, and nebular lines 
   such as [Ar III], [S III], [S IV], [Ne II], [Ne III], [Ne V], 
   and a very strong [O IV] line. The 16--18\,$\mu$m 
   feature and the broad 16--24 feature are visible as 
   well (see Section~\ref{subsec:manchester_method} 
   for details). This set of lines indicate an object with 
   high excitation PN \citep{Stanghellini:2007aa}. The 
   Spitzer spectrum of \textbf{SMP~LMC~27} is 
   classified by \cite{Stanghellini:2007aa} as a 
   featureless intermediate-excitation PN, which means 
   that there is a lack of visible dust features, except for 
   a clear 30\,$\mu$m feature. The Spitzer spectrum 
   shows [S III], [S IV], and [Ne III] nebular lines. 
   \textbf{SMP~LMC~34} is also classified by 
   \cite{Stanghellini:2007aa} as a featureless and 
   intermediate-excitation PN spectrum like 
   SMP~LMC~27. The 30\,$\mu$m feature is again the 
   only indicator of the carbon-rich nature of this object. 
   The most obvious nebular lines in the Spitzer 
   spectrum are: [S III], [S IV], [Ne II], and [Ne III]. The 
   Spitzer spectrum of \textbf{SMP~LMC~71} shows 
   clearly visible carbon-rich dust features: PAHs, 
   together with the 8, 12\,$\mu$m plateau 
   features, and the 16--18\,$\mu$m feature. The 
   [Ar III], [S III], [S IV], [Ne V], [Ne III], and [O IV] 
   nebular lines are visible in the spectrum, which imply 
   a very high excitation by a star with 
   T$_{\rm eff}$ $\sim$80\,000\,K, and 
   ${\rm log}\,{\rm L/L}_{\odot}$ < 4.27 \citep{Villaver:2003aa}. 
   The Spitzer spectrum of \textbf{SMP~LMC~79} is 
   dominated by PAH bands, together with the 8 and 12 
   $\mu$m plateau features, and the 16--18\,$\mu$m 
   feature. Another interesting dust emission feature 
   seen in the spectrum is a broad 16--24\,$\mu$m 
   feature. The nebular lines visible in the spectrum are 
   [S III], [S IV], [Ne III], and [O IV], which are 
   characteristic for high excitation PN 
   \citep{Stanghellini:2007aa}.
   
   Another four LMC PNe were taken from 
   \citet{Bernard-Salas:2009aa}. According to the 
   authors, all of them are the carbon-rich sources. 
   The Spitzer spectrum of the first object, 
   \textbf{SMP~LMC~36}, shows very strong PAH 
   features as in the spectrum of SMP~LMC~79, 
   including the 8 and 12\,$\mu$m plateau features, and 
   the 16-18\,$\mu$m feature, but with a lack of the 
   broad 16-24\,$\mu$m feature. The nebular lines 
   visible in the spectrum are [Ar III], [S III], [S IV], 
   [Ne III], [Ne V], and [O IV], which indicate a high 
   excitation PN. \textbf{SMP~LMC~38} has a [WR] 
   central star and its spectrum shows strong PAHs with 
   the 8 and 12\,$\mu$m plateau features, and the 
   16-18\,$\mu$m feature. The nebular [S III], [S IV], 
   and [Ne III] lines are visible in the spectrum; thus we 
   are dealing with an intermediate-excitation PN. 
   \textbf{SMP~LMC~61} has also a [WR] central star 
   and its spectrum shows strong PAHs with the 8 and 
   12\,$\mu$m plateau features, the 16--18\,$\mu$m 
   feature, and nebular lines typical for 
   intermediate-excitation PN: [Ar III], [S III], [S IV], 
   [Ne II], and [Ne III]. The next object in this group is 
   \textbf{SMP~LMC~78}. The Spitzer spectrum shows 
   the easily visible PAHs with the 8 and 12\,$\mu$m 
   plateau features, the 16-18\,$\mu$m feature, the 
   broad 16-24 feature, and nebular lines typical for a 
   high excitation PN: [Ar III], [S III], [S IV], [Ne II], 
   [Ne III], [Ne V], and [O IV].
   
   The last two PNe that we have added have been 
   analysed by \cite{Matsuura:2014fu}. The spectrum 
   of \textbf{SMP~LMC~52} does not show any sign of 
   PAHs. However, it has a very clearly visible 
   30\,$\mu$m feature, which is the only indicator of 
   carbon-rich nature of this object. The presence of 
   the nebular lines of [Ar III], [S III], [S IV], [Ne III], 
   [Ne V], and [O IV] indicates that this object is a high 
   excitation PN. The spectrum of 
   \textbf{SMP~LMC~75} contains strong PAHs, the 8 
   and 12 plateau features, and the 16-18\,$\mu$m 
   feature. In the spectrum the nebular [S III], [S IV], 
   [Ne II] and [Ne III] lines are visible, which 
   corresponds to this being an intermediate-excitation 
   PN.
   
   We rejected two objects from the \cite{Sloan:2014fj} 
   LMC sample. \textbf{SMP~LMC~11} is a member of 
   their `red group' for which the only requirement is a 
   red continuum. It shows weak PAH features, but we 
   conclude that the spectrum does not contain a 
   30\,$\mu$m feature. \textbf{SMP~SMC~25} is also a 
   member of the red group, but we rejected this object, 
   because crystalline silicates are visible.
   
   In addition, the \cite{Sloan:2014fj} sample contains 
   two objects from the LMC with the 30\,$\mu$m 
   feature, but which do not have a well defined 
   classification: IRAS~05315-7145 and 
   IRAS~05495-7034. Both of them are assigned to 
   the red group in their paper. Here, we classify them 
   from the literature. The first object, 
   \textbf{IRAS~05315-7145}, is classified by 
   \citet{Gruendl:2008aa} as an ERO. The bolometric 
   luminosity of our target is 9\,500\,${\rm L}_{\odot}$, 
   whereas the estimated mass-loss rate is 
   $1.7\times 10^{-4}\,{\rm M}_{\odot}\,{\rm yr}^{-1}$. 
   \cite{Woods:2011aa} classify it as a carbon-rich 
   AGB star. However, we do not find any information 
   about the long period variability, which is 
   characteristic of AGB stars. The Spitzer spectrum 
   shows the 30\,$\mu$m feature, but the SiC is not 
   well visible and there is a lack of C$_{2}$H$_{2}$ 
   absorption. On the basis of the collected information, 
   we assume this is a carbon-rich AGB star. The 
   second object, \textbf{IRAS~05495-7034} is 
   assigned by \citet{Gruendl:2008aa} to the ERO 
   group, with a bolometric luminosity 
   11\,100\,${\rm L}_{\odot}$ and an estimated mass-loss 
   rate of $2.3\times 10^{-4}\,{\rm M}_{\odot}\,{\rm yr}^{-1}$. 
   \cite{Woods:2011aa} mark this object as a 
   carbon-rich AGB star. The SED of the source in 
   our catalogue (see 
   Section~\ref{subsec:torun_catalogues}) is typical 
   of an AGB star. Because of the classification 
   of IRAS~05315-7145 and IRAS~05495-7034 as 
   carbon-rich AGB stars, these objects have been 
   moved from the post-AGB sample of 
   \citet{Sloan:2014fj}, and treated as the additions 
   to the carbon-rich AGB sample of 
   \citet{Sloan:2016aa}.
   
   We re-examined the sample from 
   \citet{Sloan:2016aa}, which consists of 184 
   carbon-rich AGB stars from the MCs, and we 
   selected the 89 objects showing the 30\,$\mu$m 
   feature. Our sample contains almost all the objects 
   for which \citet{Sloan:2016aa} reported the values 
   of the strength of the 30\,$\mu$m feature (see 
   Section~\ref{subsec:strength_central_wav} 
   for the definition of this parameter). However, after 
   re-analysing the spectra, we excluded 
   \textbf{MSX~LMC~036} and 
   \textbf{OGLE~J051306.52-690946.4} because we 
   did not detect the 30\,$\mu$m feature (see 
   Section~\ref{subsec:strength_central_wav} for details). 
   Moreover, our sample contains two objects 
   (\textbf{IRAS~05026-6809} and 
   \textbf{MSX~LMC~494}) for which the values of the 
   strength of the feature were previously not 
   determined. However, we consider them as true 
   30\,$\mu$m sources, because the features are 
   measurable, and the end of the Spitzer spectrum 
   is turning down.
   
   We found four carbon-rich AGB stars in the Sgr dSph 
   galaxy, which were analysed by 
   \cite{Lagadec:2009aa}: \textbf{Sgr~3} 
   (IRAS~F18413-3040), \textbf{Sgr~7} 
   (IRAS~F18436-2849), 
   \textbf{Sgr~15} (IRAS~F18555-3001), and 
   \textbf{Sgr~18} (IRAS~19013-3117). All of them are 
   long-period pulsators with periods between 417 and 
   485 days. Their Spitzer spectra show typical 
   carbon-rich dust feature of SiC, and molecular 
   absorptions of C$_{2}$H$_{2}$ at 7.5 and 
   13.7\,$\mu$m. We found only one PN in this galaxy 
   with a Spitzer spectrum, \textbf{Wray16-423}, which 
   was analysed in detail by \cite{Otsuka:2015aa}. 
   This spectrum shows the weak 8 and 12\,$\mu$m 
   plateau features as well as the broad 
   16--24\,$\mu$m feature.
   
   The Galactic sample analysed here consists of 
   59 objects (carbon-rich AGBs, post-AGBs, and 
   PNe). The part of our sample containing the 
   carbon-rich AGB objects is different from the work 
   by \cite{Sloan:2016aa}, because their list is based 
   on ISO spectra. We found ten S-type stars, which are 
   in the transition phase between M-type and C-type 
   AGB stars \citep{Smolders:2012aa}: 
   \textbf{CSS~987}, \textbf{GY~Lac}, \textbf{CSS~1}, 
   \textbf{CSS~466}, \textbf{CSS~657}, 
   \textbf{CSS~380}, \textbf{CSS~661}, 
   \textbf{CSS~749}, \textbf{CSS~438}, and 
   \textbf{AO~Gem}. \citet{Lagadec:2012aa} analysed 
   seven objects towards the Galactic halo: 
   \textbf{IRAS~04188+0122}, \textbf{IRAS~08427+0338}, 
   \textbf{Lyng\aa~7~V1}, \textbf{IRAS~16339-0317}, 
   \textbf{IRAS~18120+4530}, \textbf{IRAS~18384-3310}, 
   and \textbf{IRAS~19074-3233}. These objects' 
   membership of the Galactic halo or the Galactic thick 
   disc is established by 
   \citep{Lagadec:2010aa, Lagadec:2012aa}. According 
   to the authors, IRAS~04188, IRAS~08427, and 
   Lyng\aa~7~V1 are members of the Galactic thick disc 
   population, whereas IRAS~16339, IRAS~18120, 
   IRAS~18384, and IRAS~19074 belong to the Galactic 
   halo. In addition, IRAS~18384 and IRAS~19074 are 
   included to the analysis of carbon-rich AGB stars in 
   the Sgr dSph galaxy \citep{Lagadec:2009aa}, but 
   their radial velocity study of the optical spectra 
   shows that they clearly belong to the Milky Way 
   (i.e.~they are in the foreground). Furthermore, the 
   Spitzer spectrum of Lyng\aa~7~V1 is first published 
   by \cite{Sloan:2010aa} after their observations of 
   Galactic globular clusters. It is not clear whether this 
   object is a true member of the globular cluster or not.
   
   The most numerous sources in the Milky Way sample 
   are carbon-rich post-AGB objects. Seven of them 
   were analysed by \cite{Hrivnak:2009aa}: 
   \textbf{IRAS~05113+1347}, \textbf{IRAS~05341+0852}, 
   \textbf{IRAS~06530-0213}, \textbf{IRAS~07430+1115}, 
   \textbf{IRAS~19477+2401}, \textbf{IRAS~22574+6609}, 
   and \textbf{IRAS~23304+6147}. All of them have high 
   resolution spectra, and they possess the 21\,$\mu$m 
   and 30\,$\mu$m features. The short-high part of the 
   Spitzer spectrum of IRAS~19477 is unusable (probably 
   due to a bad positioning; see \cite{Hrivnak:2009aa} 
   for details). IRAS~23304 has been previously 
   observed by the ISO mission, and it is included in 
   the original paper reporting the discovery of the 
   21\,$\mu$m feature \citep{Kwok:1989aa}. The 
   discovery list contains one more object in the current 
   sample -- \textbf{IRAS~04296+3429}. The Spitzer 
   spectrum of this source was investigated by 
   \citet{Zhang:2010aa}, as well as the spectra of the 
   carbon-rich post-AGB objects 
   \textbf{IRAS~22223+4327} and 
   \textbf{IRAS~01005+7910}. In the case of 
   IRAS~01005 the 21\,$\mu$m is not visible, though 
   the broad 16-24\,$\mu$m feature appears in the 
   spectrum, reported by \cite{Zhang:2010aa}. This 
   object is also the first example of a carbon-rich 
   post-AGB source in which fullerene C$_{\rm 60}$ 
   was found \citep{Zhang:2011aa}. Another three 
   carbon-rich post-AGB objects were analysed by 
   \cite{Cerrigone:2011aa} and show the 
   21\,$\mu$m feature: \textbf{IRAS~13245-5036}, 
   \textbf{IRAS~14429-4539}, and 
   \textbf{IRAS~15482-5741}. In the low-resolution 
   Spitzer spectrum of IRAS~13245, a big shift 
   between the SL and LL segments is visible. 
   Seven additional carbon-rich post-AGB objects 
   were included after inspection of the Spitzer 
   spectra from Spitzer General Observer program 
   30258 (PI: P. Garcia-Lario). They are: 
   \textbf{IRAS~11339-6004}, \textbf{IRAS~15038-5533}, 
   \textbf{IRAS~15229-5433}, \textbf{IRAS~15531-5704}, 
   \textbf{IRAS~16296-4507}, \textbf{IRAS~18533+0523}, 
   and \textbf{IRAS~21525+5643}. There are two 
   21\,$\mu$m sources among this group: IRAS~11339, 
   which is a new detection in our work, and IRAS~18533. 
   The inspection of the Spitzer spectra from program 
   50116 (PI: G. Fazio) provided us with another three 
   carbon-rich post-AGB objects: 
   \textbf{IRAS~12145-5834}, 
   \textbf{IRAS~14325-4628}, and 
   \textbf{IRAS 21546+4721}. One object in this group, 
   IRAS~21546, shows the emission of fullerene 
   \citep{Raman:2017aa}.
   
   The list of carbon-rich Galactic PNe is based on 15 
   objects observed by \citet{Stanghellini:2012aa}: 
   \textbf{PN~K3-19}, \textbf{PN~K3-31}, 
   \textbf{PN~M1-71}, \textbf{PN~K3-37}, 
   \textbf{PN~K3-39}, \textbf{PN~K3-54}, 
   \textbf{PN~K3-62}, \textbf{PN~M1-5}, 
   \textbf{PN~M1-12}, \textbf{PN~PB~2}, 
   \textbf{Hen~2-5}, \textbf{Hen~2-26}, 
   \textbf{Hen~2-41}, \textbf{Hen~2-68}, 
   and \textbf{Hen~2-115}. Four objects in this group 
   show fullerene emission: PN~K3-54, PN~K3-62, 
   PN~M1-12, and Hen~2-68. In addition, 
   \citet{Otsuka:2013aa} report the presence of the 
   16--24\,$\mu$m feature in the spectrum of 
   PN~M1-12. The presence of this feature in the 
   spectrum that we have analysed is not obvious, thus 
   we have not removed it from further analysis (see 
   Section \ref{subsec:manchester_method}). The 
   sample of PNe is supplemented by two carbon-rich 
   objects from \citet{Perea-Calderon:2009aa}, who 
   observed PNe in the Galactic bulge: 
   \textbf{PN~M1-20} and \textbf{PN~Tc~1}. The 
   planetary nebula Tc~1 is the best example of a 
   fullerene source \citep{Cami:2010aa}. The 
   spectrum of PN~M1-20 also shows fullerene 
   emission. In addition the list of the 
   fullerene-containing PNe includes the carbon-rich 
   object \textbf{PN~M1-11}, studied in detail by 
   \citet{Otsuka:2013aa}. The authors also report the 
   presence of the broad 16-24\,$\mu$m feature in this 
   object. The last carbon-rich PN in the Galactic 
   sample is \textbf{PN~K3-60}, which was included 
   after inspection of data from the program 
   30430 (PI: H. Dinerstein).

%-------------------------------------------------------------
% ----------  APPENDIX - ALL SPECTRA ----------
%-------------------------------------------------------------
\section{Whole spectra}
\label{sec:appendix_all_spectra}

      \Cref{appfig:all_spectra_agb_1,appfig:all_spectra_agb_2,%
      appfig:all_spectra_agb_3,appfig:all_spectra_agb_4,%
      appfig:all_spectra_agb_5,appfig:all_spectra_agb_6,%
      appfig:all_spectra_pagb_1,appfig:all_spectra_pagb_2,%
      appfig:all_spectra_pn_1,appfig:all_spectra_pn_2,%
      appfig:all_spectra_pn_3} show the Spitzer spectra, 
      which have been analysed in this work.

%*************************************************************************************************************************
%*************************************************************************************************************************
\begin{figure*}
\includegraphics[width=17.5cm]{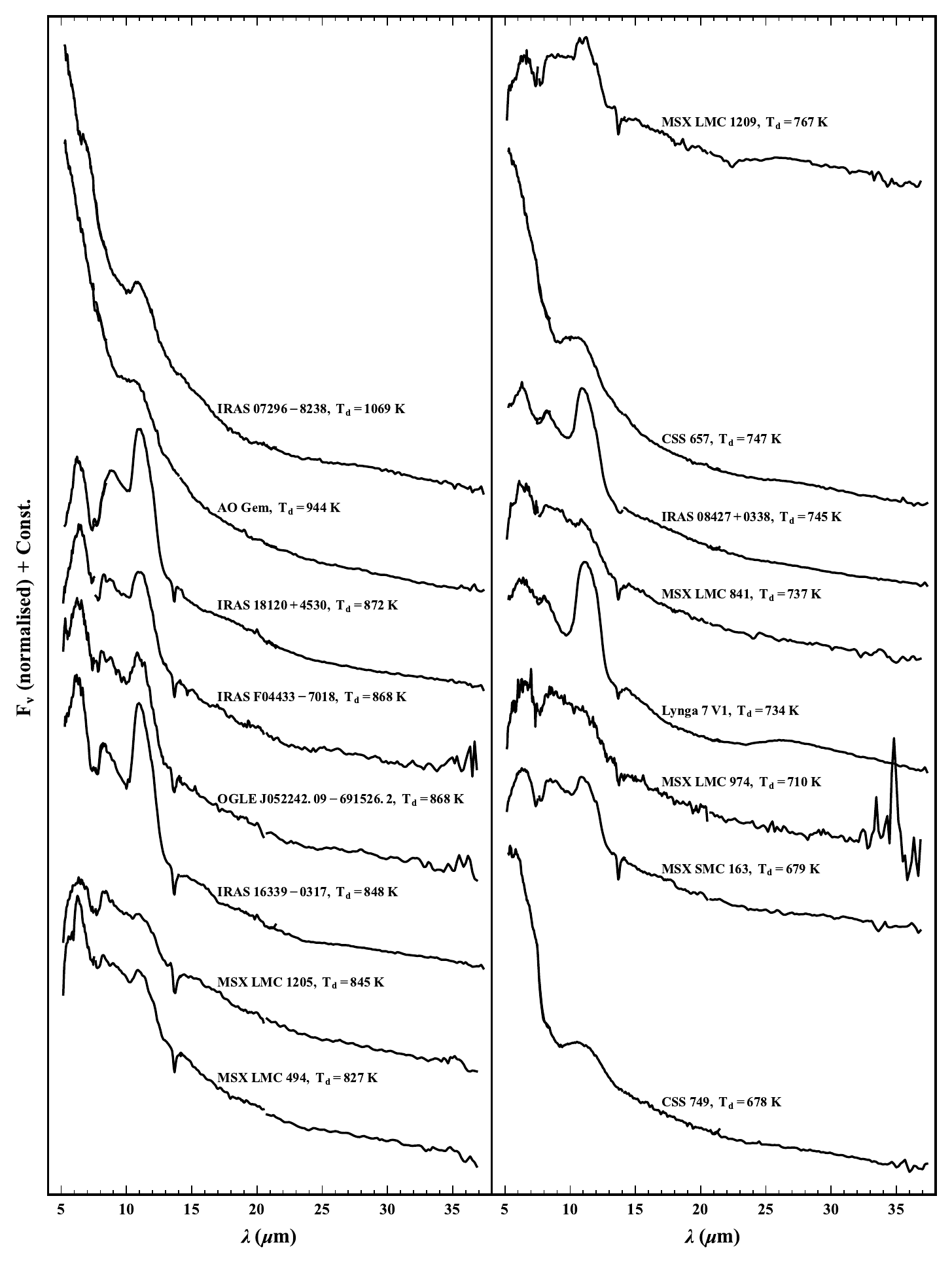}
  \caption{Overview of the Spitzer spectra (black solid 
  lines) of the carbon-rich AGB stars with the 
  30\,$\mu$m feature. The spectra are ordered 
  according to the T$_{\rm d}$ from high to low 
  temperature, top to bottom, and left to right, and 
  are normalised to the flux density at 18\,$\mu$m and 
  offset for clarity. The names of objects and values 
  of the T$_{\rm d}$ are shown above the spectra.}
  \label{appfig:all_spectra_agb_1}
\end{figure*}
%*************************************************************************************************************************
%*************************************************************************************************************************

%*************************************************************************************************************************
%*************************************************************************************************************************
\begin{figure*}
\includegraphics[width=17.5cm]{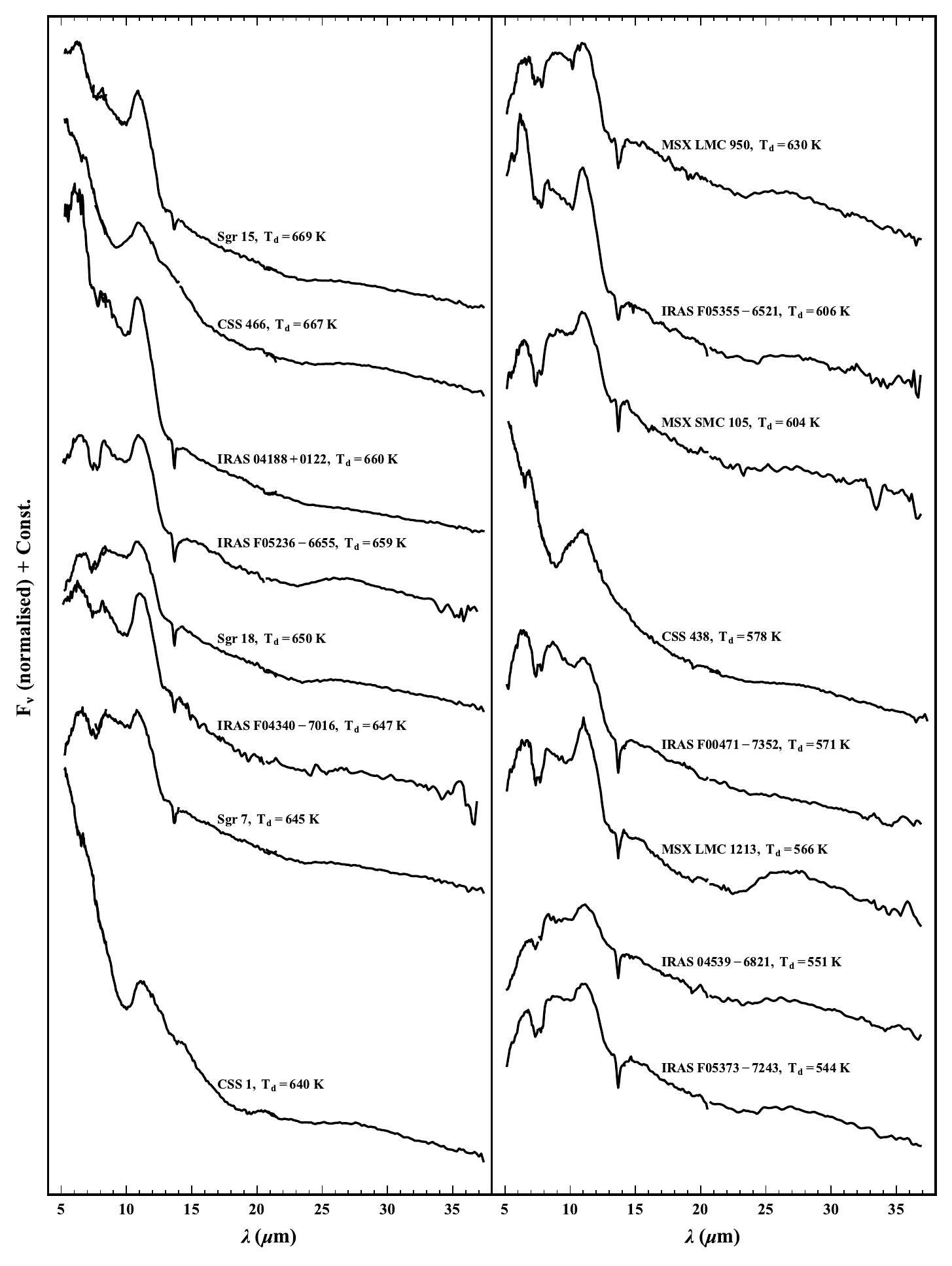}
  \caption{Overview of the Spitzer spectra (black solid 
  lines) of the carbon-rich AGB stars with the 
  30\,$\mu$m feature. The spectra are ordered 
  according to the T$_{\rm d}$ from high to low 
 temperature, top to bottom, and left to right, and 
 are normalised to the flux density at 18\,$\mu$m and 
 offset for clarity. The names of objects and values 
 of the T$_{\rm d}$ are shown above the spectra.}
  \label{appfig:all_spectra_agb_2}
\end{figure*}
%*************************************************************************************************************************
%*************************************************************************************************************************

%*************************************************************************************************************************
%*************************************************************************************************************************
\begin{figure*}
\includegraphics[width=17.5cm]{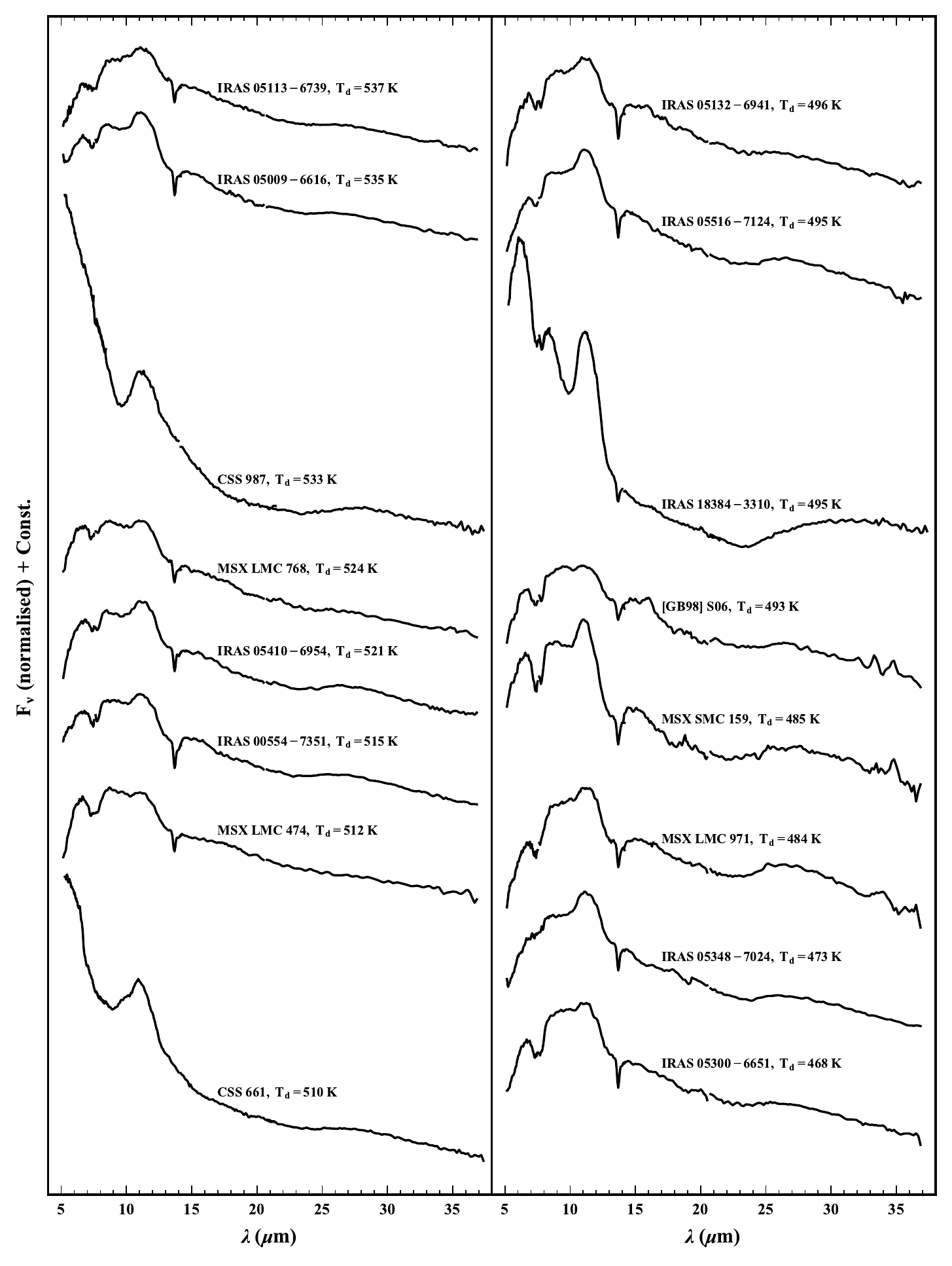}
  \caption{Overview of the Spitzer spectra (black solid 
  lines) of the carbon-rich AGB stars with the 
  30\,$\mu$m feature. The spectra are ordered 
  according to the T$_{\rm d}$ from high to low 
  temperature, top to bottom, and left to right, and 
  are normalised to the flux density at 18\,$\mu$m and 
  offset for clarity. The names of objects and values 
  of the T$_{\rm d}$ are shown above the spectra.}
  \label{appfig:all_spectra_agb_3}
\end{figure*}
%*************************************************************************************************************************
%*************************************************************************************************************************

\clearpage % the lack of this causes that a compilation becomes impossible (too many figures in the paper)
%*************************************************************************************************************************
%*************************************************************************************************************************
\begin{figure*}
\includegraphics[width=17.5cm]{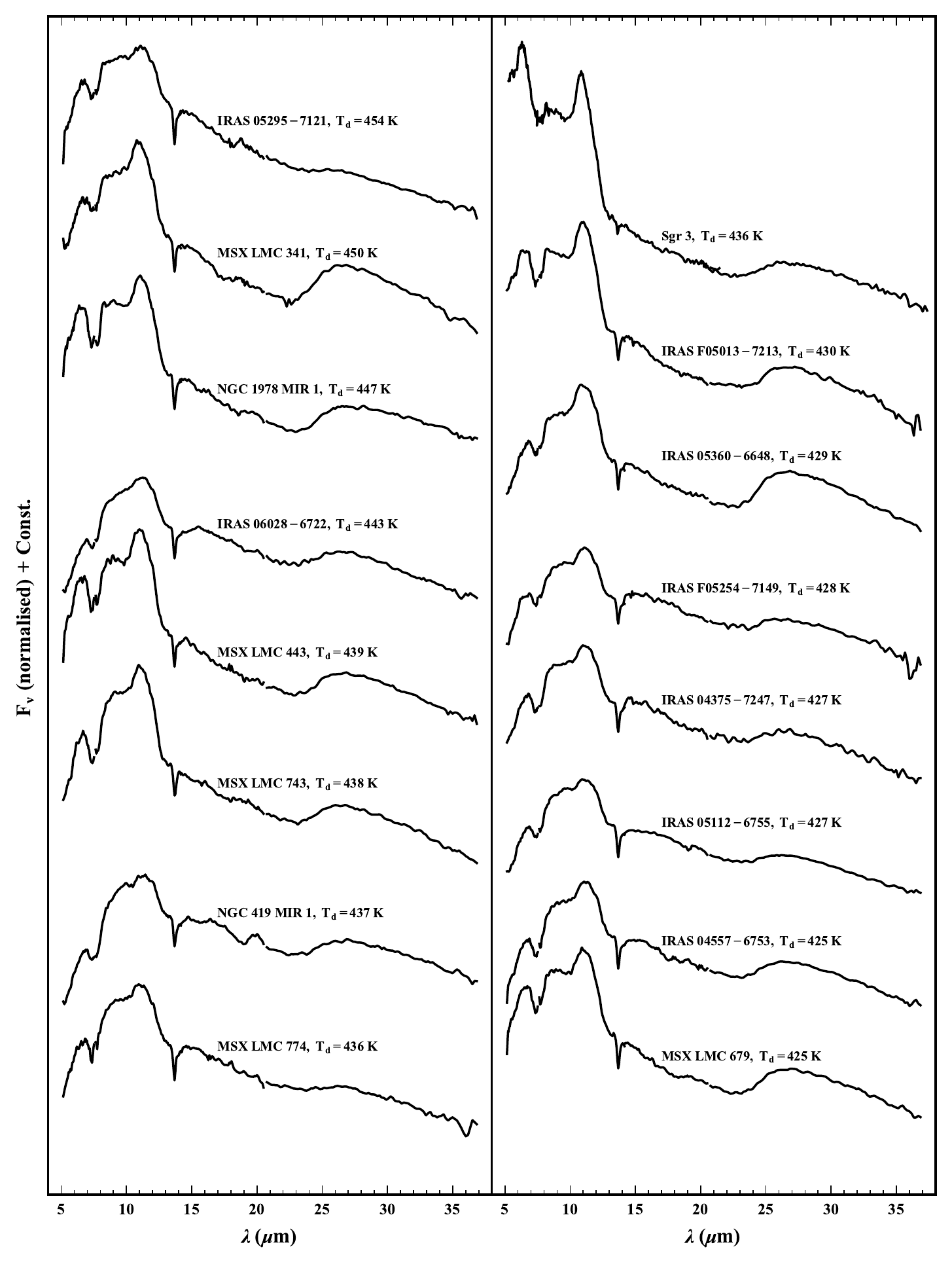}
  \caption{Overview of the Spitzer spectra (black solid 
  lines) of the carbon-rich AGB stars with the 
  30\,$\mu$m feature. The spectra are ordered 
  according to the T$_{\rm d}$ from high to low 
 temperature, top to bottom, and left to right, and 
 are normalised to the flux density at 18\,$\mu$m and 
 offset for clarity. The names of objects and values 
 of the T$_{\rm d}$ are shown above the spectra.}
  \label{appfig:all_spectra_agb_4}
\end{figure*}
%*************************************************************************************************************************
%*************************************************************************************************************************

%*************************************************************************************************************************
%*************************************************************************************************************************
\begin{figure*}
\includegraphics[width=17.5cm]{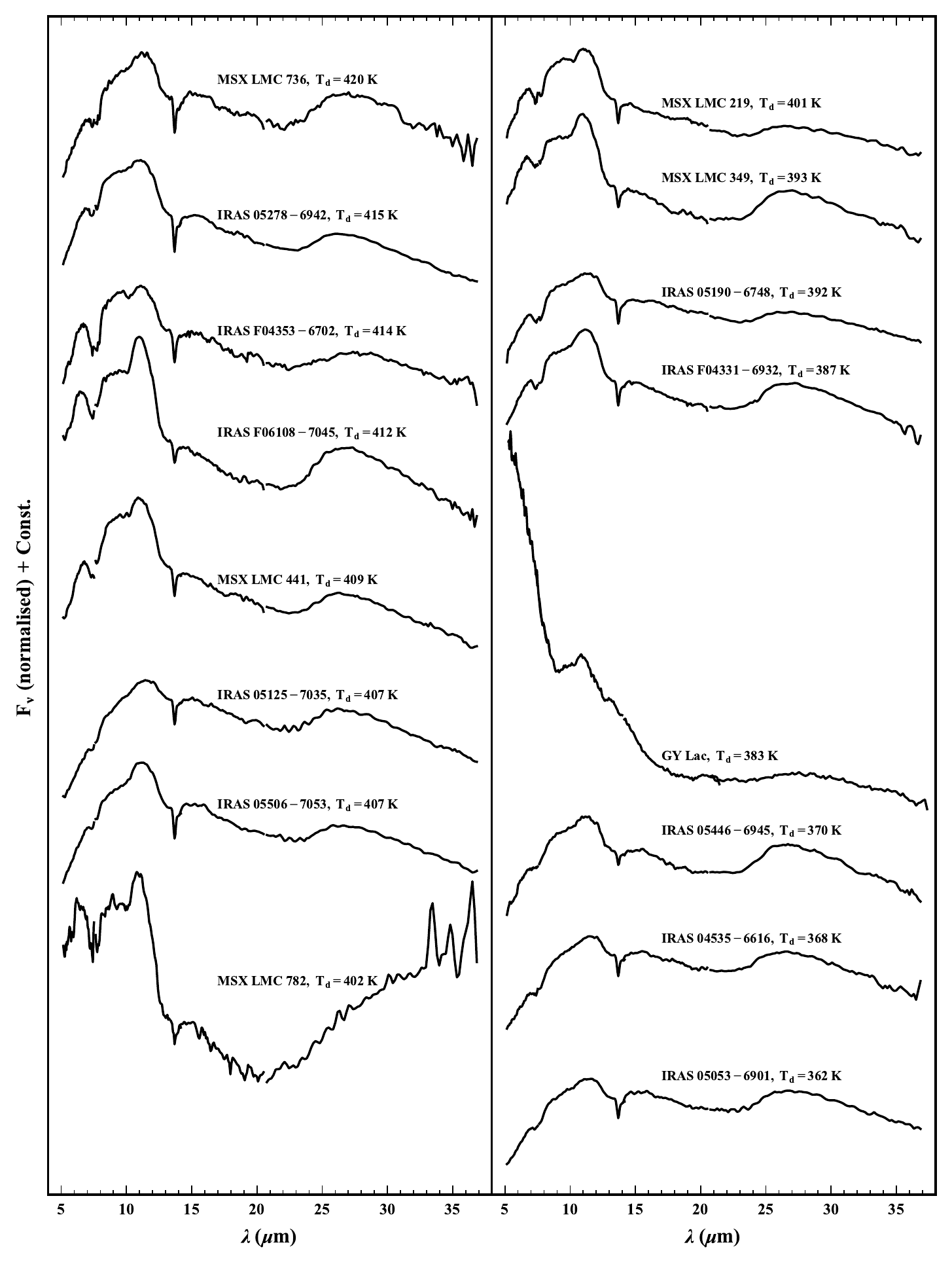}
  \caption{Overview of the Spitzer spectra (black solid 
  lines) of the carbon-rich AGB stars with the 
  30\,$\mu$m feature. The spectra are ordered 
  according to the T$_{\rm d}$ from high to low 
  temperature, top to bottom, and left to right, and 
  are normalised to the flux density at 18\,$\mu$m and 
  offset for clarity. The names of objects and values 
  of the T$_{\rm d}$ are shown above the spectra.}
  \label{appfig:all_spectra_agb_5}
\end{figure*}
%*************************************************************************************************************************
%*************************************************************************************************************************

%*************************************************************************************************************************
%*************************************************************************************************************************
\begin{figure*}
\includegraphics[width=17.5cm]{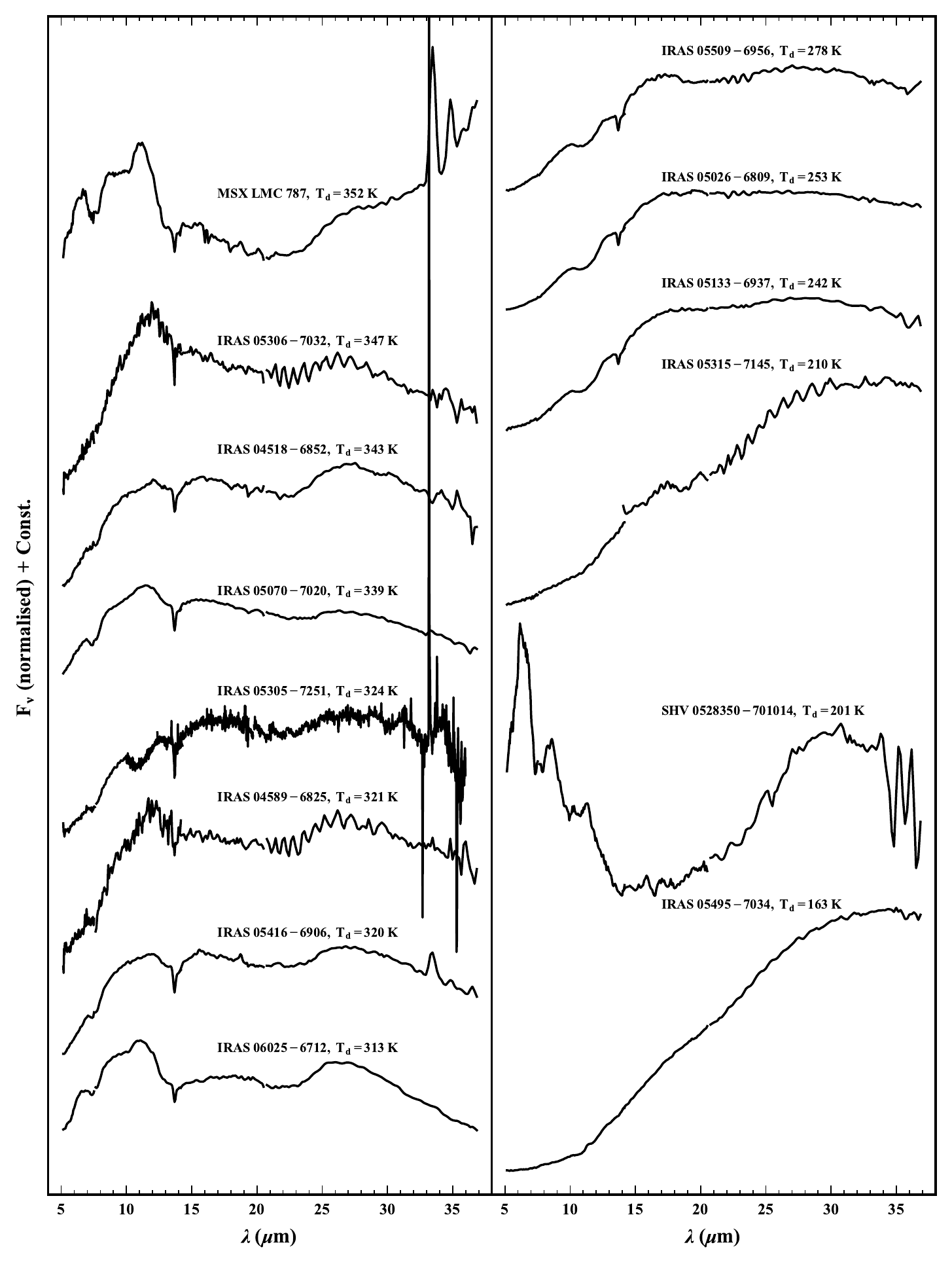}
  \caption{Overview of the Spitzer spectra (black solid 
  lines) of the carbon-rich AGB stars with the 
  30\,$\mu$m feature. The spectra are ordered 
  according to the T$_{\rm d}$ from high to low 
  temperature, top to bottom, and left to right, and 
  are normalised to the flux density at 18\,$\mu$m and 
  offset for clarity. The names of objects and values 
  of the T$_{\rm d}$ are shown above the spectra.}
  \label{appfig:all_spectra_agb_6}
\end{figure*}
%*************************************************************************************************************************
%*************************************************************************************************************************

%*************************************************************************************************************************
%*************************************************************************************************************************
\begin{figure*}
\includegraphics[width=17.5cm]{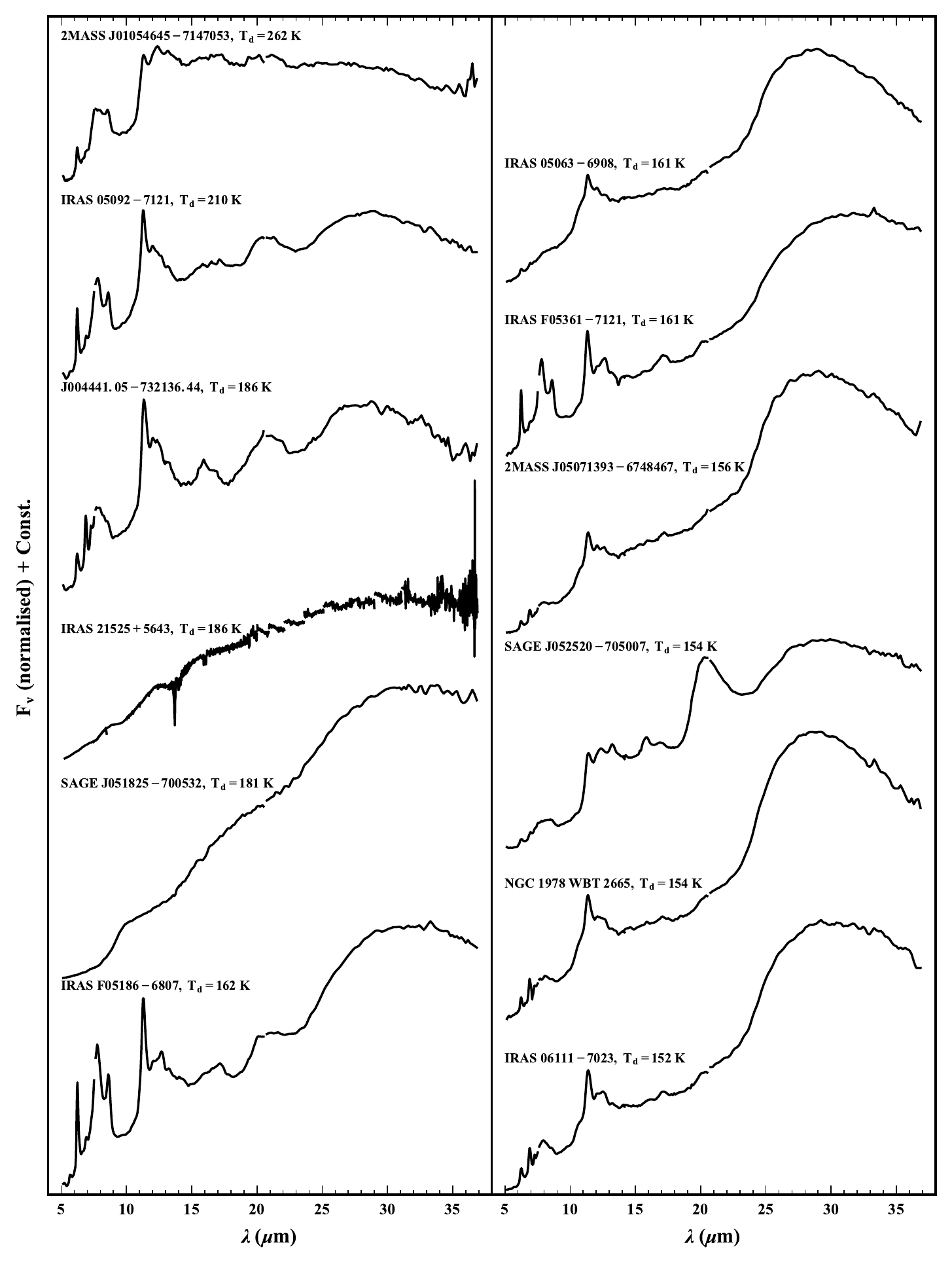}
  \caption{Overview of the Spitzer spectra (black solid 
  lines) of the carbon-rich post-AGB stars with the 
  30\,$\mu$m feature. The spectra are ordered 
  according to the T$_{\rm d}$ from high to low 
  temperature, top to bottom, and left to right, and 
  are normalised to the flux density at 18\,$\mu$m and 
  offset for clarity. The names of objects and values 
  of the T$_{\rm d}$ are shown above the spectra.}
  \label{appfig:all_spectra_pagb_1}
\end{figure*}
%*************************************************************************************************************************
%*************************************************************************************************************************

%*************************************************************************************************************************
%*************************************************************************************************************************
\begin{figure*}
\includegraphics[width=17.5cm]{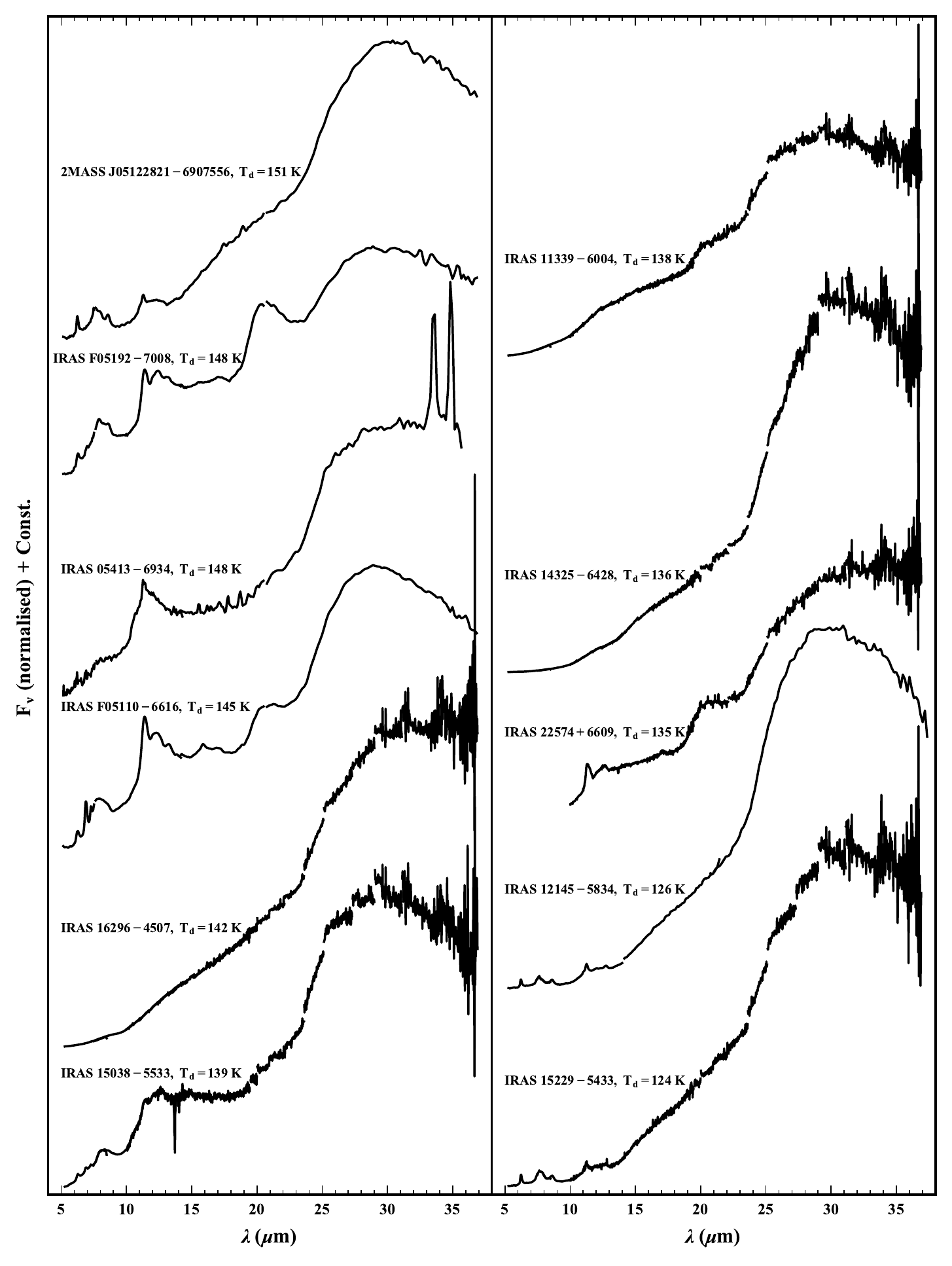}
  \caption{Overview of the Spitzer spectra (black solid 
  lines) of the carbon-rich post-AGB stars with the 
  30\,$\mu$m feature. The spectra are ordered 
  according to the T$_{\rm d}$ from high to low 
  temperature, top to bottom, and left to right, and 
  are normalised to the flux density at 18\,$\mu$m and 
  offset for clarity. The names of objects and values 
  of the T$_{\rm d}$ are shown above the spectra.}
  \label{appfig:all_spectra_pagb_2}
\end{figure*}
%*************************************************************************************************************************
%*************************************************************************************************************************

%*************************************************************************************************************************
%*************************************************************************************************************************
\begin{figure*}
\includegraphics[width=17.5cm]{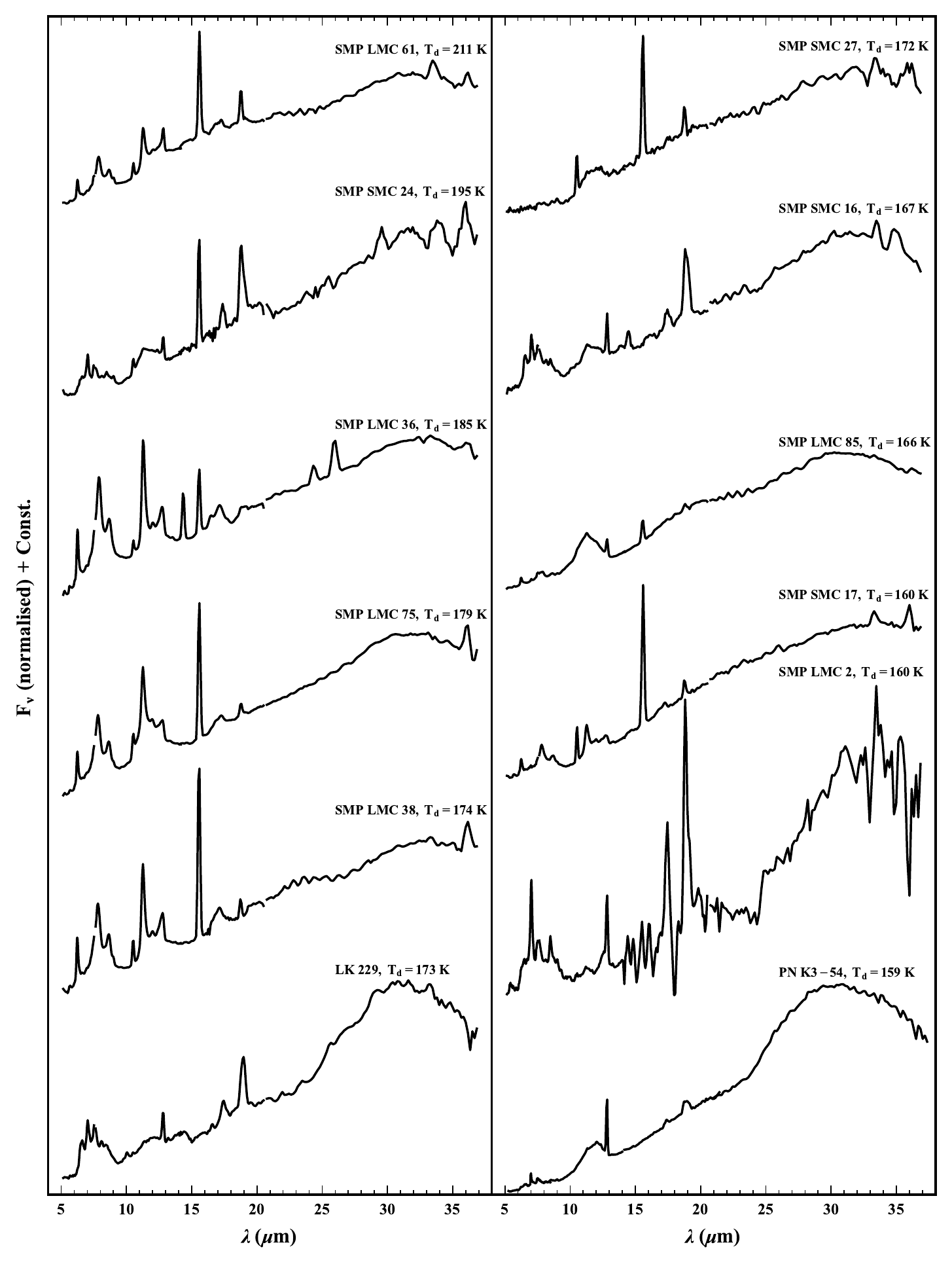}
  \caption{Overview of the Spitzer spectra (black solid 
  lines) of the carbon-rich PNe with the 
  30\,$\mu$m feature. The spectra are ordered 
  according to the T$_{\rm d}$ from high to low 
  temperature, top to bottom, and left to right. They 
  are also normalised to the flux density at 18\,$\mu$m 
  (in case of SMP~LMC~2 it is 18.4\,$\mu$m) and 
  offset for clarity. The names of objects and values 
  of the T$_{\rm d}$ are shown above spectra.}
  \label{appfig:all_spectra_pn_1}
\end{figure*}
%*************************************************************************************************************************
%*************************************************************************************************************************

%*************************************************************************************************************************
%*************************************************************************************************************************
\begin{figure*}
\includegraphics[width=17.5cm]{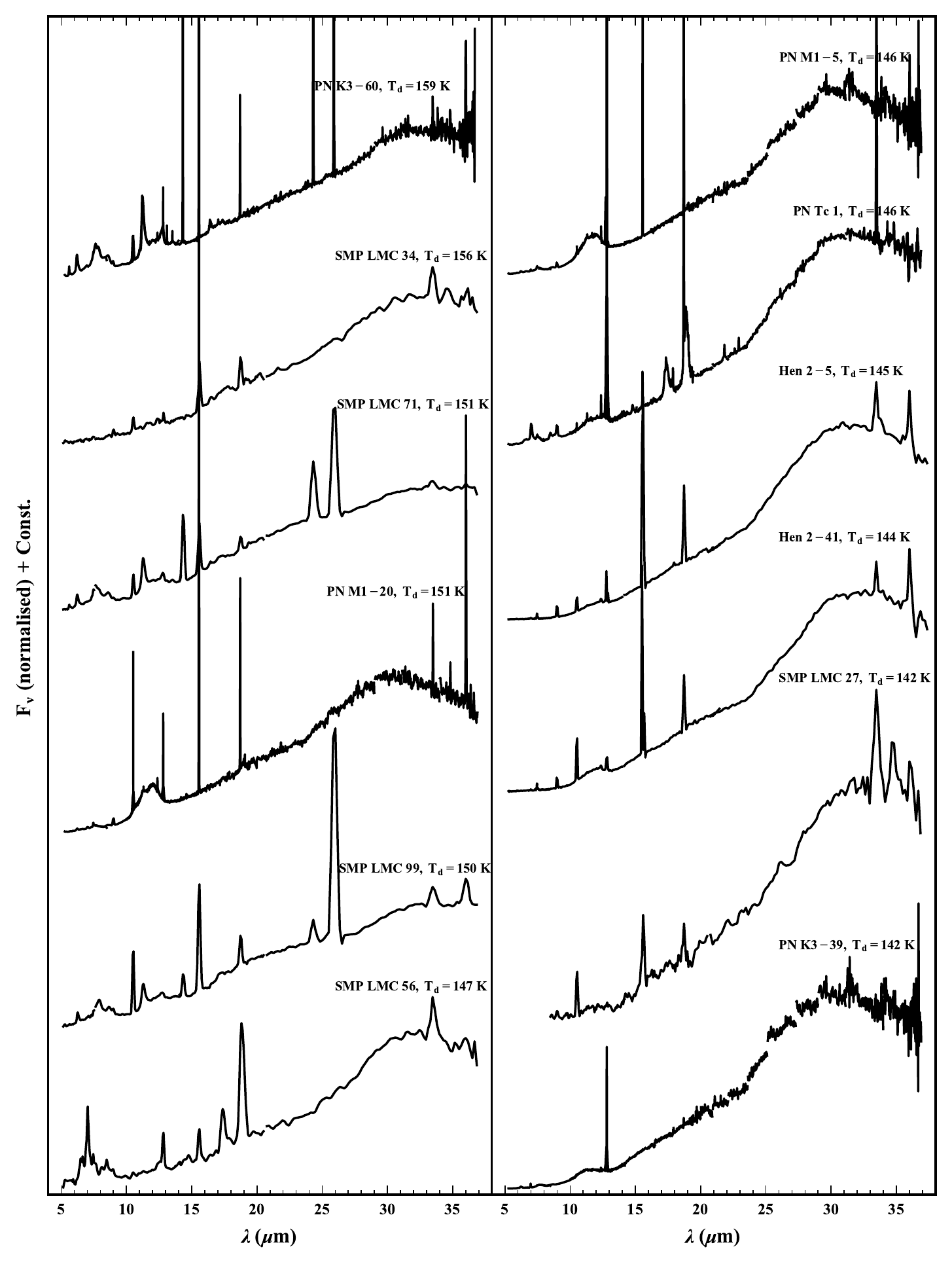}
  \caption{Overview of the Spitzer spectra (black solid 
  lines) of the carbon-rich PNe with the 
  30\,$\mu$m feature. The spectra are ordered 
  according to the T$_{\rm d}$ from high to low 
  temperature, top to bottom, and left to right, and are 
  normalised to the flux density at 18\,$\mu$m and 
  offset for clarity. In case of SMP~LMC~27, the part 
  of spectrum between around 5 -- 8.5\,$\mu$m is not 
  shown because it is very noisy and makes other 
  spectra less readable in this region. The names of 
  objects and values of the T$_{\rm d}$ are shown 
  above the spectra.}
  \label{appfig:all_spectra_pn_2}
\end{figure*}
%*************************************************************************************************************************
%*************************************************************************************************************************

%*************************************************************************************************************************
%*************************************************************************************************************************
\begin{figure*}
\includegraphics[width=17.5cm]{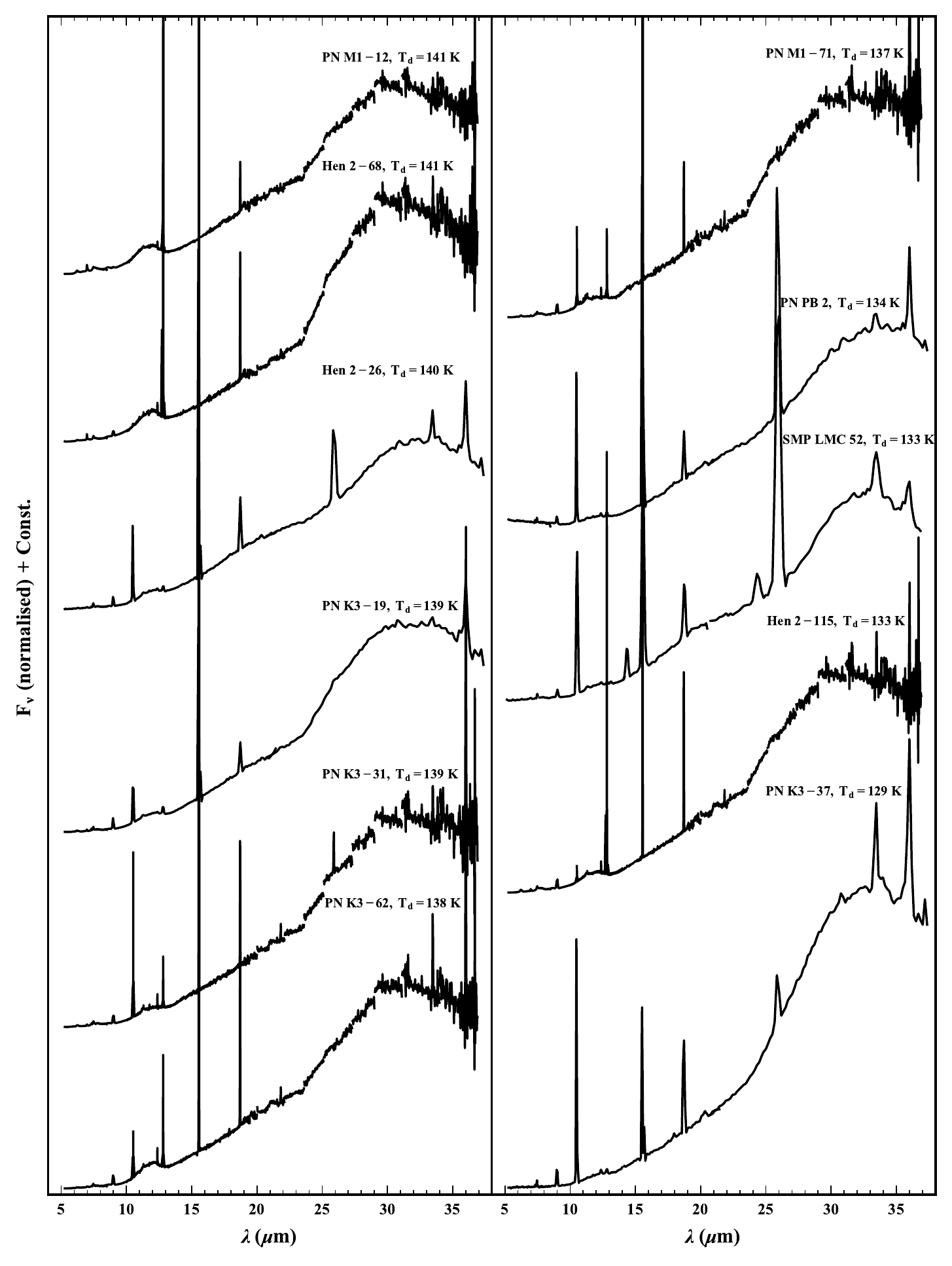}
  \caption{Overview of the Spitzer spectra (black solid lines) of the
 carbon-rich PNe with the 30\,$\mu$m feature. The spectra 
 are ordered according to the T$_{\rm d}$ from high to low 
 temperature, top to bottom, and left to right, and are
 normalised to the flux density at 18\,$\mu$m and offset 
 for clarity. The names of objects and values of the 
 T$_{\rm d}$ are shown above the spectra.}
  \label{appfig:all_spectra_pn_3}
\end{figure*}
%*************************************************************************************************************************
%*************************************************************************************************************************

\clearpage % the lack of this causes that a compilation becomes impossible (too many figures in the paper)

%-------------------------------------------------------------
% ------- APPENDIX - MEDIAN PROFILES ------
%-------------------------------------------------------------
\section{Normalised median profiles} % Second appendix
\label{sec:appendix_profiles}

      In Section~\ref{subsec:profiles_of_the_30_um_feature} 
      we compare the normalised median profiles of the 
      30\,$\mu$m feature for the AGB stars, post-AGB 
      objects, and PNe 
      (see \Cref{fig:agb_norm_profiles_t,fig:pagb_norm_profiles_t,fig:pne_norm_profiles_t}). 
      In \Cref{appfig:agb_norm_profiles_smc_400-600,appfig:agb_norm_profiles_lmc_200-400,%
      appfig:agb_norm_profiles_lmc_400-600,appfig:agb_norm_profiles_lmc_600-800,%
      appfig:agb_norm_profiles_lmc_800-1100,appfig:agb_norm_profiles_sgr_600-800,%
      appfig:agb_norm_profiles_gal_400-600,appfig:agb_norm_profiles_gal_600-800,%
      appfig:agb_norm_profiles_gal_800-1200,appfig:pagb_norm_profiles_lmc,%
      appfig:pagb_norm_profiles_gal,appfig:pn_norm_profiles_smc,appfig:pn_norm_profiles_lmc,%
      appfig:pn_norm_profiles_only_low_res_gal,appfig:pn_norm_profiles_gal} 
      we present the individual normalised profiles of the 
      30\,$\mu$m feature, which are a part of every 
      median profile shown in 
      \Cref{fig:agb_norm_profiles_t,fig:pagb_norm_profiles_t,%
      fig:pne_norm_profiles_t} (Section~\ref{subsec:profiles_of_the_30_um_feature}). 
      The profiles for the AGB stars are illustrated in 
      \Cref{appfig:agb_norm_profiles_smc_400-600,%
      appfig:agb_norm_profiles_lmc_200-400,appfig:agb_norm_profiles_lmc_400-600,%
      appfig:agb_norm_profiles_lmc_600-800,appfig:agb_norm_profiles_lmc_800-1100,%
      appfig:agb_norm_profiles_sgr_600-800,appfig:agb_norm_profiles_gal_400-600,%
      appfig:agb_norm_profiles_gal_600-800,appfig:agb_norm_profiles_gal_800-1200}, 
      whereas for post-AGB objects and PNe are shown in 
      \Cref{appfig:pagb_norm_profiles_lmc,appfig:pagb_norm_profiles_gal,%
      appfig:pn_norm_profiles_smc,appfig:pn_norm_profiles_lmc,%
      appfig:pn_norm_profiles_only_low_res_gal,appfig:pn_norm_profiles_gal}, 
      respectively. Furthermore, 
      \Cref{appfig:pagb_norm_profiles_gal,appfig:pn_norm_profiles_gal} 
      present the profiles of the 30\,$\mu$m feature for 
      the Galactic post-AGB objects and PNe derived 
      from the high resolution Spitzer spectra. In each of 
      Figures presented here, we show the individual 
      normalised profiles of the 30\,$\mu$m feature by 
      the solid grey lines, whereas the result median of 
      all normalised profiles is presented on top by the 
      coloured solid line with a colour consistent to the 
      analysed galaxy (lime -- SMC, red -- LMC, gold -- 
      Sgr dSph, black -- Milky Way).

%*************************************************************************************************************************
%*************************************************************************************************************************
   \begin{figure}[h!]
   \centering
   \includegraphics[width=\hsize]{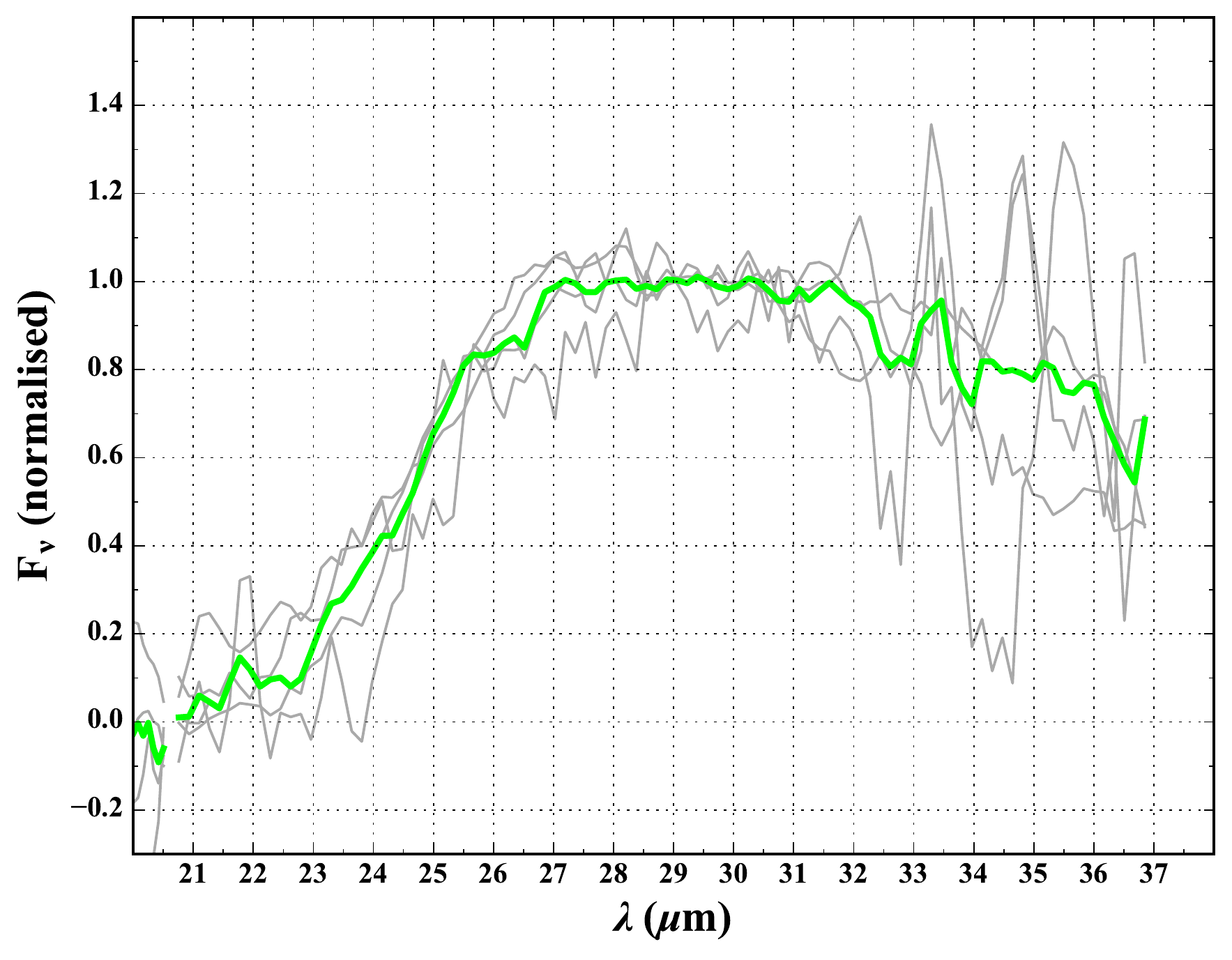}
      \caption{Normalised profiles of the 30\,$\mu$m 
      feature (grey solid lines) for the AGB stars in the 
      SMC for which the obtained T$_{\rm d}$ is between 
      400 and 600\,K. The normalised median profile of 
      the feature is shown by the solid lime line.}
         \label{appfig:agb_norm_profiles_smc_400-600}
   \end{figure}
%*************************************************************************************************************************
%*************************************************************************************************************************

%*************************************************************************************************************************
%*************************************************************************************************************************
   \begin{figure}[h!]
   \centering
   \includegraphics[width=\hsize]{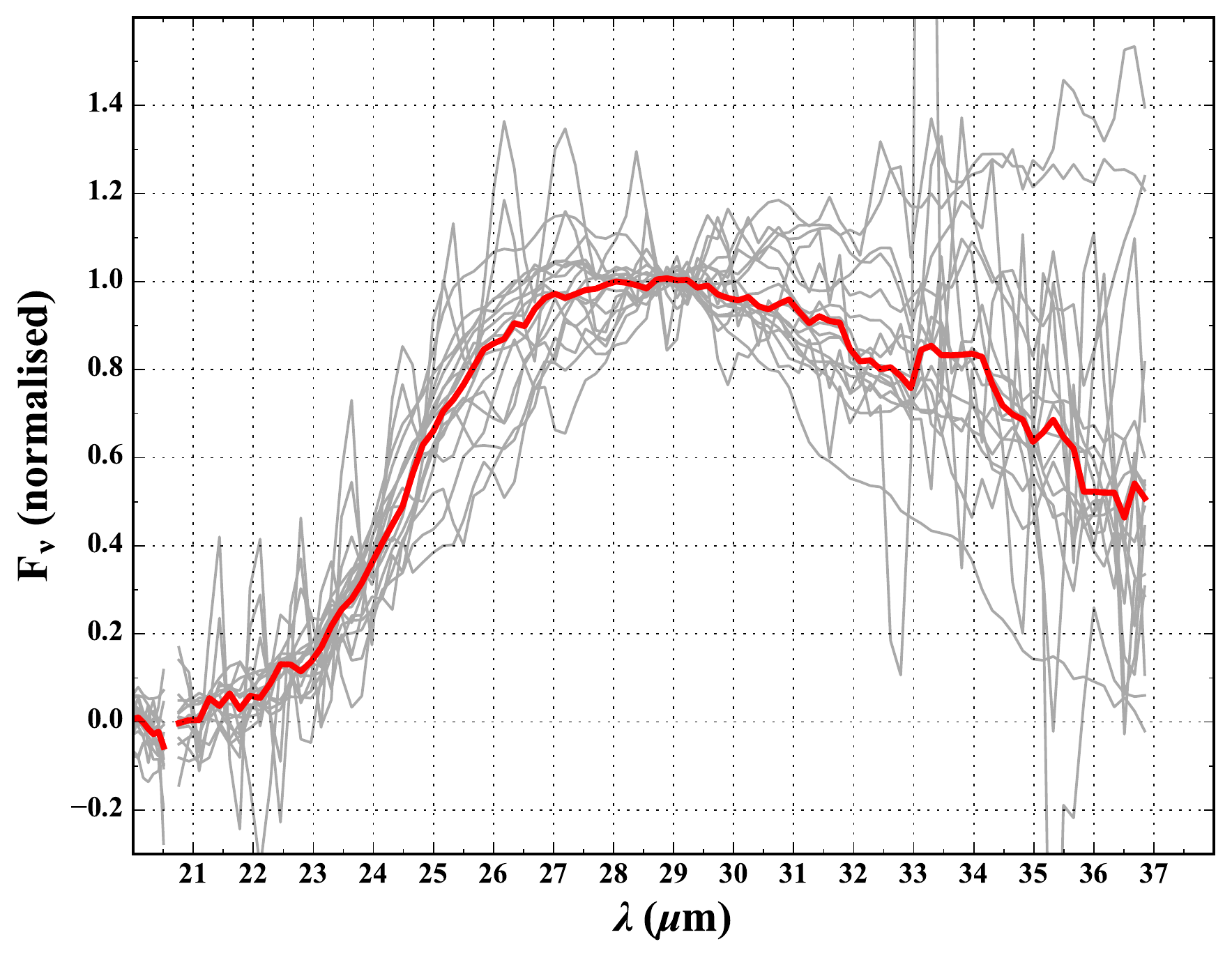}
      \caption{Normalised profiles of the 30\,$\mu$m 
      feature (grey solid lines) for the AGB stars in the 
      LMC for which the obtained T$_{\rm d}$ is between 
      200 and 400\,K. The normalised median profile of 
      the feature is shown by the solid red line.}
         \label{appfig:agb_norm_profiles_lmc_200-400}
   \end{figure}
%*************************************************************************************************************************
%*************************************************************************************************************************

%*************************************************************************************************************************
%*************************************************************************************************************************
   \begin{figure}[h!]
   \centering
   \includegraphics[width=\hsize]{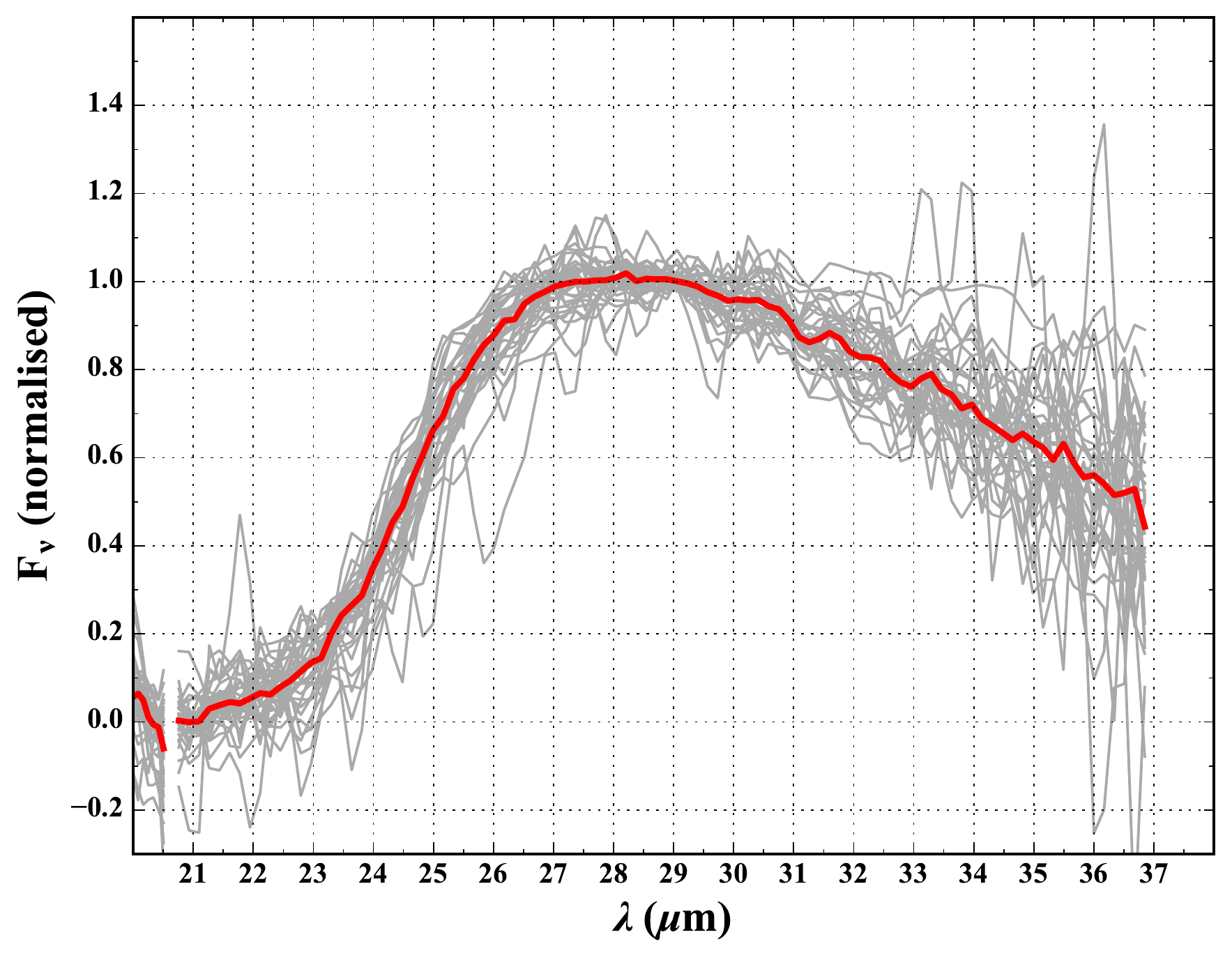}
      \caption{Normalised profiles of the 30\,$\mu$m 
      feature (grey solid lines) for the AGB stars in the 
      LMC for which the obtained T$_{\rm d}$ is between 
      400 and 600\,K. The normalised median profile of 
      the feature is shown by the solid red line.}
         \label{appfig:agb_norm_profiles_lmc_400-600}
   \end{figure}
%*************************************************************************************************************************
%*************************************************************************************************************************

%*************************************************************************************************************************
%*************************************************************************************************************************
   \begin{figure}[h!]
   \centering
   \includegraphics[width=\hsize]{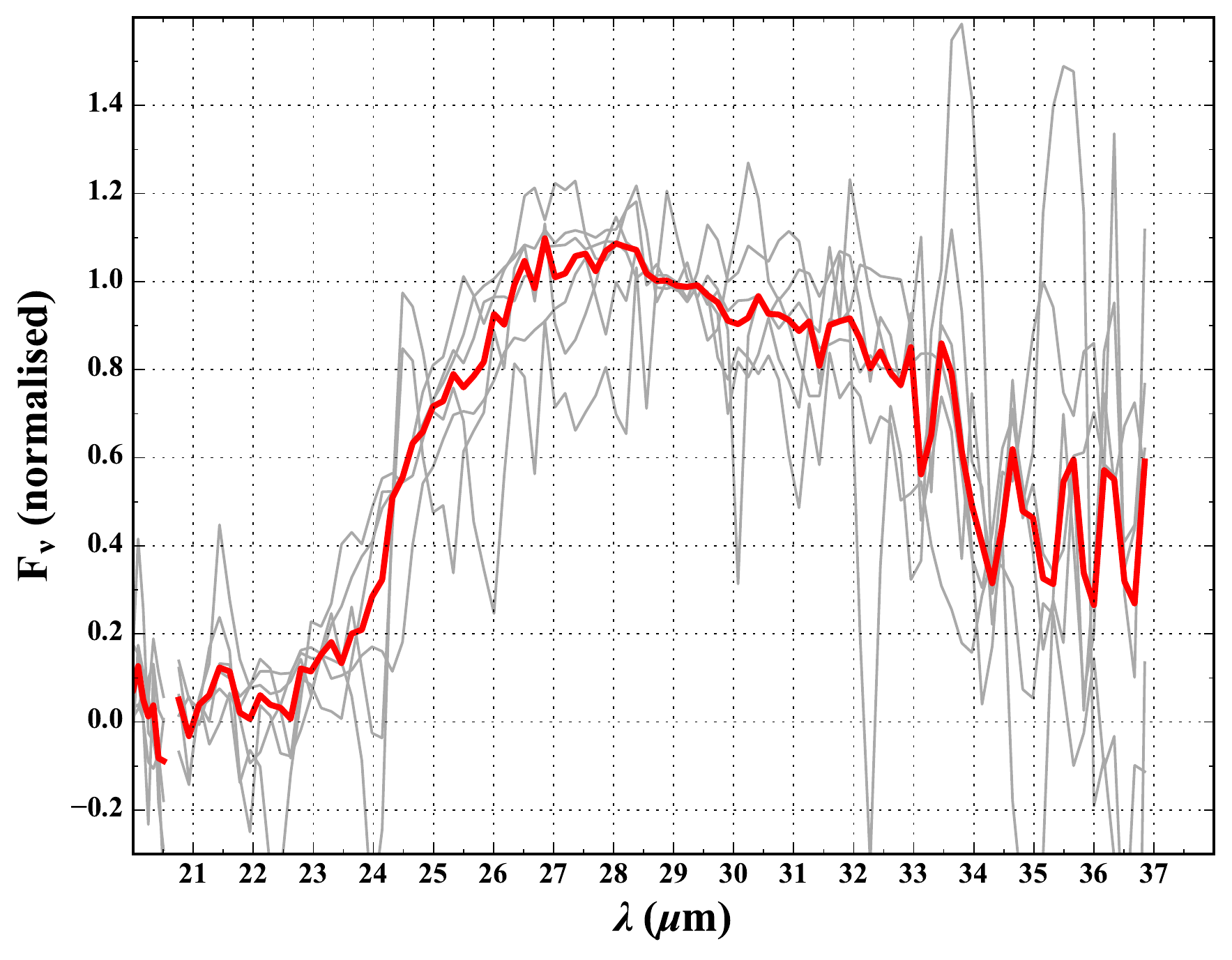}
      \caption{Normalised profiles of the 30\,$\mu$m 
      feature (grey solid lines) for the AGB stars in the 
      LMC for which the obtained T$_{\rm d}$ is between 
      600 and 800\,K. The normalised median profile of 
      the feature is shown by the solid red line.}
         \label{appfig:agb_norm_profiles_lmc_600-800}
   \end{figure}
%*************************************************************************************************************************
%*************************************************************************************************************************

%*************************************************************************************************************************
%*************************************************************************************************************************
   \begin{figure}[h!]
   \centering
   \includegraphics[width=\hsize]{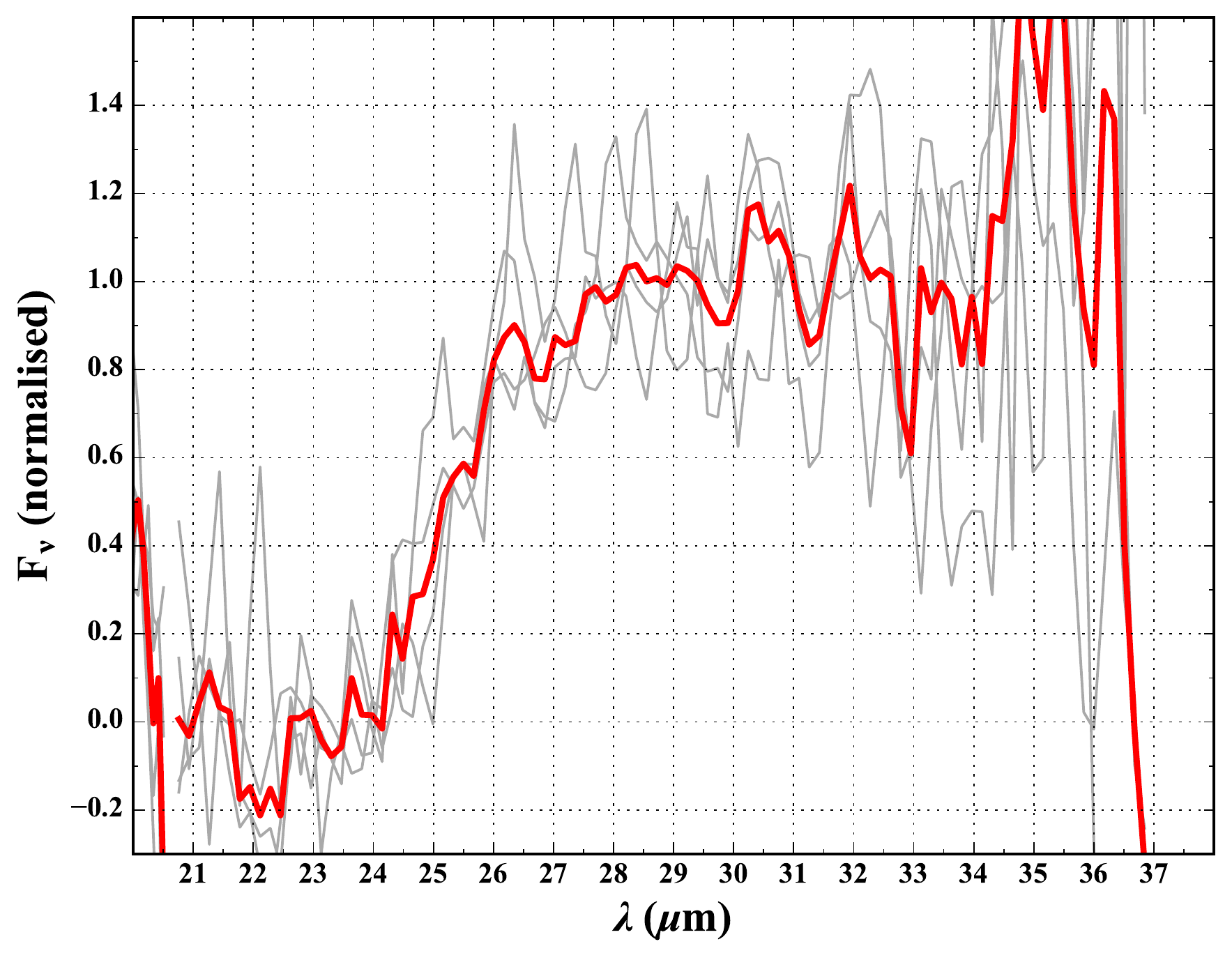}
      \caption{Normalised profiles of the 30\,$\mu$m 
      feature (grey solid lines) for the AGB stars in the 
      LMC for which the obtained T$_{\rm d}$ is higher 
      than 800\,K. The normalised median profile of the 
      feature is shown by the solid red line.}
         \label{appfig:agb_norm_profiles_lmc_800-1100}
   \end{figure}
%*************************************************************************************************************************
%*************************************************************************************************************************

%*************************************************************************************************************************
%*************************************************************************************************************************
   \begin{figure}[h!]
   \centering
   \includegraphics[width=\hsize]{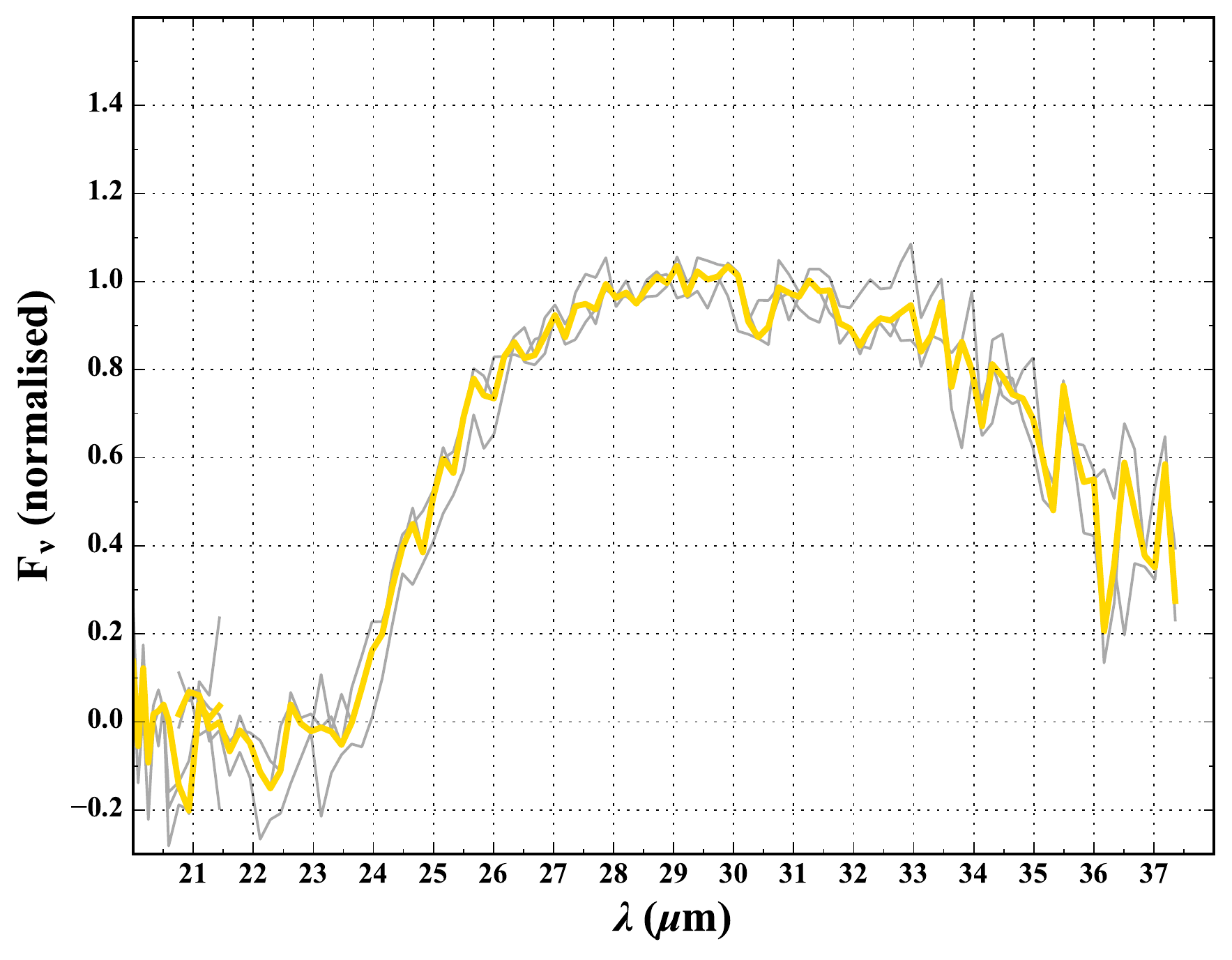}
      \caption{Normalised profiles of the 30\,$\mu$m 
      feature (grey solid lines) for the AGB stars in the 
      Sgr dSph galaxy for which the obtained T$_{\rm d}$ 
      is between 600 and 800\,K. The normalised median 
      profile of the feature is shown by the solid gold line.}
         \label{appfig:agb_norm_profiles_sgr_600-800}
   \end{figure}
%*************************************************************************************************************************
%*************************************************************************************************************************

%*************************************************************************************************************************
%*************************************************************************************************************************
   \begin{figure}[h!]
   \centering
   \includegraphics[width=\hsize]{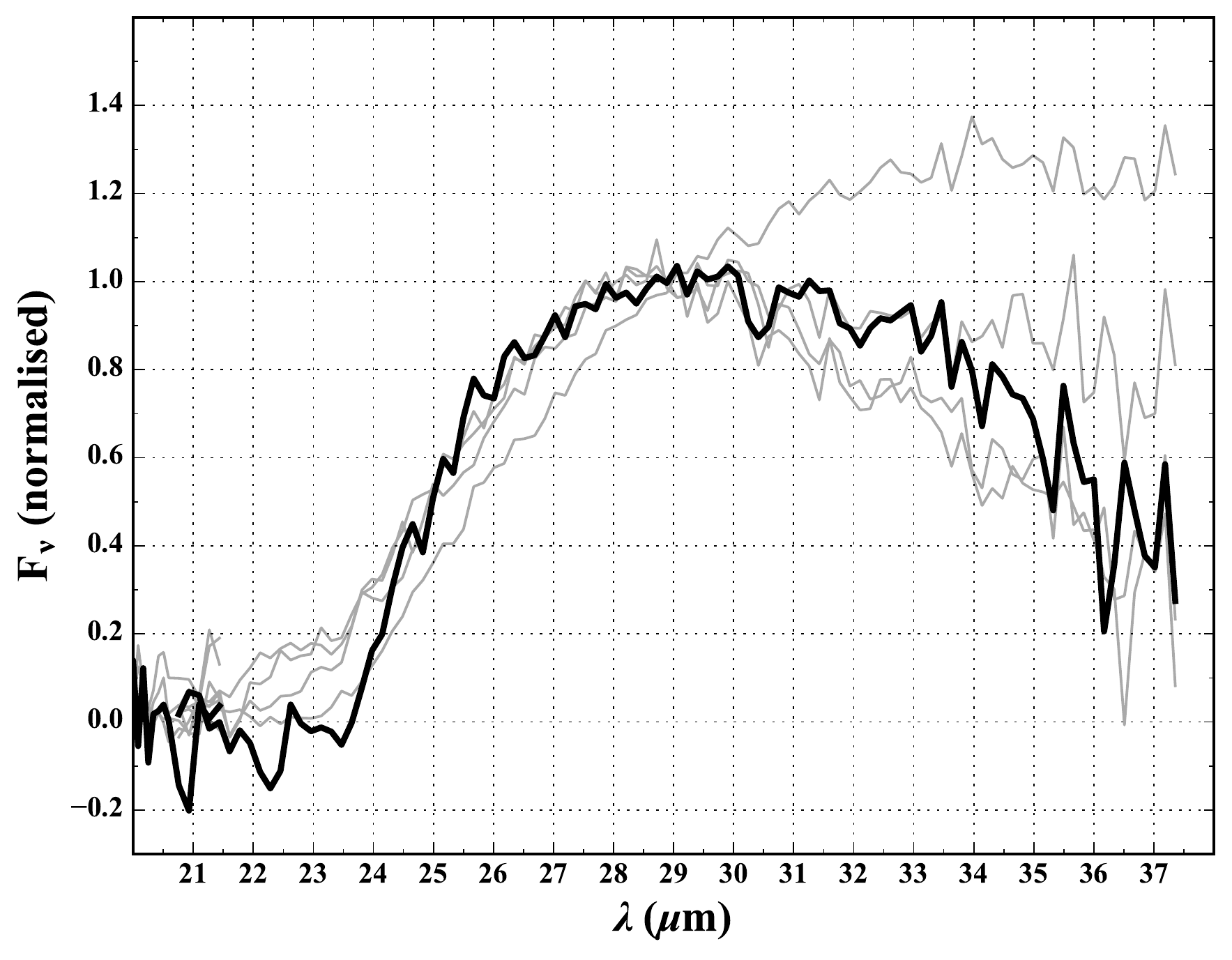}
      \caption{Normalised profiles of the 30\,$\mu$m 
      feature (grey solid lines) for the AGB stars in the 
      Milky Way for which the obtained T$_{\rm d}$ 
      is between 400 and 600\,K. The normalised median 
      profile of the feature is shown by the solid black line.}
         \label{appfig:agb_norm_profiles_gal_400-600}
   \end{figure}
%*************************************************************************************************************************
%*************************************************************************************************************************

%*************************************************************************************************************************
%*************************************************************************************************************************
   \begin{figure}[h!]
   \centering
   \includegraphics[width=\hsize]{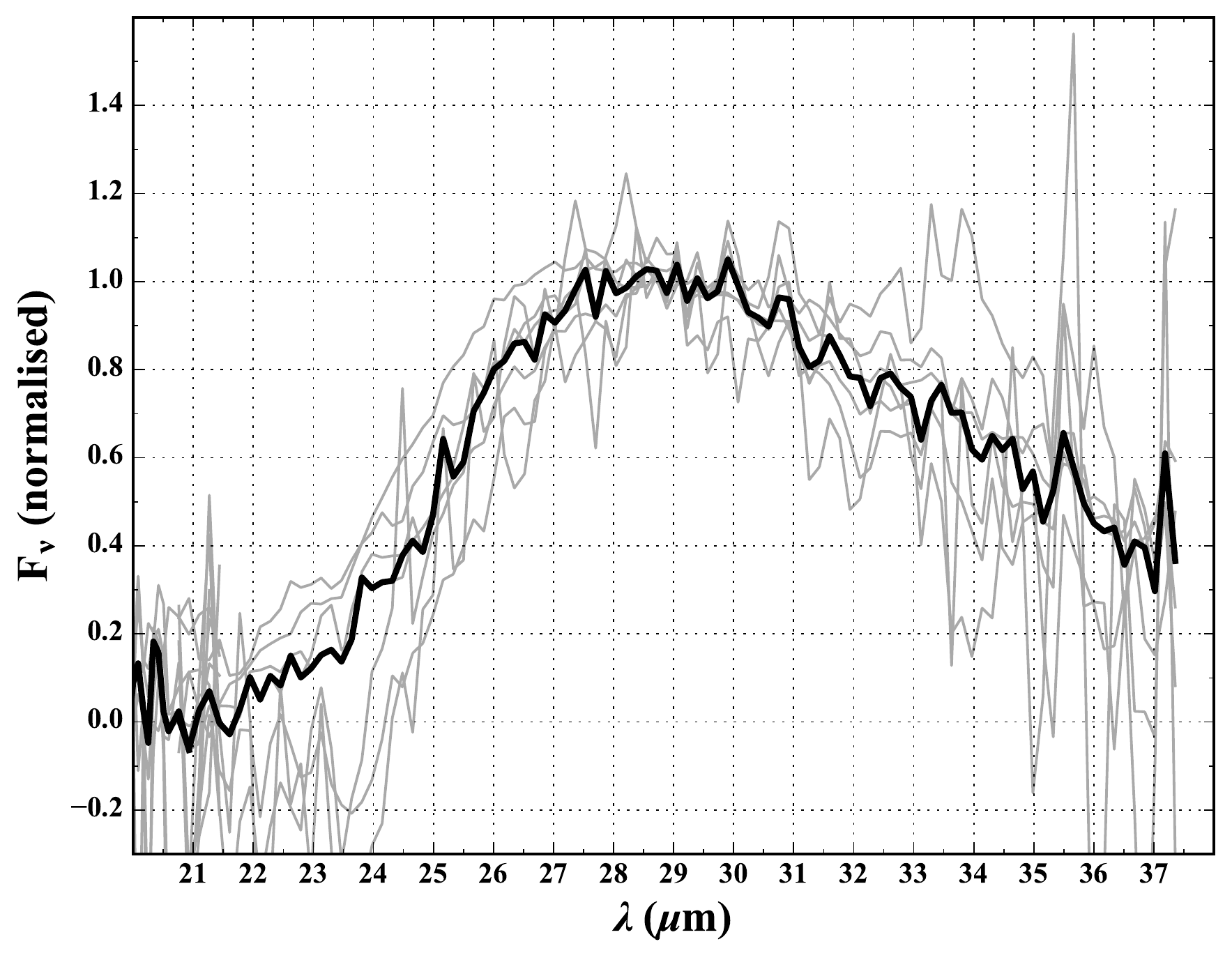}
      \caption{Normalised profiles of the 30\,$\mu$m 
      feature (grey solid lines) for the AGB stars in the 
      Milky Way for which the obtained T$_{\rm d}$ 
      is between 600 and 800\,K. The normalised median 
      profile of the feature is shown by the solid black line.}
         \label{appfig:agb_norm_profiles_gal_600-800}
   \end{figure}
%*************************************************************************************************************************
%*************************************************************************************************************************

%*************************************************************************************************************************
%*************************************************************************************************************************
   \begin{figure}[h!]
   \centering
   \includegraphics[width=\hsize]{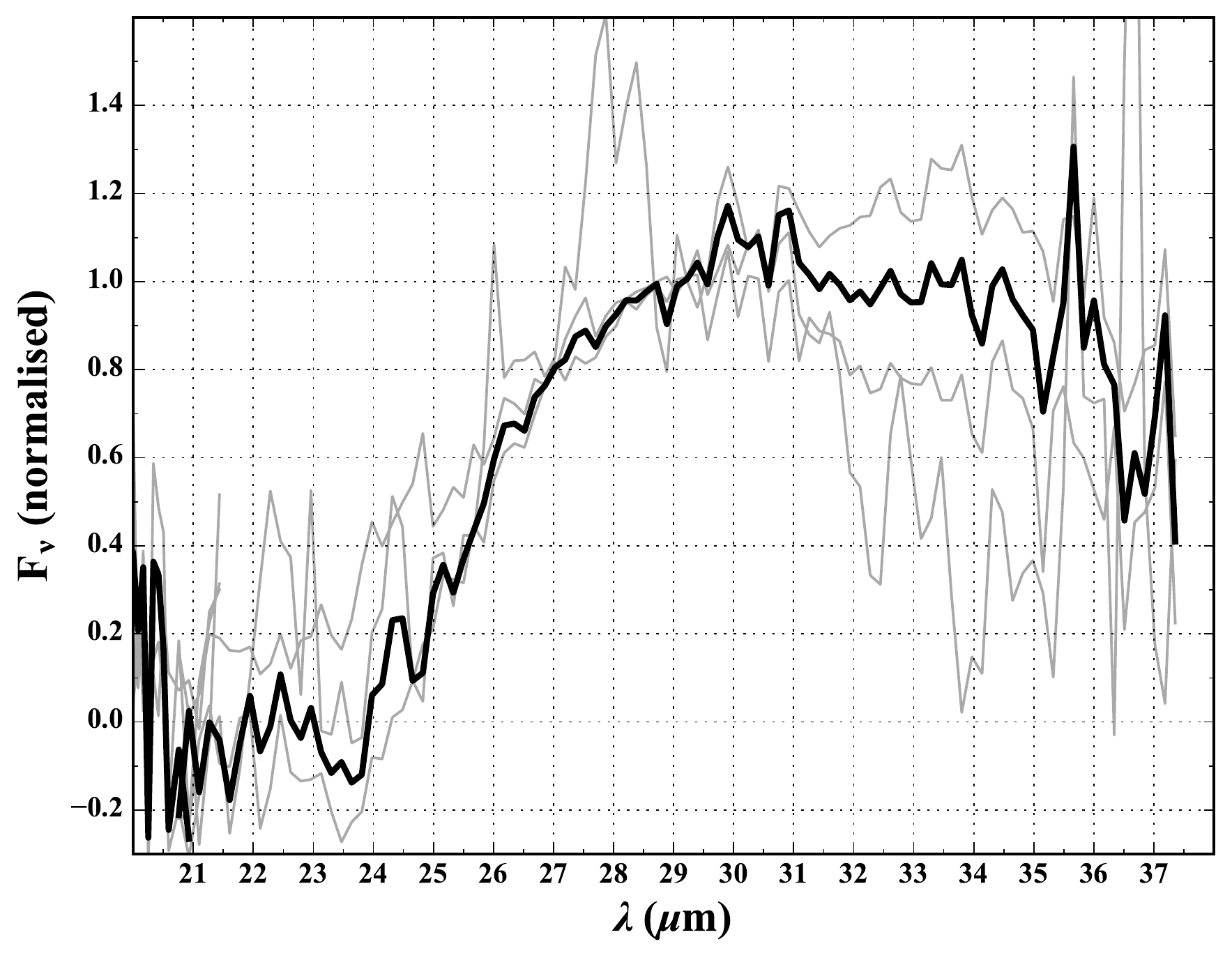}
      \caption{Normalised profiles of the 30\,$\mu$m 
      feature (grey solid lines) for the AGB stars in the 
      Milky Way for which the obtained T$_{\rm d}$ 
      is higher than 800\,K. The normalised median 
      profile of the feature is shown by the solid black line.}
         \label{appfig:agb_norm_profiles_gal_800-1200}
   \end{figure}
%*************************************************************************************************************************
%*************************************************************************************************************************

%*************************************************************************************************************************
%*************************************************************************************************************************
   \begin{figure}[h!]
   \centering
   \includegraphics[width=\hsize]{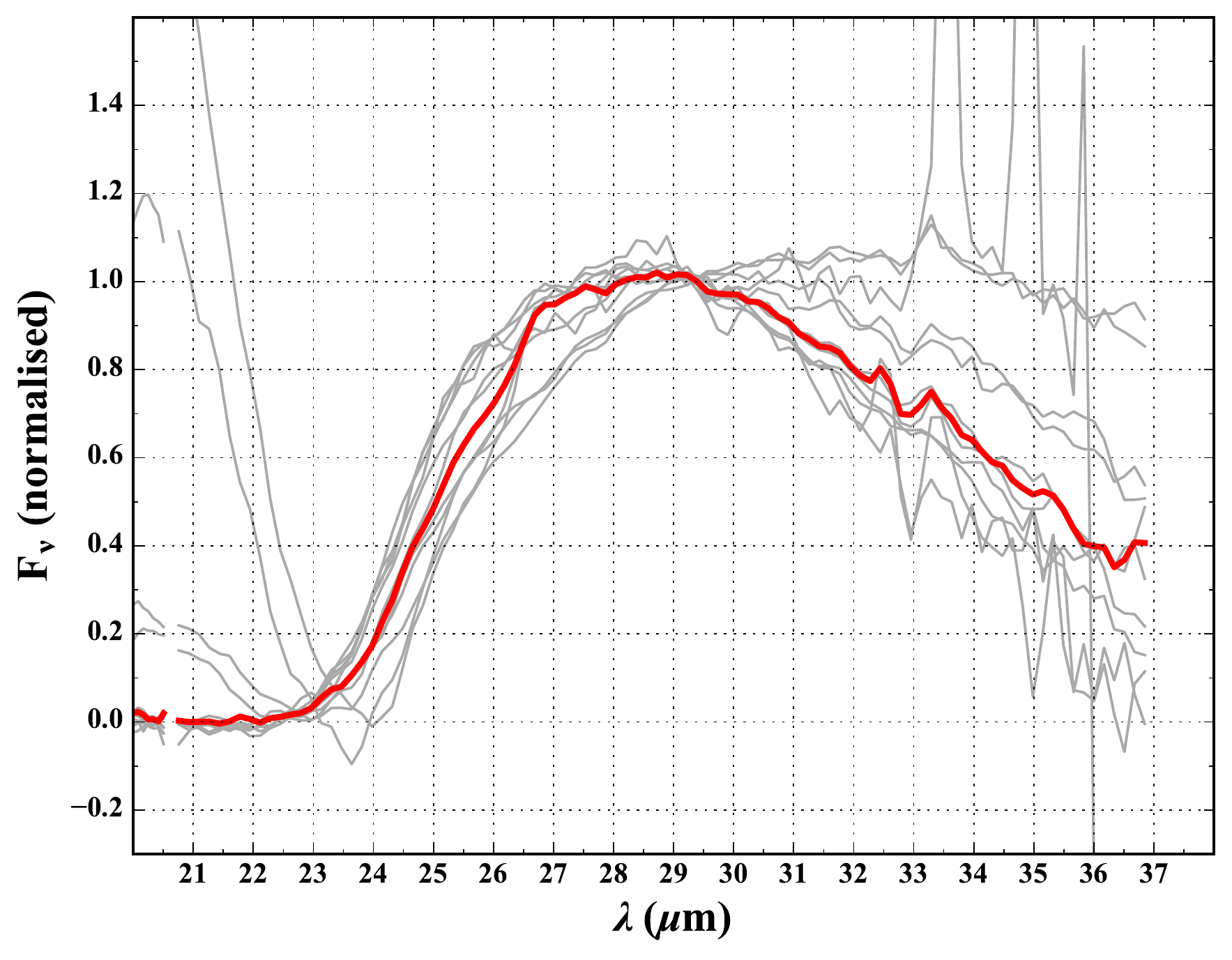}
      \caption{Normalised profiles of the 30\,$\mu$m 
      feature (grey solid lines) for the post-AGB stars in the 
      LMC. The normalised median profile of the feature is 
      shown by the solid red line.}
         \label{appfig:pagb_norm_profiles_lmc}
   \end{figure}
%*************************************************************************************************************************
%*************************************************************************************************************************

%*************************************************************************************************************************
%*************************************************************************************************************************
   \begin{figure}[h!]
   \centering
   \includegraphics[width=\hsize]{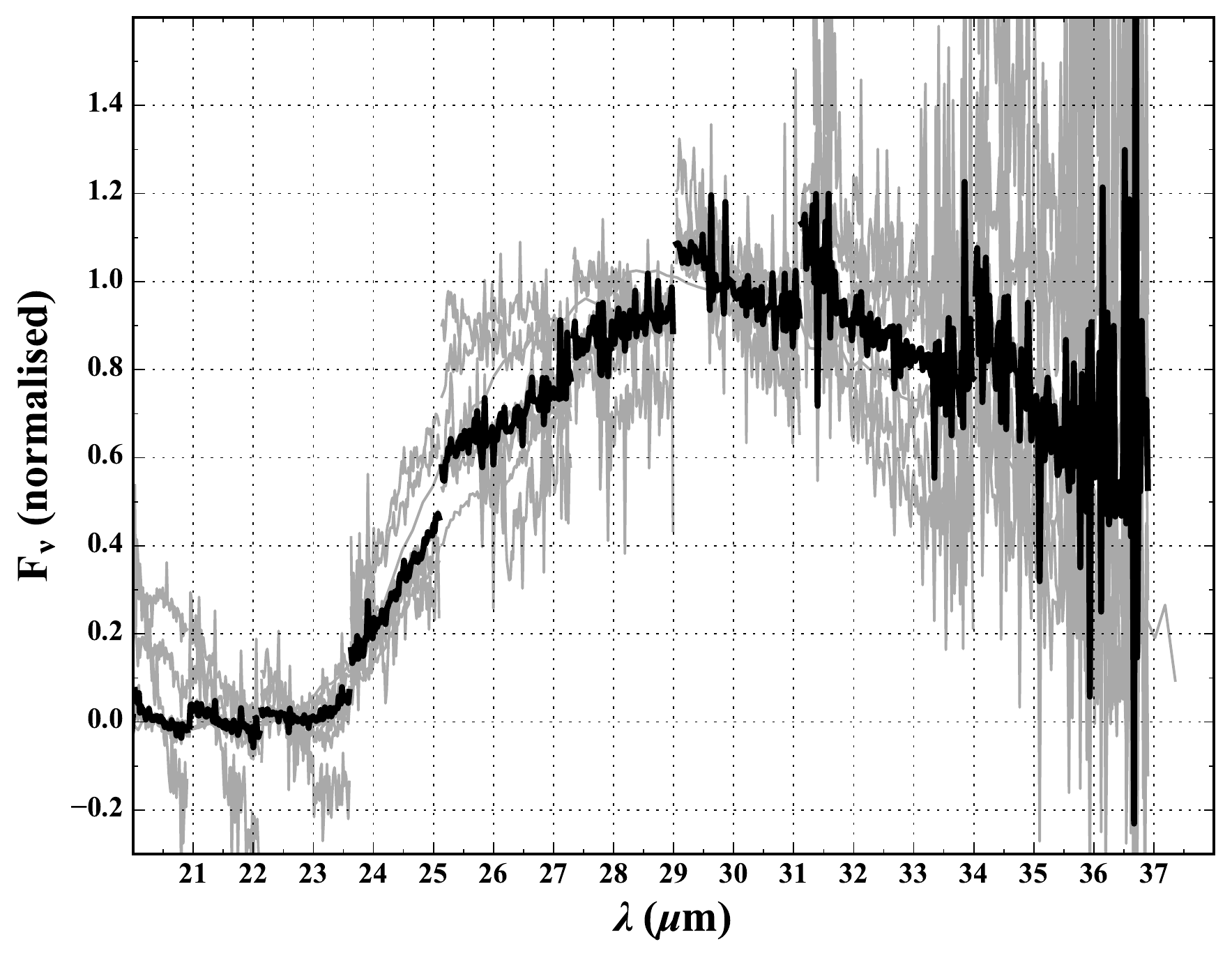}
      \caption{Low and high resolution normalised profiles 
      of the 30\,$\mu$m feature (grey solid lines) for the 
      post-AGB stars in the Milky Way. The normalised 
      median profile of the feature is shown by the solid 
      black line. The low resolution normalised profiles 
      are resampled to the high resolution wavelength 
      grid using the linear interpolation.}
         \label{appfig:pagb_norm_profiles_gal}
   \end{figure}
%*************************************************************************************************************************
%*************************************************************************************************************************

%*************************************************************************************************************************
%*************************************************************************************************************************
   \begin{figure}[h!]
   \centering
   \includegraphics[width=\hsize]{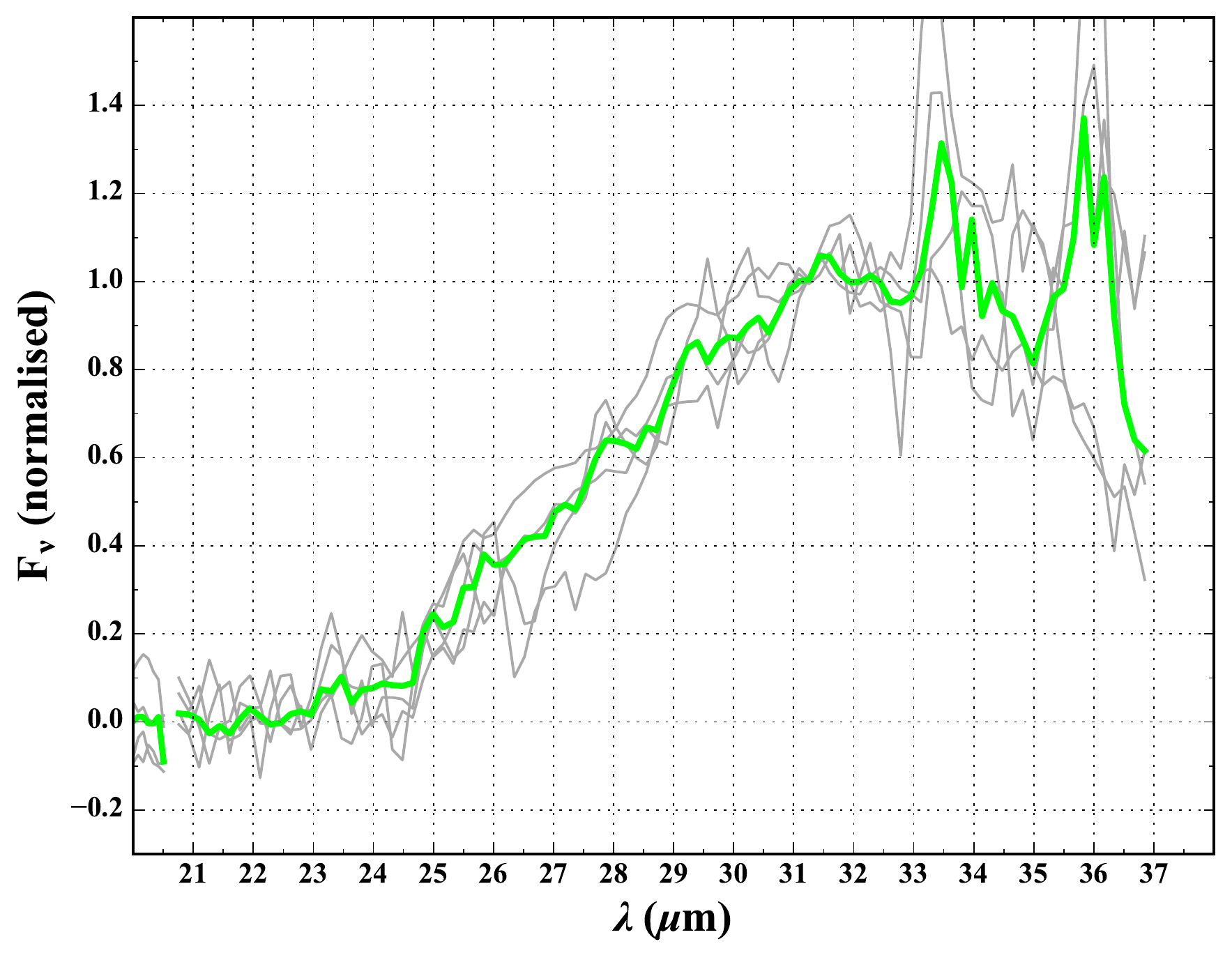}
      \caption{Normalised profiles of the 30\,$\mu$m 
      feature (grey solid lines) for the PNe in the SMC. 
      The normalised median profile of the feature is 
      shown by the solid lime line.}
         \label{appfig:pn_norm_profiles_smc}
   \end{figure}
%*************************************************************************************************************************
%*************************************************************************************************************************

%*************************************************************************************************************************
%*************************************************************************************************************************
   \begin{figure}[h!]
   \centering
   \includegraphics[width=\hsize]{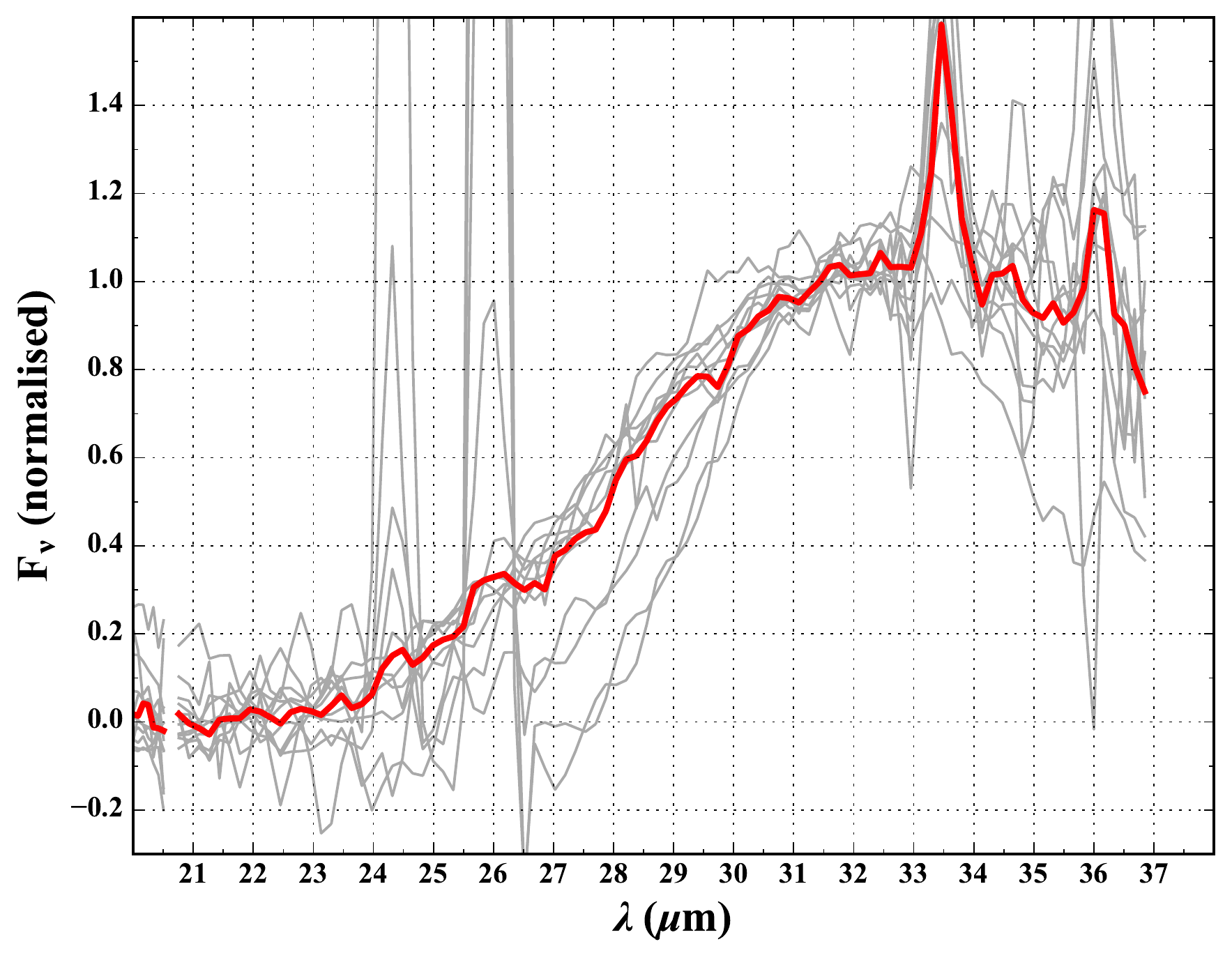}
      \caption{Normalised profiles of the 30\,$\mu$m 
      feature (grey solid lines) for the PNe in the LMC. 
      The normalised median profile of the feature is 
      shown by the solid red line.}
         \label{appfig:pn_norm_profiles_lmc}
   \end{figure}
%*************************************************************************************************************************
%*************************************************************************************************************************

%*************************************************************************************************************************
%*************************************************************************************************************************
   \begin{figure}[h!]
   \centering
   \includegraphics[width=\hsize]{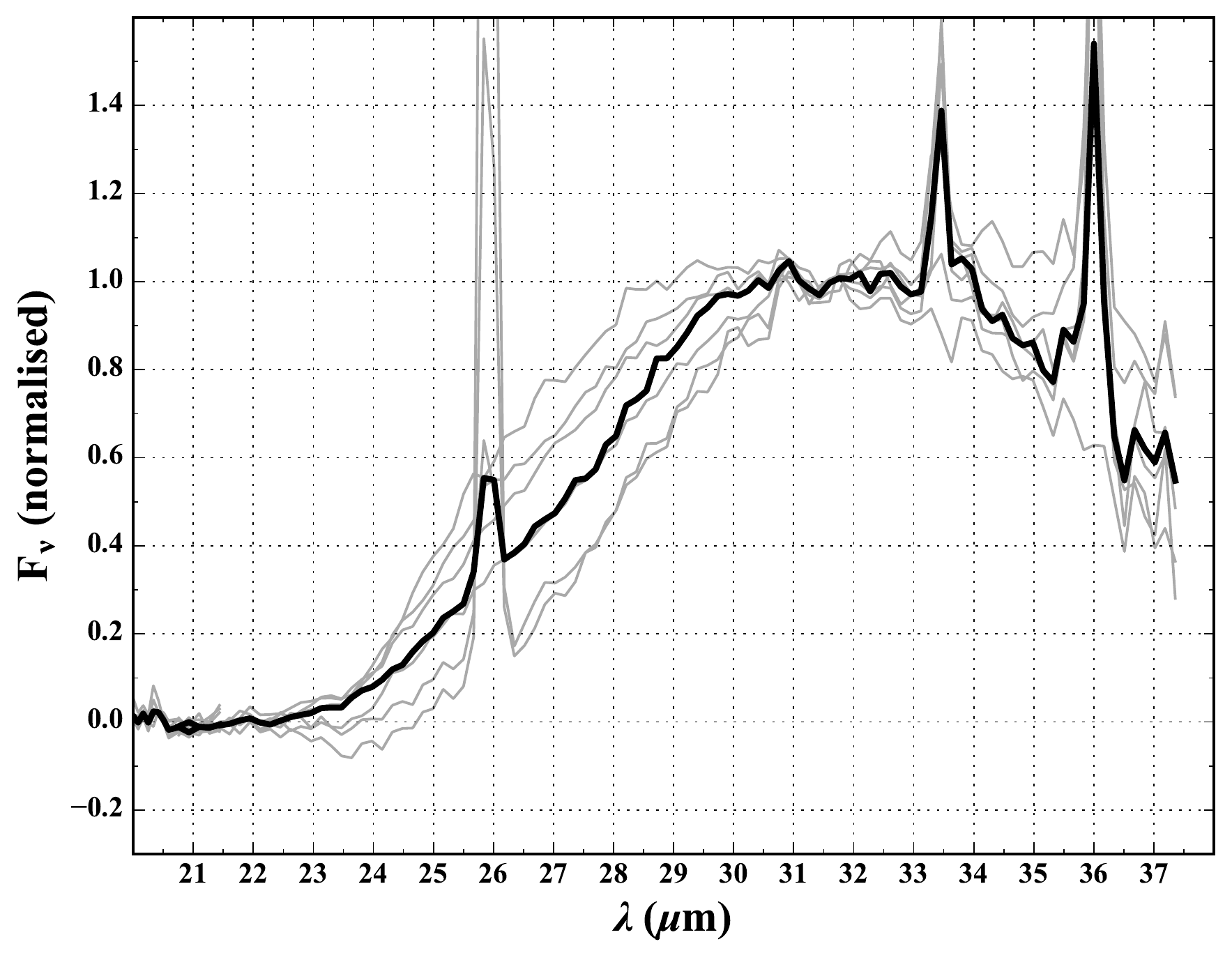}
      \caption{Low resolution normalised profiles of the 
      30\,$\mu$m feature (grey solid lines) for the PNe 
      in the Milky Way. The normalised median profile of 
      the feature is shown by the solid black line.}
         \label{appfig:pn_norm_profiles_only_low_res_gal}
   \end{figure}
%*************************************************************************************************************************
%*************************************************************************************************************************

%*************************************************************************************************************************
%*************************************************************************************************************************
   \begin{figure}[h!]
   \centering
   \includegraphics[width=\hsize]{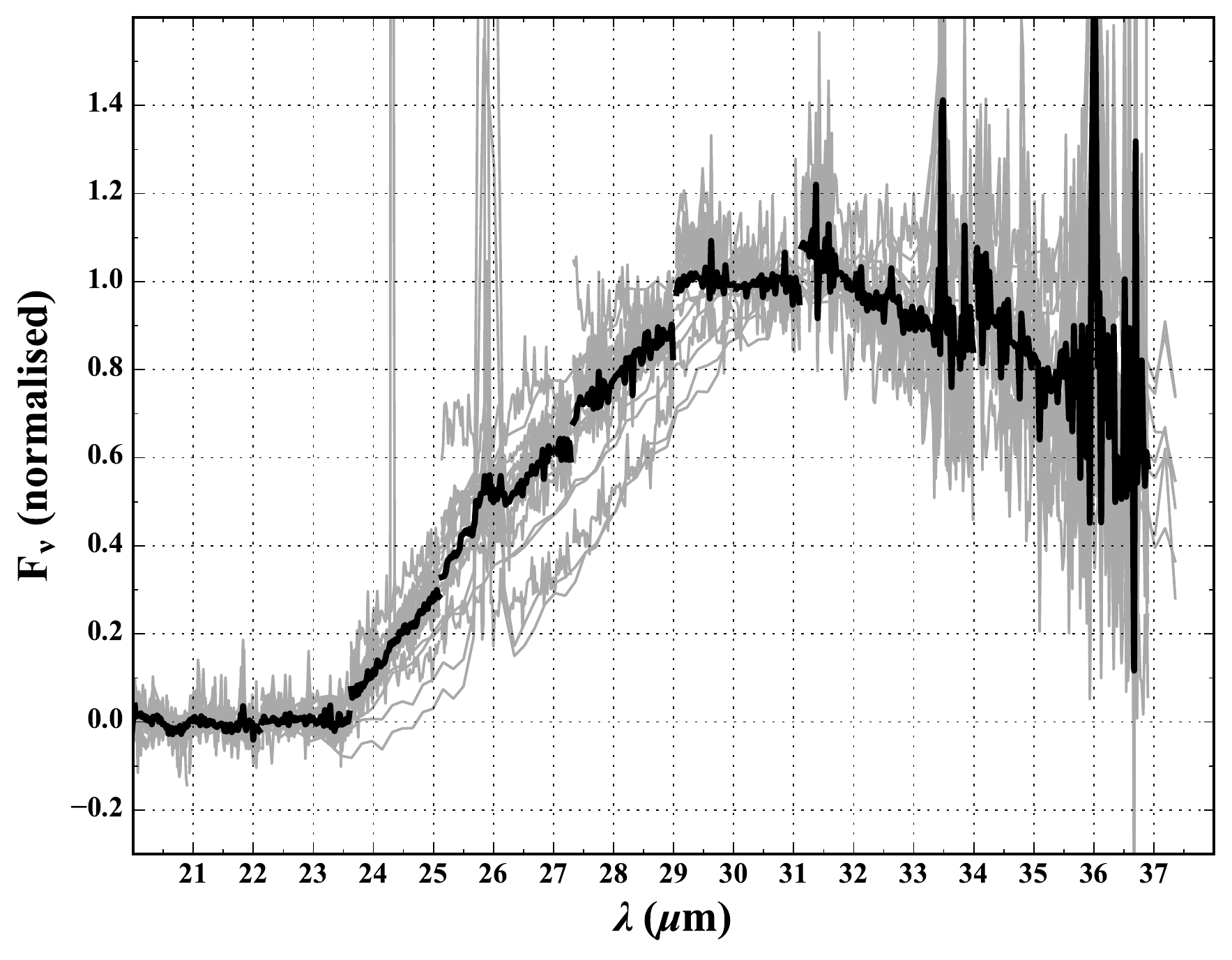}
      \caption{Low and high resolution normalised profiles of the 
      30\,$\mu$m feature (grey solid lines) for the PNe 
      in the Milky Way. The normalised median profile of 
      the feature is shown by the solid black line. The low 
      resolution normalised profiles are resampled to the 
      high resolution wavelength grid using the linear 
      interpolation.}
         \label{appfig:pn_norm_profiles_gal}
   \end{figure}
%*************************************************************************************************************************
%*************************************************************************************************************************

\clearpage
%-------------------------------------------------------------
% --------------- APPENDIX - TABLES --------------
%-------------------------------------------------------------
\section{Tables}
\label{sec:appendix_tables}

      In Section~\ref{subsec:target_selection} we present 
      Table~\ref{tab:sgr_basic_info} with basic information 
      about objects in the Sgr dSph. Here, we list 
      \Cref{app_tab:smc_basic_info,%
      app_tab:lmc_basic_info,app_tab:gal_basic_info}
      with basic information about objects in the 
      remaining galaxies: the SMC, LMC, and Milky 
      Way. A brief description is also given in 
      Section~\ref{subsec:target_selection}.
      
      In Section~\ref{subsec:strength_central_wav} we 
      show Table~\ref{tab:sgr_spectral_results} with the 
      results of the spectroscopic analysis for the objects 
      in the Sgr dSph. Here, we list 
      \Cref{app_tab:smc_spectral_results,%
      app_tab:lmc_spectral_results,app_tab:gal_spectral_results} 
      with the spectroscopic results for the objects in the 
      SMC, LMC, and Milky Way. A brief description is 
      also given in Section~\ref{subsec:strength_central_wav}.

%*************************************************************************************************************************
%*************************************************************************************************************************
\begin{landscape}
\begin{table*}
\scriptsize
\caption{The SMC sample: names, coordinates, reference position, 
classification of spectra (C-AGB: carbon-rich AGB star, C-pAGB: carbon-rich post-AGB star, C-PN: carbon-rich planetary nebula), period with reference, program identifier, and AOR key.}
\label{app_tab:smc_basic_info}
\centering
\begin{tabular}{llccclccrr}
\hline
\hline
                                                &                                       &       \multicolumn{2}{c}{(J2000.0)}   \\\cmidrule{3-4}
Name                                    &       Other name              &       RA                         &       Dec                     &Position               &               Class                           &       Period  &Period         &Program ID      & AOR key\\
                                                &                                       &       (deg)                   &         (deg)                   &reference      &                                                       &       (days)  &reference      &                       &\\
\hline
\textbf{SMP~SMC~1}              &               \ldots          &       005.994         &       $-$73.634               &       IRAC    &       C-PN, 16-24\,$\mu$m feat.\tablefootmark{b}, fullerenes\tablefootmark{f}       &       \ldots  &       \ldots  &       103             &       4953088 \\
\textbf{IRAS~00350-7436}                &               \ldots          &       009.248         &       $-$74.331               &       IRAC    &       C-pAGB, gr. I                                                                                           &       142             &       (5)             &       50240   &       27517184        \\
\textbf{SMP~SMC~6}              &               \ldots          &       010.366         &       $-$73.785               &       IRAC    &       C-PN, 16-24\,$\mu$m feat.\tablefootmark{b}                                            &       \ldots  &       \ldots  &       103             &       4954112 \\
\textbf{MSX~SMC~054}            &               \ldots          &       010.775         &       $-$73.361               &       IRAC    &       C-AGB                                                                                                   &       396             &       (4)             &       3277            &       10657280        \\      
LK~229                                  &           LIN 49                      &       010.975         &       $-$72.921               &       IRAC    &       C-PN, fullerenes\tablefootmark{f}                                                             &       \ldots  &       \ldots  &       50240   &       27537664        \\
J004441.05-732136.44            &               \ldots          &       011.171         &       $-$73.360               &       IRAC    &       C-pAGB, gr. III, 21\,$\mu$m\tablefootmark{a}                                            &       96              &       (5)     &       50240   &       27525120        \\
MSX~SMC~105                     &               \ldots          &       011.259         &       $-$72.873               &       IRAC    &       C-AGB                                                                                                   &       668             &       (4)             &       3277            &       10658816        \\
$[$GB98$]$~S06                  &       MSX~SMC~060     &       011.668         &       $-$73.280               &       IRAC    &       C-AGB                                                                                                   &       435             &       (4)             &       3277            &       10657792        \\
IRAS~F00471-7352                        &               \ldots          &       012.248         &       $-$73.594               &       IRAC    &       C-AGB                                                                                                   &       685             &       (4)             &       50240   &       27523840        \\
\textbf{SMP~SMC~13}             &               \ldots          &       012.465         &       $-$73.739               &       IRAC    &       C-PN, 16-24\,$\mu$m feat.\tablefootmark{e}, fullerenes\tablefootmark{e}       &       \ldots  &       \ldots  &       20443   &       14706176        \\
MSX~SMC~163                     &               \ldots          &       012.753         &       $-$72.422               &       IRAC    &       C-AGB                                                                                                   &       665             &       (4)             &       3277            &       10661120        \\
\textbf{SMP~SMC~15}             &               \ldots          &       012.781         &       $-$73.960               &       IRAC    &       C-PN, 16-24\,$\mu$m feat.\tablefootmark{e}, fullerenes\tablefootmark{e}       &       \ldots  &       \ldots  &       20443   &       14706688        \\
SMP~SMC~16                              &               \ldots          &       012.863         &       $-$72.437               &       IRAC    &       C-PN, fullerenes\tablefootmark{c}                                                             &       \ldots  &       \ldots  &       20443   &       14706944        \\
SMP~SMC~17                              &               \ldots          &       012.985         &       $-$71.412               &       IRAC    &       C-PN, 16-18\,$\mu$m feat.                                                                     &       \ldots  &       \ldots  &       20443   &       14707200        \\
\textbf{SMP~SMC~18}             &               \ldots          &       012.992         &       $-$73.342               &       IRAC    &       C-PN, 16-24\,$\mu$m feat.\tablefootmark{e}, fullerenes\tablefootmark{e}       &       \ldots  &       \ldots  &       20443   &       14707456        \\
MSX~SMC~159                     &               \ldots          &       013.593         &       $-$72.725               &       IRAC    &       C-AGB                                                                                                   &       560             &       (2)             &       3277            &       10660608        \\
\textbf{SMP~SMC~20}             &               \ldots          &       014.022         &       $-$70.324               &       IRAC    &       C-PN, 16-24\,$\mu$m feat.\tablefootmark{e}                                            &       \ldots  &       \ldots  &       20443   &       14707968        \\
IRAS~00554-7351                 &               \ldots          &       014.266         &       $-$73.587               &       IRAC    &       C-AGB                                                                                                   &       800             &       (1)             &       3505            &       12936192        \\
SMP~SMC~24                              &               \ldots          &       014.817         &       $-$72.033               &       IRAC    &       C-PN, fullerenes\tablefootmark{e}                                                             &       \ldots  &       \ldots  &       103             &       15901952        \\
2MASS~J01054645-7147053 &               \ldots          &       016.443         &       $-$71.785               &       IRAC    &       C-pAGB, gr. II, 21\,$\mu$m\tablefootmark{a}                                             &$\sim$10-30    &       (5)             &       50240   &       27518464        \\
NGC~419~MIR~1                   &               \ldots          &       017.073         &       $-$72.886               &       IRAC   &       C-AGB                                                                                                   &       738             &       (3)             &       3505            &       12939776        \\
SMP~SMC~27                              &               \ldots          &       020.294         &       $-$73.243               &       IRAC    &       C-PN                                                                                                    &       \ldots  &       \ldots  &       20443   &       14708992        \\
\hline
\end{tabular}

\tablebib{(1)~\citet{Whitelock:1989aa}, (2)~\citet{Sloan:2006kx}, (3)~\citet{Kamath:2010aa}, (4)~\citet{Soszynski:2011aa}, (5)~\citet{Hrivnak:2015aa}.\\
(a)~\citet{Volk:2011lr}, (b)~\citet{Bernard-Salas:2009aa}, (c)~\citet{Garcia-Hernandez:2010aa}, (d)~\citet{Bernard-Salas:2012aa}, (e)~\citet{Garcia-Hernandez:2012qy}, (f)~\citet{Sloan:2014fj}.
}

\end{table*}
\end{landscape}
%*************************************************************************************************************************
%*************************************************************************************************************************

\clearpage
\onecolumn

%*************************************************************************************************************************
%*************************************************************************************************************************
%\longtab{
\begin{landscape}
\scriptsize
\begin{longtable}{llccclccrr}
\caption{The LMC sample: names, coordinates, reference position, classification of spectra (C-AGB: carbon-rich AGB star, C-pAGB: carbon-rich post-AGB star, C-PN: carbon-rich planetary nebula), period with reference, program identifier, and AOR key.}
\label{app_tab:lmc_basic_info}\\
\hline
\hline
                                                &                                                       &       \multicolumn{2}{c}{(J2000.0)}   \\\cmidrule{3-4}
Name                                    &               Other name                      &       RA                         &       Dec             &Position               &               Class                           &       Period  &Period         &Program ID      & AOR key\\
                                                &                                                       &       (deg)                   &         (deg)           &reference      &                                                       &       (days)  &reference      &                       &\\
\hline
\endfirsthead
\caption{Continued.} \\
\hline
                                                &                                                       &       \multicolumn{2}{c}{(J2000.0)}   \\\cmidrule{3-4}
Name                                    &               Other name                      &       RA                         &       Dec             &Position               &               Class                           &       Period  &Period         &Program ID      & AOR key\\
                                                &                                                       &       (deg)                   &         (deg)           &reference      &                                                       &       (days)  &reference      &                       &\\
\hline
\endhead
\hline
\endfoot
\hline
\endlastfoot
\textbf{IRAS~04286-6937}                &       MSX~LMC~1007                    &       067.126         &       $-$69.514       &       IRAC    &       C-AGB                                                                                                   &       662             &       (1)             &       3426            &       10972672        \\
IRAS~F04331-6932                        &       MSX~LMC~1008                    &       068.239         &       $-$69.443       &       IRAC    &       C-AGB                                                                                                   &       \ldots  &       \ldots  &       30788   &       19012352        \\
IRAS~F04340-7016                 &       MSX~LMC~1077                    &       068.432         &       $-$70.164       &       IRAC    &       C-AGB                                                                                                   &       442             &       (3)             &       30788   &       19012864        \\
IRAS~F04353-6702                        &       MSX~LMC~1017                    &       068.850         &       $-$66.947       &       2MASS   &       C-AGB                                                                                                   &       \ldots  &       \ldots  &       30788   &       19012608        \\
IRAS~F04375-7247                        &       MSX~LMC~1067                    &       069.185         &       $-$72.700       &       2MASS   &       C-AGB                                                                                                   &       \ldots  &       \ldots  &       30788   &       19008768        \\
\end{longtable}

\tablebib{(1)~\citet{Whitelock:2003aa}, (2)~\citet{Groenewegen:2009aa}, (3)~\citet{Soszynski:2009aa}, (4)~\citet{Kamath:2010aa}, (5)~\citet{Hrivnak:2015aa}, (6)~\citet{Sloan:2016aa}.\\
(a)~\citet{Bernard-Salas:2009aa}, (b)~\citet{Volk:2011lr}, (c)~\citet{Bernard-Salas:2012aa}, (d)~\citet{Garcia-Hernandez:2012qy}, (e)~\citet{Matsuura:2014fu}, (f)~\citet{Sloan:2014fj}, (g)~\citet{Otsuka:2015aa}.
}

\end{landscape}
%} % End longtab
%*************************************************************************************************************************
%*************************************************************************************************************************

%*************************************************************************************************************************
%*************************************************************************************************************************
%\longtab{
\begin{landscape}
\scriptsize
\begin{longtable}{llccclccrr}
\caption{The Galactic sample: names, coordinates, reference position, classification of spectra (C-AGB: carbon-rich AGB star, C-pAGB: carbon-rich post-AGB star, C-PN: carbon-rich planetary nebula), period with reference, program identifier, and AOR key.}
\label{app_tab:gal_basic_info}\\
\hline
\hline
                                                &                               &       \multicolumn{2}{c}{(J2000.0)}   \\\cmidrule{3-4}
Name                                    &       Other name      &       RA                         &       Dec                     &       Position        &               Class                           &       Period  &Period         &Program ID      & AOR key\\
                                                &                               &       (deg)                   &         (deg)                   &       reference       &                                                       &       (days)  &reference      &                       &\\
\hline
\endfirsthead
\caption{Continued.} \\
\hline
                                                &                               &       \multicolumn{2}{c}{(J2000.0)}   \\\cmidrule{3-4}
Name                                    &       Other name      &       RA                         &       Dec                     &       Position        &               Class                           &       Period  &Period         &Program ID      & AOR key\\
                                                &                               &       (deg)                   &         (deg)                   &       reference       &                                                       &       (days)  &reference      &                       &\\
\hline
\endhead
\hline
\endfoot
\hline
\endlastfoot
CSS~1                                   &                                       &       001.609         &       +68.992         &       2MASS   &       C-AGB, S-type star\tablefootmark{i}                                                                            &       \ldots  &       \ldots  &       30737   &       18917632        \\
\textbf{IRAS~01005+7910}        &                                       &       016.190         &       +79.446         &       2MASS   &       C-pAGB, gr. I, 16-24\,$\mu$m feat.\tablefootmark{f}, fullerenes\tablefootmark{h}        &       \ldots  &       \ldots  &       30036   &       17505792        \\
IRAS~04188+0122                 &                                       &       065.364         &       +01.487         &       2MASS   &       C-AGB                                                                                                                   &       226             &       (4)             &       50835   &       29037824        \\
\textbf{IRAS~04296+3429}        &                                       &       068.237         &       +34.603         &       2MASS   &       C-pAGB, gr. III, 21\,$\mu$m\tablefootmark{a}                                                            &       71              &       (1)             &       45              &       4116736 \\
\textbf{IRAS~05113+1347}        &                                       &       078.532         &       +13.841         &       2MASS   &       C-pAGB, gr. II, 21\,$\mu$m\tablefootmark{b}                                                             &       133             &       (1)             &       93, 20208   &       4900352, 14399232       \\
\end{longtable}

\tablebib{(1)~\citet{Hrivnak:2010aa}, (2)~\citet{Smolders:2012aa}, (3)~\citet{Mauron:2014aa}, (4)~\citet{Samus:2017aa}.\\
(a)~\citet{Kwok:1989aa}, (b)~\citet{Hrivnak:2009aa}, (c)~\citet{Bernard-Salas:2009aa}, (d)~\citet{Cami:2010aa}, (e)~\citet{Garcia-Hernandez:2010aa}, (f)~\citet{Zhang:2010aa}, (g)~\citet{Cerrigone:2011aa}, (h)~\citet{Zhang:2011aa}, (i)~\citet{Smolders:2012aa}, (j)~\citet{Garcia-Hernandez:2012qy}, (k)~\citet{Otsuka:2013aa}, (l)~\citet{Otsuka:2014eu}, (m)~\citet{Otsuka:2015aa}, (n)~\citet{Raman:2017aa}.
}

\end{landscape}

%}% End longtab
%*************************************************************************************************************************
%*************************************************************************************************************************

%*************************************************************************************************************************
%*************************************************************************************************************************
%*************************************************************************************************************************
%*************************************************************************************************************************
%*************************************************************************************************************************
%*************************************************************************************************************************

%*************************************************************************************************************************
%*************************************************************************************************************************
\begin{landscape}
\begin{table*}
\scriptsize
\caption{Spectroscopic results for the SMC objects: names, six colours, dust temperature (T$_{\rm d}$), central wavelength ($\lambda_{\rm c}$), and strength of the 30\,$\mu$m feature (F/Cont).}
\label{app_tab:smc_spectral_results}
\centering
\begin{tabular}{lccccccccc} % 9 columns
\hline
\hline
Target                                  &                       [5.8]$-$[9.3]           &               [6.4]$-$[9.3]                   &               [16.5]$-$[21.5]                         &               [18.4]$-$[22.45]                        &               [18.4]$-$[22.75]                        &       [17.95]$-$[23.2]                &                       T$_{\rm d}$     &       $\lambda_{\rm c}$ (30\,$\mu$m)& F/Cont (30\,$\mu$m)\\
                                                &                       (mag)                   &               (mag)                           &               (mag)                           &                 (mag)                           &               (mag)                           &       (mag)                   &                       (K)                     &       ($\mu$m)                                        &                                       \\
\hline          
\textbf{SMP~SMC~1}              &       2.407   $\pm$   0.028   &                       \ldots                  &       1.078   $\pm$   0.021   &                       \ldots                  &                       \ldots                  &                       \ldots                  &                       \ldots          &                       \ldots                  &                       \ldots  \\
\textbf{IRAS~00350-7436}                &       0.954   $\pm$   0.004   &                       \ldots                  &                       \ldots                  &       0.514   $\pm$   0.005   &                       \ldots                  &                       \ldots                  &                       \ldots          &                       \ldots                  &                       \ldots  \\
\textbf{SMP~SMC~6}              &       2.513   $\pm$   0.029   &                       \ldots                  &                       \ldots                  &                       \ldots                  &                       \ldots                  &                       \ldots                  &                       \ldots          &                       \ldots                  &                       \ldots  \\
\textbf{MSX~SMC~054}            &                       \ldots                  &       0.756   $\pm$   0.004   &       0.101   $\pm$   0.016   &                       \ldots                  &                       \ldots                  &                       \ldots                  &                       \ldots          &                       \ldots                  &                       \ldots  \\
LK~229                                  &       1.622   $\pm$   0.017   &                       \ldots                  &       0.993   $\pm$   0.022   &                       \ldots                  &                       \ldots                  &                       \ldots                  &       173             $\pm$   3       &       30.291  $\pm$   0.062   &       0.607   $\pm$   0.018   \\
J004441.05-732136.44            &       2.754   $\pm$   0.014   &                       \ldots                  &                       \ldots                  &                       \ldots                  &                       \ldots                  &       0.803   $\pm$   0.009   &       186             $\pm$   2       &       28.617  $\pm$   0.158   &       0.171   $\pm$   0.012   \\
MSX~SMC~105                     &                       \ldots                  &       0.866   $\pm$   0.003   &       0.226   $\pm$   0.022   &                       \ldots                  &                       \ldots                  &                       \ldots                  &       604             $\pm$   48      &       28.971  $\pm$   0.327   &       0.254   $\pm$   0.029   \\
$[$GB98$]$~S06                  &                       \ldots                  &       0.969   $\pm$   0.003   &       0.284   $\pm$   0.020   &                       \ldots                  &                       \ldots                  &                       \ldots                  &       493             $\pm$   28      &       29.160  $\pm$   0.128   &       0.358   $\pm$   0.025   \\
IRAS~F00471-7352                        &                       \ldots                  &       0.701   $\pm$   0.008   &       0.240   $\pm$   0.009   &                       \ldots                  &                       \ldots                  &                       \ldots                  &       571             $\pm$   18      &       29.124  $\pm$   0.234   &       0.120   $\pm$   0.011   \\
\textbf{SMP~SMC~13}             &       2.334   $\pm$   0.041   &                       \ldots                  &                       \ldots                  &                       \ldots                  &                       \ldots                  &                       \ldots                  &                       \ldots          &                       \ldots                  &                       \ldots  \\
MSX~SMC~163                     &                       \ldots                  &       0.705   $\pm$   0.004   &       0.198   $\pm$   0.010   &                       \ldots                  &                       \ldots                  &                       \ldots                  &       679             $\pm$   31      &       30.167  $\pm$   0.345   &       0.143   $\pm$   0.014   \\
\textbf{SMP~SMC~15}             &       2.504   $\pm$   0.033   &                       \ldots                  &                       \ldots                  &                       \ldots                  &                       \ldots                  &                       \ldots                  &                       \ldots          &                       \ldots                  &                       \ldots  \\
SMP~SMC~16                              &       1.662   $\pm$   0.040   &                       \ldots                  &       1.037   $\pm$   0.018   &                       \ldots                  &                       \ldots                  &                       \ldots                  &       167             $\pm$   2       &       30.545  $\pm$   0.135   &       0.251   $\pm$   0.014   \\
SMP~SMC~17                              &       2.346   $\pm$   0.058   &                       \ldots                  &                       \ldots                  &                       \ldots                  &                       \ldots                  &                       \ldots                  &       160             $\pm$   3       &       31.564  $\pm$   0.166   &       0.101   $\pm$   0.017   \\
\textbf{SMP~SMC~18}             &       2.319   $\pm$   0.025   &                       \ldots                  &       1.000   $\pm$   0.016   &                       \ldots                  &                       \ldots                  &                       \ldots                  &                       \ldots          &                       \ldots                  &                       \ldots  \\
MSX~SMC~159                     &                       \ldots                  &       0.878   $\pm$   0.005   &       0.289   $\pm$   0.012   &                       \ldots                  &                       \ldots                  &                       \ldots                  &       485             $\pm$   17      &       29.193  $\pm$   0.185   &       0.502   $\pm$   0.017   \\
\textbf{SMP~SMC~20}             &       2.605   $\pm$   0.039   &                       \ldots                  &       0.671   $\pm$   0.009   &                       \ldots                  &                       \ldots                  &                       \ldots                  &                       \ldots          &                       \ldots                  &                       \ldots  \\
IRAS~00554-7351                 &                       \ldots                  &       0.906   $\pm$   0.003   &       0.270   $\pm$   0.011   &                       \ldots                  &                       \ldots                  &                       \ldots                  &       515             $\pm$   18      &       28.670  $\pm$   0.080   &       0.266   $\pm$   0.014   \\
SMP~SMC~24                              &       1.876   $\pm$   0.022   &                       \ldots                  &       0.860   $\pm$   0.024   &                       \ldots                  &                       \ldots                  &                       \ldots                  &       195             $\pm$   4       &       30.805  $\pm$   0.158   &       0.523   $\pm$   0.020   \\
2MASS~J01054645-7147053 &       3.524   $\pm$   0.095   &                       \ldots                  &                       \ldots                  &                       \ldots                  &       0.434   $\pm$   0.004   &                       \ldots                  &       262             $\pm$   3       &       28.971  $\pm$   0.274   &       0.121   $\pm$   0.010   \\
NGC~419~MIR~1                   &                       \ldots                  &       1.358   $\pm$   0.011   &       0.326   $\pm$   0.005   &                       \ldots                  &                       \ldots                  &                       \ldots                  &       437             $\pm$   6       &       29.262  $\pm$   0.046   &       0.424   $\pm$   0.006   \\
SMP~SMC~27                              &       2.963   $\pm$   0.183   &                       \ldots                  &       0.999   $\pm$   0.019   &                       \ldots                  &                       \ldots                  &                       \ldots                  &       172             $\pm$   3       &       31.014  $\pm$   0.229   &       0.202   $\pm$   0.015   \\               
\hline
\end{tabular}
\end{table*}
\end{landscape}
%
%*************************************************************************************************************************
%*************************************************************************************************************************

%*************************************************************************************************************************
%*************************************************************************************************************************
%\longtab{
\begin{landscape}
\scriptsize
\begin{longtable}{lccccccccc}
\caption{Spectroscopic results for the LMC objects: names, six colours, dust temperature (T$_{\rm d}$), central wavelength ($\lambda_{\rm c}$), and strength of the 30\,$\mu$m feature (F/Cont).}
\label{app_tab:lmc_spectral_results}\\
\hline
\hline
Target                                  &                       [5.8]$-$[9.3]           &               [6.4]$-$[9.3]                   &               [16.5]$-$[21.5]                         &               [18.4]$-$[22.45]                        &               [18.4]$-$[22.75]                        &               [17.95]$-$[23.2]                        &                               T$_{\rm d}$     &       $\lambda_{\rm c}$ (30\,$\mu$m)& F/Cont (30\,$\mu$m)\\
                                                &                       (mag)                   &               (mag)                           &               (mag)                           &                 (mag)                           &               (mag)                           &               (mag)                           &                               (K)                     &       ($\mu$m)                                        &\\
\hline
\endfirsthead
\caption{Continued.} \\
\hline
Target                                  &                       [5.8]$-$[9.3]           &               [6.4]$-$[9.3]                   &               [16.5]$-$[21.5]                         &               [18.4]$-$[22.45]                        &               [18.4]$-$[22.75]                        &               [17.95]$-$[23.2]                        &                               T$_{\rm d}$     &       $\lambda_{\rm c}$ (30\,$\mu$m)& F/Cont (30\,$\mu$m)\\
                                                &                       (mag)                   &               (mag)                           &               (mag)                           &                 (mag)                           &               (mag)                           &               (mag)                           &                               (K)                     &       ($\mu$m)                                        &\\
\hline
\endhead
\hline
\endfoot
\hline
\endlastfoot
\textbf{IRAS~04286-6937}                &                       \ldots                  &       0.931   $\pm$   0.004   &       0.424   $\pm$   0.008   &                       \ldots                  &                       \ldots                  &                       \ldots                  &                               \ldots          &                       \ldots                  &                       \ldots  \\
IRAS~F04331-6932                        &                       \ldots                  &       1.153   $\pm$   0.008   &       0.376   $\pm$   0.007   &                       \ldots                  &                       \ldots                  &                       \ldots                  &               387             $\pm$   6       &       28.586  $\pm$   0.038   &       0.577   $\pm$   0.009   \\
IRAS~F04340-7016                 &                       \ldots                  &       0.608   $\pm$   0.008   &       0.209   $\pm$   0.019   &                       \ldots                  &                       \ldots                  &                       \ldots                  &               647             $\pm$   48      &       29.164  $\pm$   0.420   &       0.262   $\pm$   0.025   \\
IRAS~F04353-6702                        &                       \ldots                  &       1.052   $\pm$   0.006   &       0.347   $\pm$   0.007   &                       \ldots                  &                       \ldots                  &                       \ldots                  &               414             $\pm$   7       &       29.293  $\pm$   0.091   &       0.457   $\pm$   0.009   \\
\end{longtable}

\end{landscape}

%}% End onllongtab
%*************************************************************************************************************************
%*************************************************************************************************************************

%*************************************************************************************************************************
%*************************************************************************************************************************
%\longtab{
\begin{landscape}
\scriptsize
\begin{longtable}{lccccccccc}
\caption{Spectroscopic results for the Galactic objects: names, six colours, dust temperature (T$_{\rm d}$), central wavelength ($\lambda_{\rm c}$), and strength of the 30\,$\mu$m feature (F/Cont).}
\label{app_tab:gal_spectral_results}\\
\hline
\hline
Target                          &                       [5.8]$-$[9.3]           &               [6.4]$-$[9.3]                   &               [16.5]$-$[21.5]                         &               [18.4]$-$[22.45]                        &               [18.4]$-$[22.75]                        &               [17.95]$-$[23.2]                        &                               T$_{\rm d}$     &       $\lambda_{\rm c}$ (30\,$\mu$m)& F/Cont (30\,$\mu$m)\\
                                        &                       (mag)                   &               (mag)                           &               (mag)                           &                 (mag)                           &               (mag)                           &               (mag)                           &                               (K)                     &       ($\mu$m)                                        &                                       \\
\hline
\endfirsthead
\caption{Continued.} \\
\hline
Target                          &                       [5.8]$-$[9.3]           &               [6.4]$-$[9.3]                   &               [16.5]$-$[21.5]                         &               [18.4]$-$[22.45]                        &               [18.4]$-$[22.75]                        &               [17.95]$-$[23.2]                        &                               T$_{\rm d}$     &       $\lambda_{\rm c}$ (30\,$\mu$m)& F/Cont (30\,$\mu$m)\\
                                        &                       (mag)                   &               (mag)                           &               (mag)                           &                 (mag)                           &               (mag)                           &               (mag)                           &                               (K)                     &       ($\mu$m)                                        &\\
\hline
\endhead
\hline
\endfoot
\hline
\endlastfoot
CSS~1                           &                       \ldots                  &       0.185   $\pm$   0.011   &       0.211   $\pm$   0.015   &                       \ldots                  &                       \ldots                  &                       \ldots                  &               640             $\pm$   39      &       28.736  $\pm$   0.178   &       0.303   $\pm$   0.018   \\              
\textbf{IRAS~01005+7910}&       2.362   $\pm$   0.011   &                       \ldots                  &                       \ldots                  &                       \ldots                  &                       \ldots                  &                       \ldots                  &                               \ldots          &                       \ldots                  &                       \ldots                  \\
IRAS~04188+0122         &                       \ldots                  &       0.365   $\pm$   0.010   &       0.204   $\pm$   0.009   &                       \ldots                  &                       \ldots                  &                       \ldots                  &               660             $\pm$   26      &       29.974  $\pm$   0.261   &       0.102   $\pm$   0.012   \\
\textbf{IRAS~04296+3429}&       2.882   $\pm$   0.021   &                       \ldots                  &                       \ldots                  &                       \ldots                  &                       \ldots                  &       1.288   $\pm$   0.005   &                               \ldots          &                       \ldots                  &                       \ldots                  \\
\end{longtable}

\end{landscape}

%}% End onllongtab
%*************************************************************************************************************************
%*************************************************************************************************************************

\end{appendix}

\end{document}